\documentclass[prx,onecolumn,nofootinbib,citeautoscript,eqsecnum,10pt,notitlepage]{revtex4-1}
\pdfoutput=1

\usepackage{dsfont}
\usepackage{float}
\usepackage{array}
\usepackage{slashed,bbold}
\usepackage{physics}

\usepackage{amsmath,amssymb,bm} 
\usepackage{graphicx}

\usepackage[tight]{subfigure} 

\usepackage{color} 
\usepackage[papersize={8.5in,11in}]{geometry}

\usepackage{color}
\definecolor{darkblue}{rgb}{0.,0.,0.4}
\definecolor{darkred}{rgb}{0.5,0.,0.}
\definecolor{BlueViolet}{RGB}{138,43,226}
\definecolor{SkyBlue}{RGB}{30,144,255}
\definecolor{DarkGreen}{RGB}{0,100,0}
\usepackage[pdftex,colorlinks=true,linkcolor=darkblue,citecolor=blue,urlcolor=darkred]{hyperref}

\def \be{\begin{equation}}
\def \ee{\end{equation}}
\def \nn{\nonumber \\}
\begin{document}

\title{Tunneling of multi-Weyl semimetals through a potential barrier under the influence of magnetic fields}

\author{Ipsita Mandal}
\affiliation{Faculty of Science and Technology, University of Stavanger, 4036 Stavanger, Norway\\
Institute of Nuclear Physics, Polish Academy of Sciences, 31-342 Krak\'{o}w, Poland}

\author{Aritra Sen}
\affiliation{Indian Institute of Technology, Gandhinagar, Gujarat 382355, India}

\begin{abstract}
{
We investigate the tunneling of the quasiparticles arising in multi-Weyl semimetals through a barrier consisting of both electrostatic and vector potentials, existing uniformly in a finite region along the transmission axis. The dispersion of a multi-Weyl semimetal is linear in one direction (say, $k_z$), and proportional to $k_\perp^J$ in the plane perpendicular to it (where $k_\perp =\sqrt{k_x^2+k_y^2}$). Hence, we study the cases when the barrier is perpendicular to $k_z$ and $k_x$, respectively. For comparison, we also state the corresponding results for the Weyl semimetal.
}
\end{abstract}
\maketitle

\tableofcontents

%%%%%%%%%%%%%%%%%%%%%%%%%%%%%%%%%%%%%%%%%%%%%%%%%

\section{Introduction}

\begin{figure}[]
\subfigure[]{\includegraphics[width = 0.15 \textwidth]{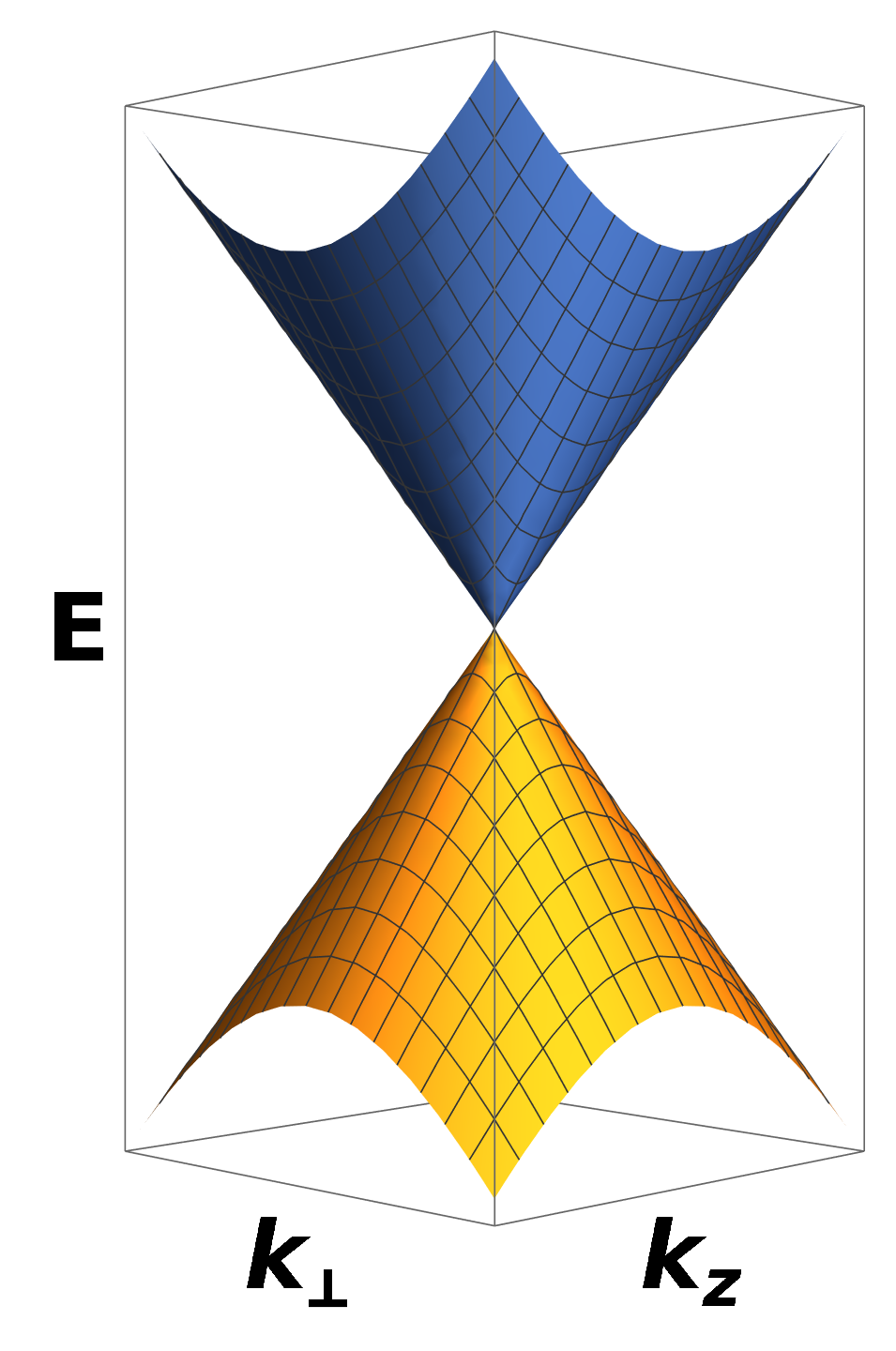}} \hspace{2 cm}
\subfigure[]{\includegraphics[width = 0.15 \textwidth]{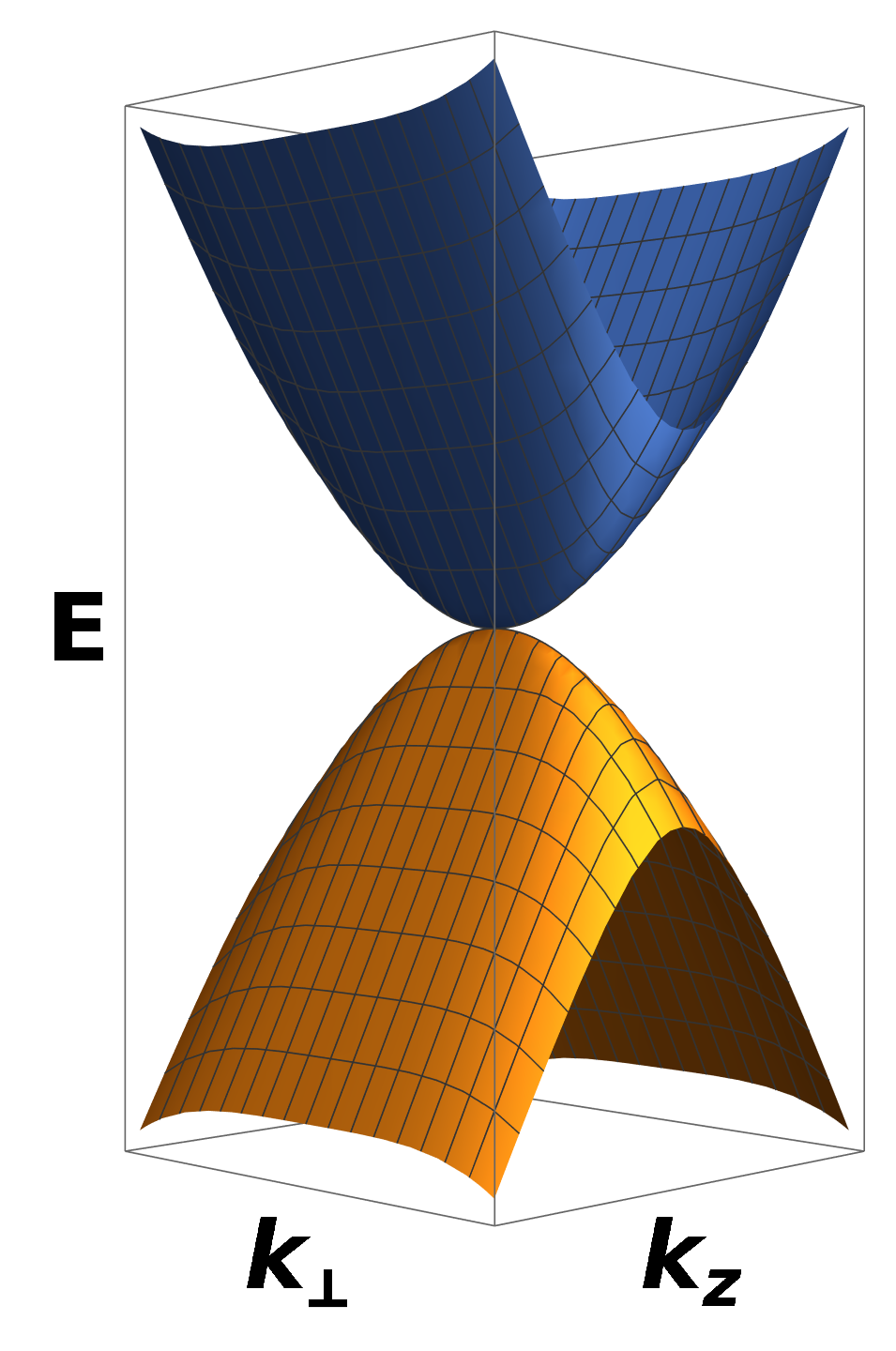}}\hspace{2 cm}
\subfigure[]{\includegraphics[width = 0.15 \textwidth]{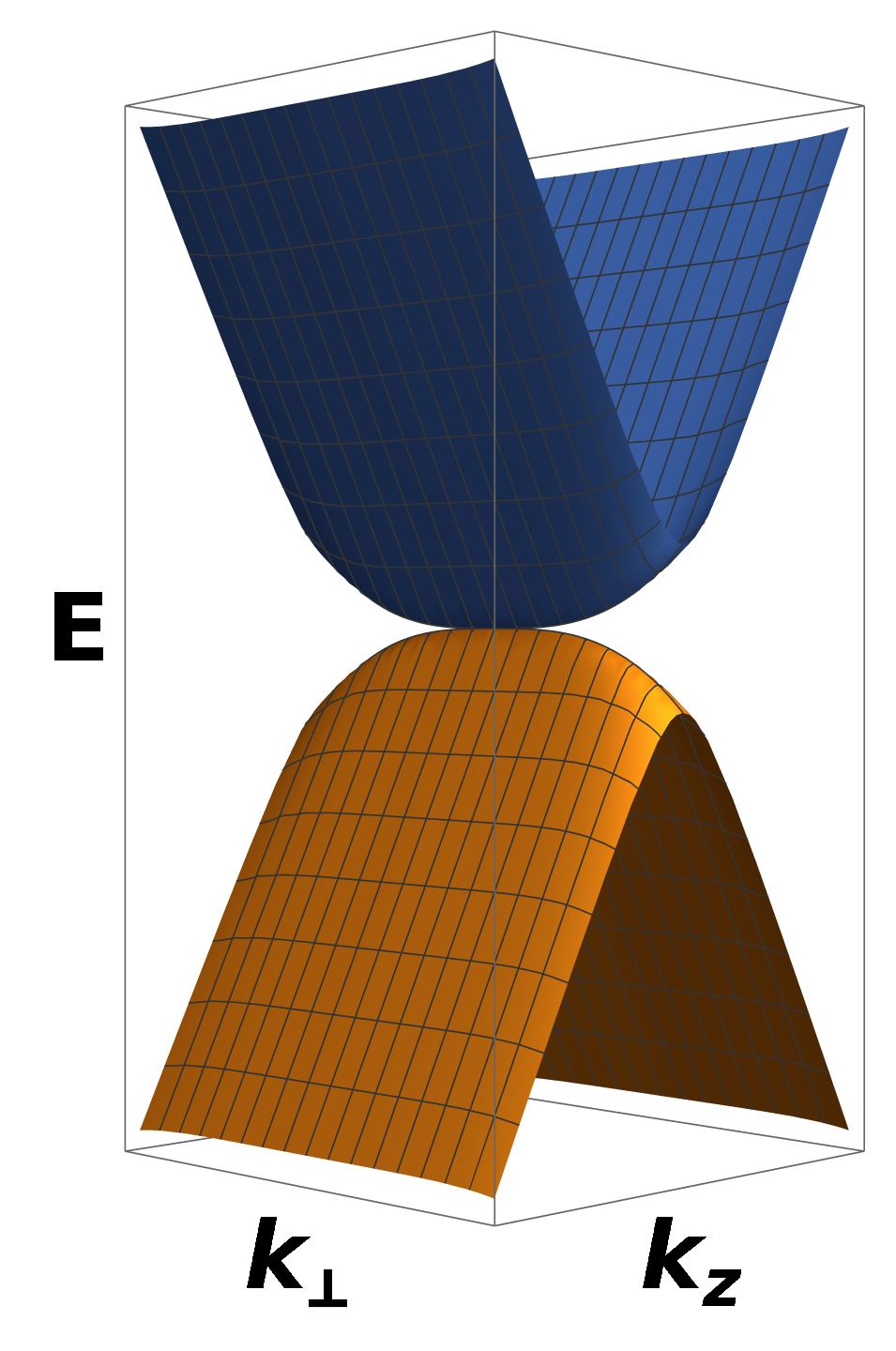}}
\subfigure[]{\includegraphics[width = 0.65 \textwidth]{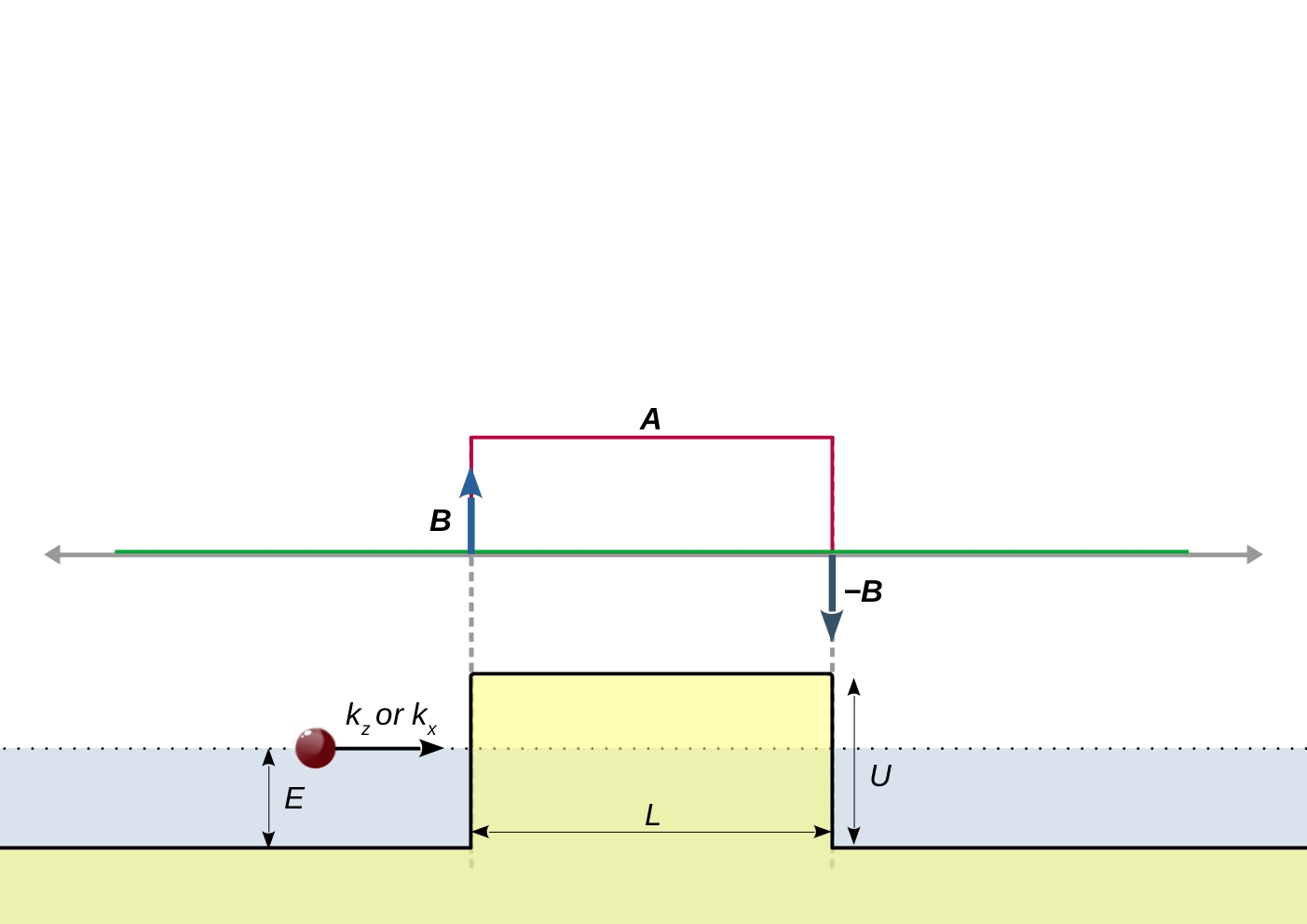}} 
\caption{Figures (a), (b), and (c) show the energy dispersion relations of Weyl and multi-Weyl semimetals with $J=1, 2,$ and $3,$ respectively.
(d) Tunneling takes place through a scalar (or electric) potential barrier of strength $V_0$, with a constant vector potential $\mathbf{A}$ superposed
in the same region. Theoretically, this vector potential can be created by applying equal and opposite delta function magnetic fields ($\mathbf{B}$ and $-\mathbf{B}$) at the edges of the barrier region, oriented perpendicular to the axis of propagation.
The lower panel represents the schematic diagram of the transport of a quasiparticle (red ball) across the potential barrier. The Fermi level is depicted by dotted lines, and lies in the conduction band outside the barrier, and in the valence band inside it. (For interpretation of the colors in the figure(s), the reader is referred to the web version of this
article.)}
\label{figbands}
\end{figure}

Recently, there has been a surge of interest in gapless topological phases that arise in multi-band crossings \cite{bernavig,bernavig2} in the Brillouin zone (BZ)
such that the bandstructures have nonzero Chern numbers. Some of these have a high-energy counterpart (e.g. Weyl semimetals), and some do not (e.g. double-Weyl and triple-Weyl semimetals). In Weyl ($J=1$) semimetals, two linearly dispersing bands in three-dimensional (3d) momentum space intersect at a point, which acts as a monopole of Berry curvature in
momentum space. A pair of such points exist which have opposite Chern numbers ($\pm 1$) and behave as the sink and source of Berry flux in momentum space. The projected images of these points are connected by topologically protected gapless Fermi arcs
as the zero-energy surface states that can be experimentally observed \cite{Inoue_2016} in Fourier-transformed scanning tunneling microscopy (STM). Analogously, double-Weyl ($J=2$) and triple-Weyl ($J=3$) semimetals have a pair of band-crossing points
\footnote{According to the  Nielsen–Ninomiya theorem, Weyl and multi-Weyl nodes always appear in pairs \cite{nielsen-ninomiya}, which are usually referred to as the valley degrees of freedom.} in 3d where the Chern numbers are $\pm 2$ and $\pm 3$, respectively \cite{bernavig2}. Consequently, the nodal points in the former and the later are connected by two and three Fermi arcs respectively.
Also important to note is the fact the dispersions in these $J>1$ semimetals
are anisotropic. The dispersion of a multi-Weyl semimetal is linear in one direction (say, $k_z$), and proportional to $k_\perp^J$ in the plane perpendicular to it (where $k_\perp =\sqrt{k_x^2+k_y^2}$). The various scenarios are depicted schematically in Fig.~\ref{figbands}(a), (b), and (c). Rotational symmetries limit the multi-Weyl systems to $J\leq 3$ in crystalline systems \cite{bernavig2}.

These exotic gapless topological band-crossings have been predicted to exist in various experimentally feasible candidate materials, based on first principles band-structure calculations and density functional theory computations. For example, Weyl semimetals have been observed in the TaAs family \cite{Huang_2015,*Xu_2015} and SrSi$_2$ \cite{Huang1180}, double-Weyl quasiparticles are expected to exist in HgCr$_2$Se$_4$ \cite{bernavig2,PhysRevLett.107.186806}, SrSi$_2$ \cite{Huang1180}, and
superconducting states of $^3$He-A \cite{volovik}, UPt$_3$ \cite{PhysRevB.92.214504}, SrPtAs \cite{PhysRevB.89.020509}, and YPtBi \cite{PhysRevB.99.054505}.
Similarly, molybdenum monochalcogenide compounds A(MoX)$_3$ (where A $ =$Na, K, Rb, In, Tl; X = S, Se, Te) are predicted \cite{PhysRevX.7.021019} to harbour triple-Weyl quasiparticles.

In this paper, we study the behavior of the transmission coefficients of the multi-Weyl semimetals through a finite rectangular potential barrier subjected to a uniform vector potential (within the barrier region) in a direction perpendicular to the direction of propagation. We try to identify the distinct features peculiar to the $J$ value.
This is shown pictorially in Fig.~\ref{figbands}(d).
%%%%%%%5
The required vector potential can be generated in real experiments
\cite{PhysRevLett.72.1518,Zhai_2008,Ramezani_Masir_2009} by placing a ferromagnetic metal strip of width $L$, deposited on the top of a thin dielectric layer placed above the semimetal, and with a magnetization parallel (or anti-parallel) to the propagation direction. The resulting fringe fields thus provide a magnetic field modulation along the current, which is assumed to be homogeneous in the perpendicular plane.
%https://arxiv.org/pdf/0805.1105.pdf
%%%%%%%%%%%%%%5
 This set-up might prove to be a tool to identify/distinguish these materials in experiments. Earlier theoretical studies \cite{ips-tunnel-linear} have investigated such a scenario for pseudospin-1 (also called Maxwell fermions \cite{PhysRevA.96.033634}) and pseudospin-3/2 (also called Rarita-Schwinger-Weyl fermions \cite{long}) quasiparticles.
 Ref.~\cite{Zhu2020,Deng2020} have computed the barrier tunneling features of the multi-Weyl quasiparticles in the absence of magnetic fields.

The paper is organized as follows. In Sec.~\ref{secmodel}, we explain the Hamiltonians of the systems under consideration, and the general set-up for carrying out the tunneling experiments. In Sec.~\ref{secperpkz} and \ref{secperpkx}, we apply the Landau-B\"{u}ttiker formalism to compute the tunneling coefficients for the cases when the propagation directions are parallel and perpendicular to the linear dispersion direction, respectively.
Finally, we end with a summary and outlook in Sec.~\ref{secsum}.

%%%%%%%%%%%%%%%%%%%%%%%%%%%%%%%%%
%%%%%%%%%%%%%%%%%%%%%%%%

%%%%%%%%%%%%%%%%%%%%%%%%%%%%%%%%%%%%%
\section{Formalism}
\label{secmodel}

The Weyl semimetal ($J=1$) Hamiltonian at the node with positive chirality is given by:
\begin{equation}
\label{a1}
\mathcal{H}_1(\mathbf{k})=  v\, \mathbf{k} \cdot \boldsymbol{\sigma}\,,
\end{equation}
%%%%%%%%%%%%%%%5
where $v$ is the isotropic Fermi velocity, and
$\boldsymbol{\sigma}$ represents the vector of the Pauli matrices. 
The energy eigenvalues are given by:
\begin{equation}
\label{a2}
\mathcal{E}_1^\pm(\mathbf{k})= \pm  v\, k\,,
\end{equation}
%%%%%%%%%%%%%%%%5
where $\pm$ correspond to the conduction and valence bands respectively. A set of normalized eigenvectors corresponding to $\mathcal{E}_1^\pm(\mathbf k)$ are given by:
%%%%%%%%%%%%%%%%%%55
\begin{align}
\label{a3}
\psi_{1}^\pm(\mathbf{k})  
=\frac{1}{\mathcal{N}_1^\pm }
\begin{pmatrix}
\frac{v^{-1}\, \mathcal{E}^\pm_1+ k_z  }
{  k_x + \mathrm{i} \,k_y }
\\ 1
\end{pmatrix},
\end{align}
where $\mathcal{N}_1^\pm$ denotes the normalization factors.
%%%%%%%%%%%%%%%%%%%%%%%%%%%%

The multi-Weyl systems are generalizations of the Weyl Hamiltonian to nodes having higher topological charges.
The effective continuum Hamiltonian for an isolated multi-Weyl
node of chirality $\chi =\pm 1$ and topological charge $J$ is given by
\cite{PhysRevB.95.201102,PhysRevB.97.045150}:
\begin{align}
\mathcal{H}_J(\mathbf{k})= \frac{ v_{\perp}
\left [  \left (k_x - \mathrm{i}\, k_y \right)^J 
\left( \sigma_x +\mathrm{i}\, \sigma_y \right)
+ \left (k_x + \mathrm{i}\, k_y \right)^J 
\left( \sigma_x -\mathrm{i}\, \sigma_y \right) \right ]}
{2\,k_0^{J-1}}    
+ \chi\, v_z\,  k_z\, \sigma_z \,,
\end{align}
where $v_z$ and $v_{\perp}$ are the Fermi velocities in the $z$ direction and $xy$-plane respectively, and $k_0$ is a system dependent parameter with the dimension of momentum.
$\mathcal{H}_1(\mathbf{k})$ can be obtained from $\mathcal{H}_J(\mathbf{k})$ by setting $v_\perp=v_z =v $.
For the sake of completeness, the explicit forms are:
\begin{align}
&    \mathcal{H}_2(\mathbf{k})= \frac{ v_{\perp} 
\left [  \left (k_y^2-k_x^2 \right)\sigma_x+ 2\, k_x\, k_y\, \sigma_y \right ]}
{k_0}    
+ \chi\, v_z\,  k_z\, \sigma_z \,,\nn
%%%%%%
&
 \mathcal{H}_3(\mathbf{k})=
\frac{ v_{\perp}\,  \left [ \left (k_x^3\, \sigma_x - k_y^3\, \sigma_y \right)
+ 3 \left (  k_x\, \sigma_y- k_y\, \sigma_x \right ) k_x\, k_y \right ]}
{k_0^2}
+ \chi\, v_z\,   k_z\, \sigma_z \,,
\end{align}

Henceforth, we will focus on the $\chi = 1$ case.
The eigenvalues are given by:
%%%%%%%%%%%%
\begin{align}
\mathcal{E}^\pm_J(\mathbf{k})=\pm
\sqrt{  \frac{v_{\perp}^2\,k_\perp^{2J}} {k_0^{2J-2}} + v_z^2\, k_z^2}\,,
\text{ where } k_\perp =\sqrt{k_x^2 +k_y^2}\,,
\end{align}
with eigenvectors
\begin{align}
\label{psij}
\psi_{J}^\pm(\mathbf{k})
=\frac{1}{\mathcal{N}^\pm_J }
\begin{pmatrix}
\frac{k_0^{J-1} 
\left ( \mathcal{E}^\pm_J+ v_z \,k_z \right )}
{ v_\perp \left(k_x + \mathrm{i} \,k_y \right )^{J}
}
\\ 1
\end{pmatrix},
\end{align}
where $\mathcal{N}_J^\pm$ denotes the normalization factors.
The labels $\pm$ denote the conduction and valence bands respectively. 
In our computations, we will set $v, v_\perp, $ and $k_0$ to unity, and $v_z$ to $0.5$.
We will follow the usual Landau-B\"{u}ttiker procedure (see, for example Refs.~\cite{salehi,beenakker,ips-qbcp-tunnel,ips-tunnel-linear}) to compute the transport coefficients. We will consider the tunneling of quasiparticles in a slab of square cross-section, with a transverse width $W$. We assume that $W$ is large enough such that the specific boundary conditions being used in the calculations are irrelevant for the bulk response. In the following, we will impose periodic boundary conditions along these transverse directions.
Our transmission problem involves semimetals with the valence and conduction bands crossing at the nodal point, and we will deal with the case when the incident particles are electron-like excitations.
In other words, the Fermi energy ($E$) is adjusted to lie in the conduction band outside the potential barrier.

Note that the energy is expressed in units of $\hbar\, v_{\perp}\, k_0$ (where we set $\hbar=1$). Lengths and magnetic vector potentials are in units of $1/k_0$ and $\hbar\, k_0/e$ (again, we set $e=1$ ), respectively.

%%%%%%%%%%%%%%%%%%%%%%%%%%%%%%%%
\section{Barrier perpendicular to $k_z$}
\label{secperpkz}

First let us consider the case when the barrier is placed perpendicular to $z$ axis, such that
\begin{equation}\label{a4}
U(x,y,z)
=\begin{cases}U&0\leq z\leq L \\0 & \text{otherwise}
\end{cases}\,.
\end{equation}
Hence the momentum components $k_x$ and $k_y$ are conserved.
On imposing periodic boundary conditions along these directions, we get the corresponding momentum components quantized as:
\begin{align} 
k_x=\frac{2\,\pi\, n_x}{W}\,,\quad
k_y=\frac{2\,\pi \,n_y}{W}\,.
\end{align}

In the next step, we subject the sample to equal and opposite magnetic fields localized at the edges of the rectangular electric potential, and directed perpendicular to the $z$-axis \cite{mansoor,kai}. This can be theoretically modeled as Dirac delta functions of opposite signs at $z=0$ and $z=L$ respectively, and gives rise to a vector potential with the components:
\begin{align}
\mathbf {A}( z ) \equiv  \lbrace a_x, a_y,0  \rbrace = \begin{cases} 
\lbrace B_y, -B_x, 0  \rbrace &  \text{ for } 0 < z < L \\
\mathbf 0 & \text{ otherwise} \,.
\end{cases}
\label{eqvecpot}
\end{align}
Note that this arises from the magnetic field $\mathbf{B}  = 
\left(  B_x\, \hat{\mathbf i}  + B_y\, \hat{\mathbf j}   \right )
\left [  \delta\left(z\right) - \delta\left( z-L \right) \right ] $.
The vector potential modifies the transverse momenta as $k_x \rightarrow k_x-e\, a_x,$ and $ k_y\rightarrow k_y-e\, a_y$, such that the effective Hamiltonians within the barrier region are given by $\mathcal{H}_J(  k_x-e\, a_x, k_y -e\, a_y, k_z) + e\,U$.

The proposed experimental set-up is depicted schematically in Fig.~\ref{figbands}.
Some possible methods to achieve this set-up in real experiments (for example, by placing ferromagnetic stripes at barrier boundaries) have been discussed
in Ref.~\cite{mansoor}.

A scattering state $\Psi_{ J,\mathbf n}(z)$, in the mode labeled by $\mathbf n =\lbrace n_x, n_y \rbrace$, is constructed from the following states:
\begin{align}
 \Psi_{J, \mathbf n} (z)=&  \begin{cases}   \phi_{J,L} & \text{ for } z <0  \\
 \phi_{J,M} & \text{ for } 0< z < L \\
  \phi_{J,R} &  \text{ for } z > L 
\end{cases} \,,\nonumber \\
%%%%%%%%%%%%%%%%%%%%%%%%%%%%%%%%%%%%%%%%
  \phi_{J,L} = & \,\frac{   
 \psi_J^+ (k_x,k_y,k_\ell)\,e^{\mathrm{i}\, k_\ell\, z}
 +r_{J,{\mathbf n}} \, \psi_J^+ (k_x,k_y,-k_\ell)\,e^{-\mathrm{i}\, k_\ell\, z}}
{\sqrt{  { \mathcal{V} }_z(k_x, k_y,k_\ell ) }}\, ,\nonumber \\
%%%%%%%%%%%%%%%%%%%%%%%%%%%%%%%%%%%%%%%%%%%%5
  \phi_{J,M}  = &\,\Big[  
 \alpha_{J,\mathbf n} \,\psi_J^+ (\tilde k_x,\tilde k_y,\tilde{k}_z) \,
 e^{\mathrm{i}\,\tilde k_z  z } 
 + 
 \beta_{J,\mathbf n} \, \psi_J^+ (\tilde k_x,\tilde k_y,-\tilde{k}_z) \,
 e^{-\mathrm{i}\,\tilde k_z z } 
\Big]   \Theta\left( E-e\,U  \right)
 \nonumber\\ & 
 + \Big[
 \alpha_{J,{\mathbf n}} \,\psi_J^- (\tilde k_x,\tilde k_y,\tilde{k}_z) \,
 e^{\mathrm{i}\,\tilde k_z  z } 
 + 
 \beta_{J,{\mathbf n}} \,\psi_J^- (\tilde k_x,\tilde k_y,-\tilde{k}_z) \,
 e^{-\mathrm{i}\,\tilde k_z  z }
   \Big]  \,\Theta\left( e\, U-E  \right),\nonumber \\
 %%%%%%%%%%%%%%%%%%%%%%%%%%%%%%%%%%%%%%%%%%%%%%%%%%%%%%%%%%%55
 \phi_{J,R} = & \,\frac{  t_{J,{\mathbf n}} \,\psi_{J}^+ ( k_x,k_y,k_{\ell}) 
 } 
{\sqrt{  { \mathcal{V} }_z  (k_x,k_y,k_\ell)}}
\, e^{\mathrm{i}\, k_\ell \left( z -L\right)}\,,\quad
%%%%%%%%%%%%%%%%%%%%%%%%%%%%%%%%% 
k_\ell = k_z = \frac{\sqrt{E^2 -  \frac{v_{\perp}^2\,k_\perp^{2J}} {k_0^{2J-2}}}
} {v_z}\,, \quad
{ \mathcal{V} }_z  (k_x,k_y,k_\ell) =   
\big |\partial_{k_\ell}  \mathcal{E}_{J}^+ (k_x,k_y,k_\ell)\big|\,,\quad
\nn
\tilde k_x = & \, k_x - e\, a_x\,,\quad 
\tilde k_y = k_y - e\, a_y\,,\quad
\tilde k_z =\frac{
\sqrt{\left(E-e\,U \right) ^2 -  
\frac{v_{\perp}^2\left( \tilde k_x^2 + \tilde k_y^2 \right)^{J}} {k_0^{2J-2}}
}} {v_z}\,,
\end{align}
where we have used the velocity $ { \mathcal{V} }_z(k_x, k_y,k_\ell) $ to normalize the incident, reflected, and transmitted plane waves. The symbol $\Theta(u)$ represents the Heaviside step function, as usual. Note that for $J=1$, we have $v_z = v_\perp =v$, which is set to unity in the numerical results.
Here $r_{J,{\mathbf n}}$ and $t_{J,{\mathbf n}}$ are the amplitudes of the reflected and transmitted waves, respectively. Altogether, we have $4$ unknown parameters $(r_{J,{\mathbf n}}, \,t_{J,{\mathbf n}}, \,\alpha_{J,{\mathbf n}},\, \beta_{J,{\mathbf n}})$, and to solve for these, we need $4$ equations which are provided by the continuity of the two components of the wavefunction at $z=0$ and $z=L$.

\subsection{Transmission coefficients}
%%%%%%%%%%%%%%%%%%%%%%%%%%%

We show below the expressions for $t_{J,{\mathbf n}}$:
\begin{align}
\frac{2\,\tilde k_z \,k_z \,e^{ \mathrm{i}\,\tilde k_z L}} 
{t_{1,{\mathbf n}}}
=   \left (e^{ 2\,\mathrm{i}\,\tilde k_z L}- 1 \right )
\Big  [ E \left (e\,U-E \right )
+k_x \left(k_x-a_x\right) + k_y \left(k_y-a_y \right)  \Big ]
+ \tilde k_z \,k_z \left (e^{ 2\,\mathrm{i}\,\tilde k_z L} + 1 \right ),
\end{align}
\begin{align}
\frac{2\,\tilde k_z \,k_z\, e^{ \mathrm{i}\,\tilde k_z L}} {t_{2,{\mathbf n}}}
&=
4  \left (e^{ 2\,\mathrm{i}\,\tilde k_z L}- 1 \right )
 \Big[
 E \left  (e\,U-E \right)
+
2 \left \lbrace k_x \left(a_y+k_x-a_x \right)-k_y \left(a_x+a_y\right)
+k_y^2 \right \rbrace 
\nn & \hspace{ 3 cm} 
\left \lbrace
a_x \left(k_y-k_x\right)-a_y \left(k_x+k_y\right)
+k_x^2+k_y^2 \right \rbrace 
 \Big]
 + \left ( e^{2\,\mathrm{i}\,\tilde k_z L}+ 1\right ) \tilde k_z \,k_z \, ,
\end{align}
\begin{align}
\frac{ 2\,\tilde k_z\, k_z\, e^{ \mathrm{i}\,\tilde k_z L}} 
{t_{3,{\mathbf n}}}
& = 4 \left ( e^{2\,\mathrm{i}\,\tilde k_z L} -1 \right )
\Bigg  [  
 E\left( e\, U-E \right)
+
\left \lbrace k_x  \left( k_x - a_x \right)
+k_y \left(k_y - a_y \right)\right \rbrace  
\bigg[
k_y^2 \left(a_y^2+2 \,k_x^2 -2 \,a_x \,k_x -3\, a_x^2 \right)
\nn & \hspace{3 cm}
-2 \,a_y \,k_x\, k_y 
\left(k_x-4 \,a_x\right)
+k_x^2 \left \lbrace \left(k_x-a_x\right)^2-3 \,a_y^2 \right \rbrace
-2 a_y k_y^3+k_y^4
\bigg]
\Bigg ]
+\left ( e^{2\,\mathrm{i}\,\tilde k_z L}+ 1\right ) \tilde k_z \,k_z \,.
\end{align}

\begin{figure}[]
\subfigure[]{\includegraphics[width=.5\textwidth]{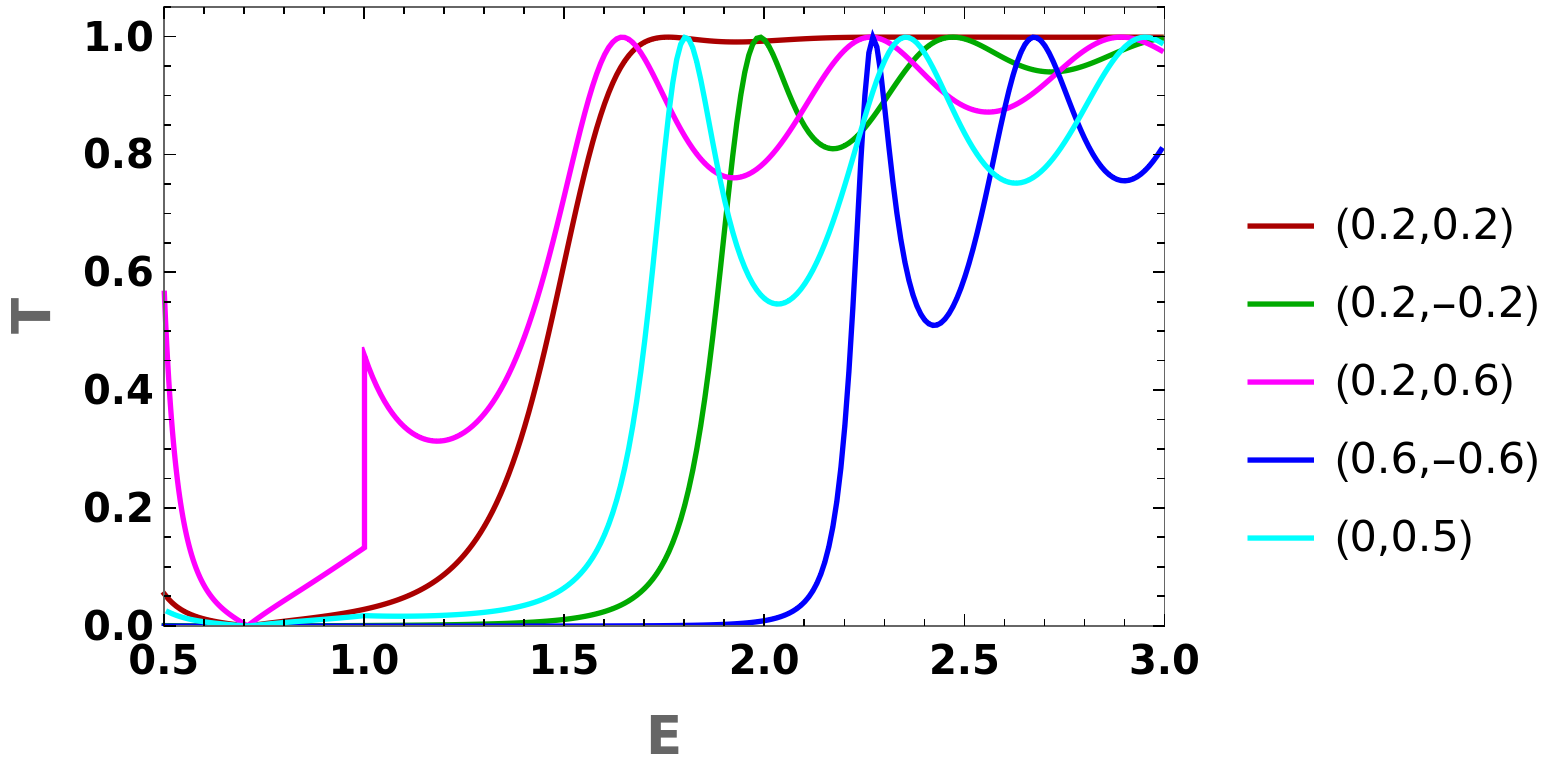}}\quad
\subfigure[]{\includegraphics[width=.46\textwidth]{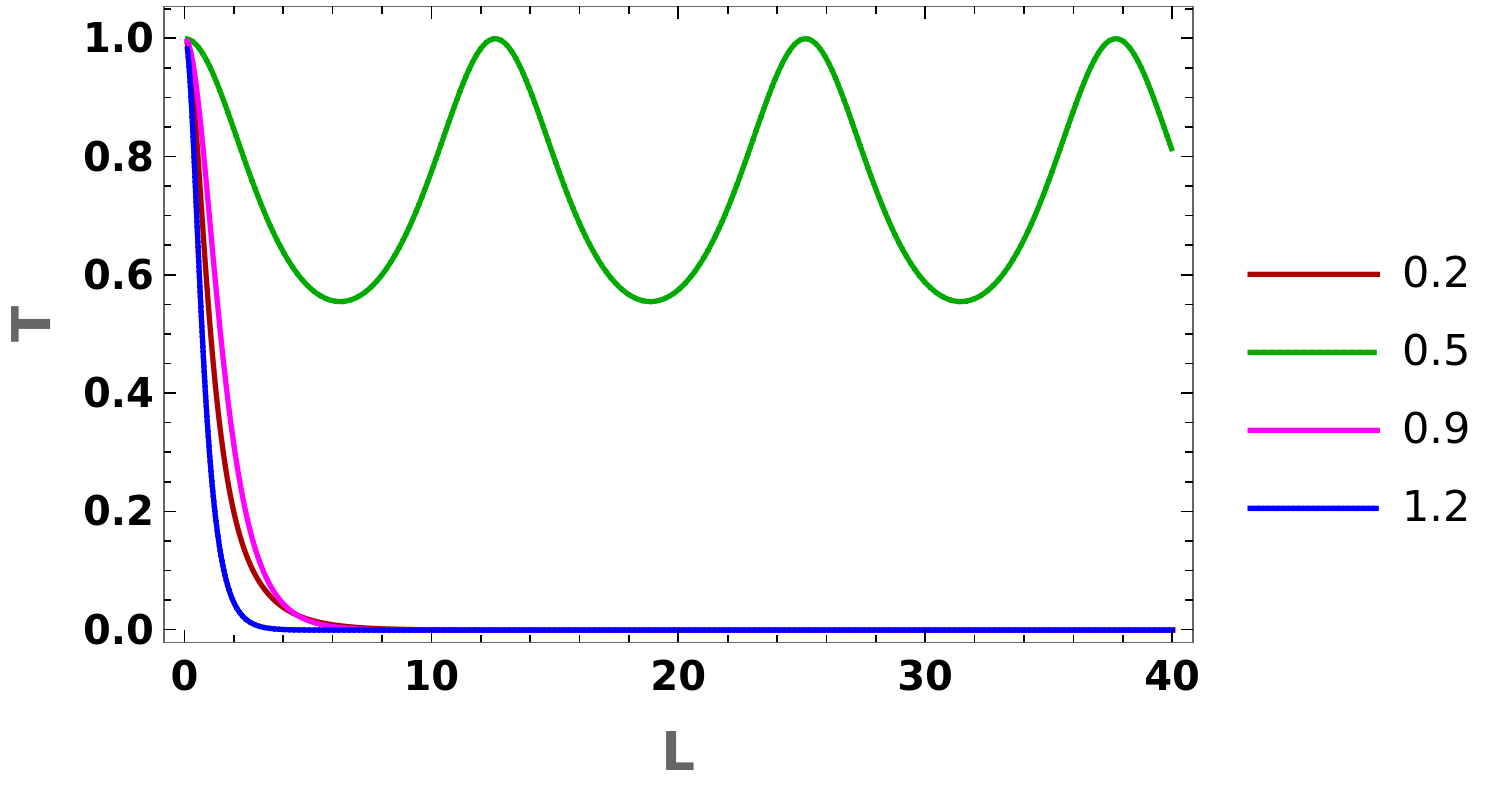}}
%\caption{$J=1$}
\subfigure[]{\includegraphics[width=.5\textwidth]{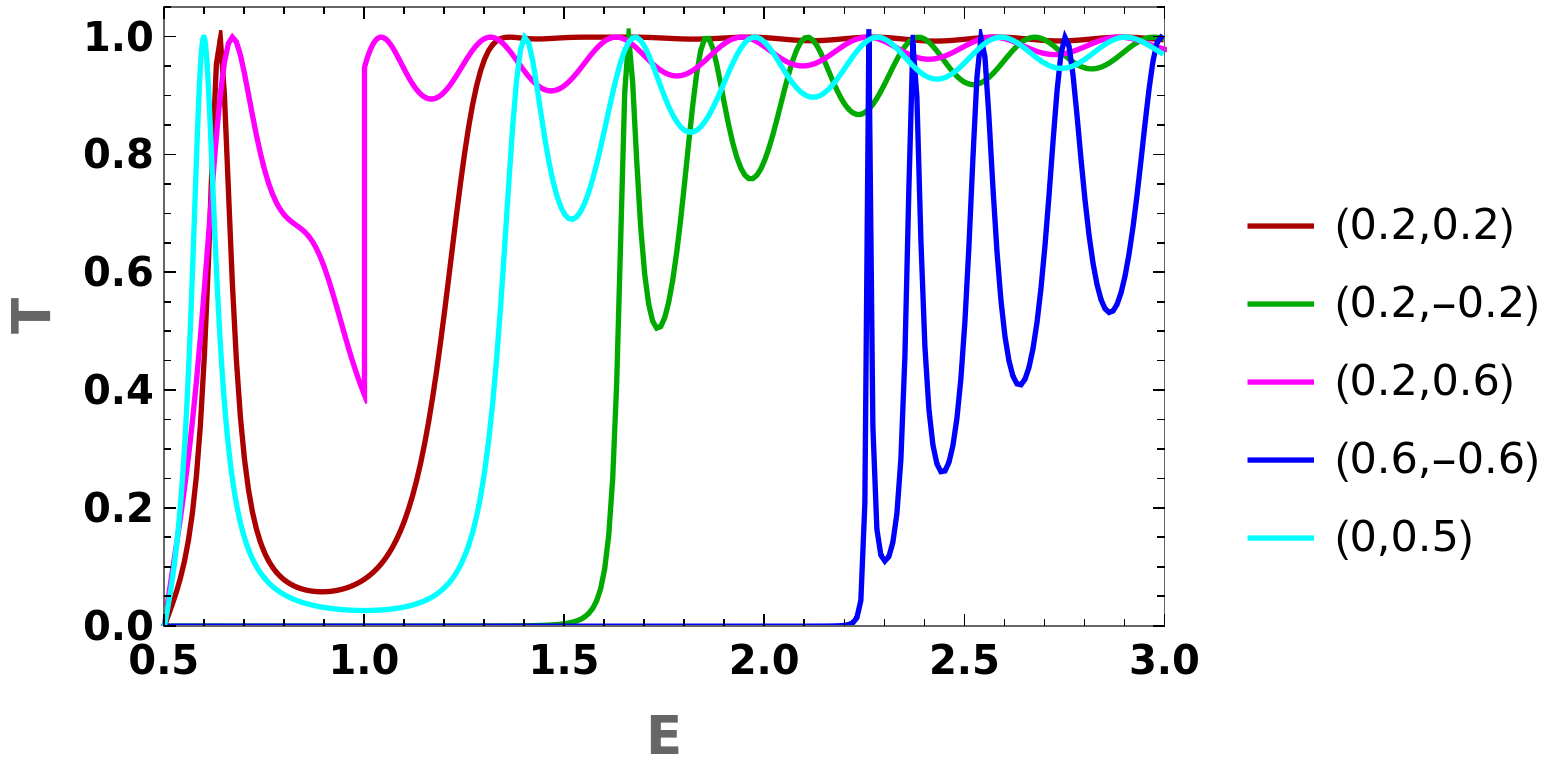}}\quad
\subfigure[]{\includegraphics[width=.46\textwidth]{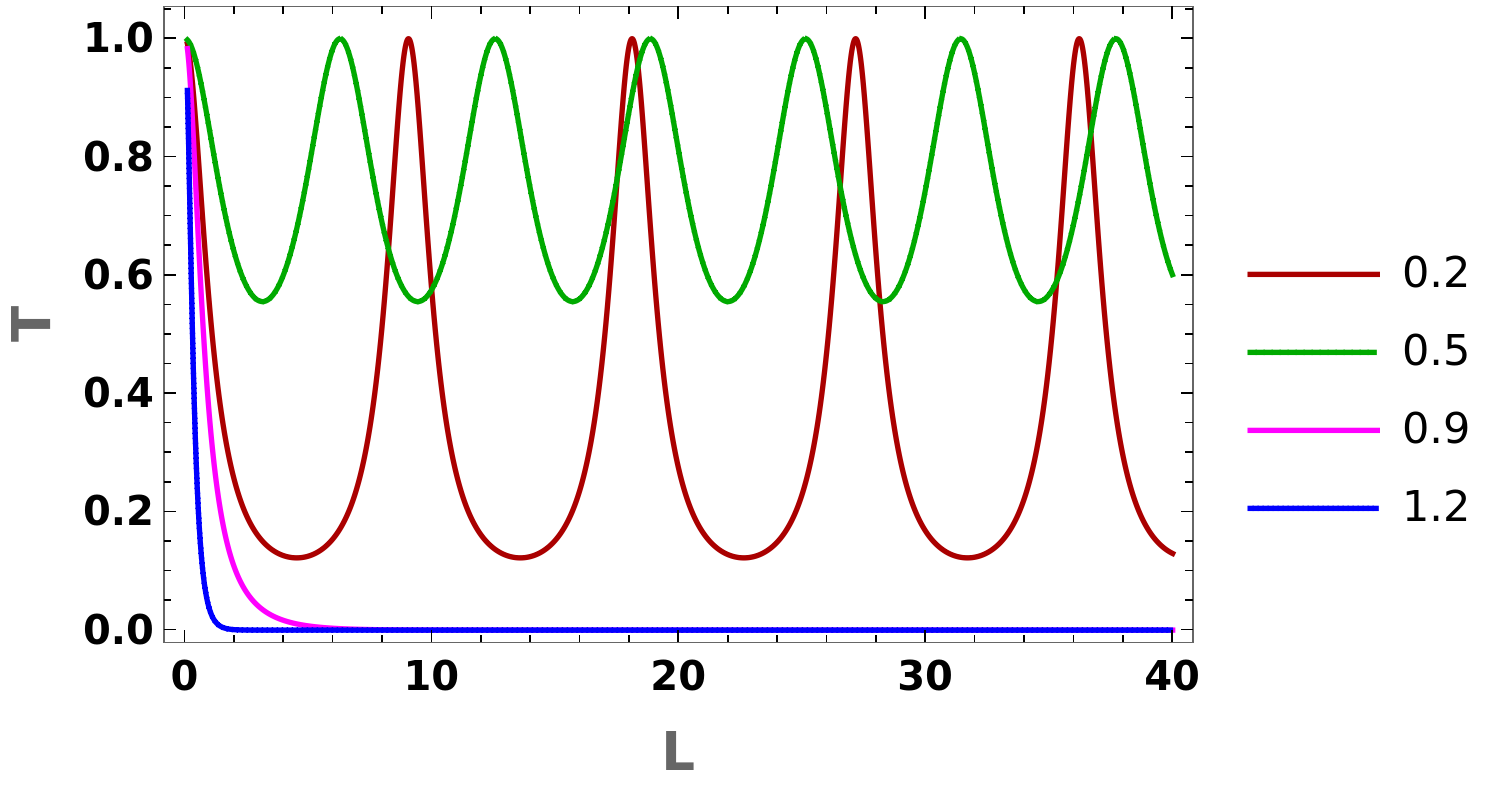}}
%\caption{$J=2$}
\subfigure[]{\includegraphics[width=.5\textwidth]{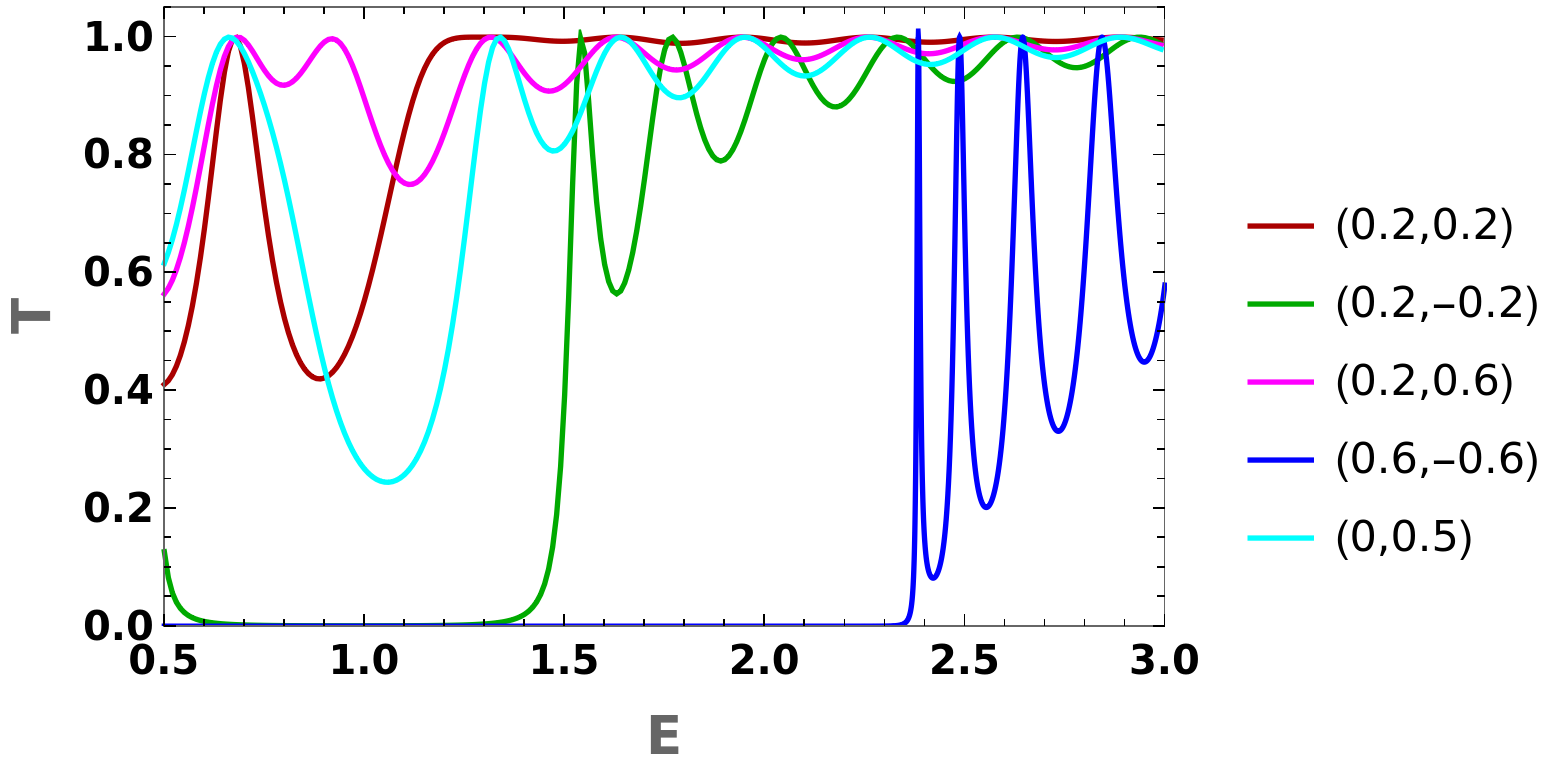}}\quad
\subfigure[]{\includegraphics[width=.46\textwidth]{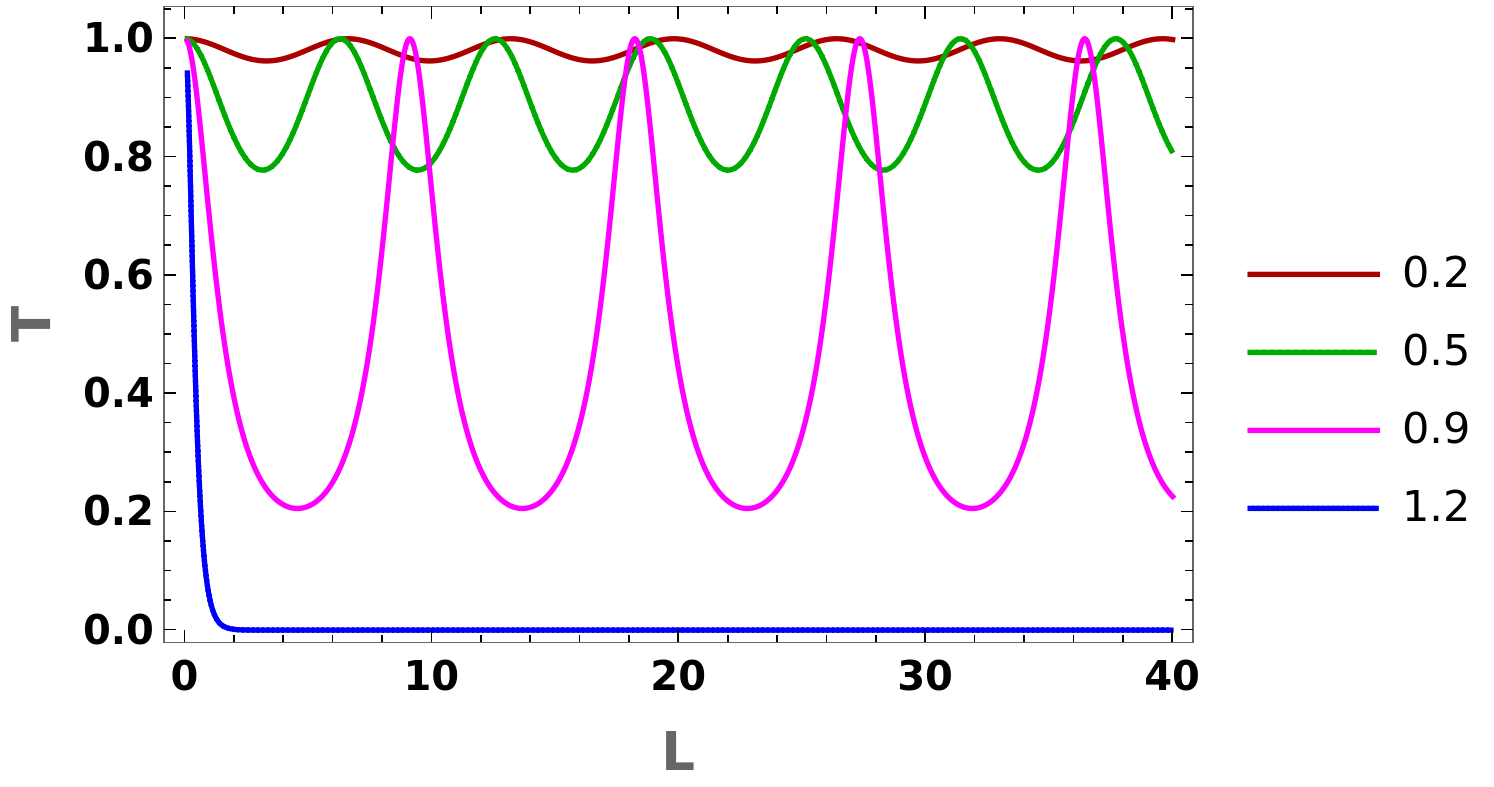}}
%\caption{$J=3$}
\caption{\label{TforKz}
Barrier perpendicular to $k_z$:
Panels (a), (c), and (e) show the transmission coefficient $T$ as a function of the Fermi energy $E$ for
$J=1,\,2,$ and $3$, respectively, with $e\,U=1$, $L=5$, $k_x=k_y =0.5$, and the $( a_x, \, a_y ) $ values indicated in the plot-legends.
Panels (b), (d), and (f) show $T$ as a function of the barrier length $L$ for
$J=1,\,2$ and $3$, respectively, with $e\,U=1$, $E=0.75$, $k_x=k_y =0.5$, and the $ a_x= a_y  $ values indicated in the plot-legends.}
\end{figure}
%%%%%%%%%%%%

The value of the transmission coefficient $T$ is obtained by taking the square of the absolute value of the corresponding transmission amplitude, i.e. $T=\left |  t_{J,{\mathbf n}}\right |^2$. For the case when $\tilde k_z$ is real
(or $\left [ \left(E-e\,U \right) ^2 -  
\frac{v_{\perp}^2\left( \tilde k_x^2 + \tilde k_y^2 \right)^{J}} {k_0^{2J-2}} \right ]
> 0$), we get:
\begin{align}
T = \begin{cases}
\frac{1}
{\frac{\sin ^2({\tilde k_z} L) (E (e U-E)+k_x {\tilde k_x}+k_y {\tilde k_y})^2}{k_z^2 {\tilde k_z}^2}+\cos ^2({\tilde k_z} L)}
& \text{ for } J=1 \\
%%%%%%%%%%%%%%%%%%%%
\frac{1}
{\frac{ 16 \sin ^2({\tilde k_z} L) 
\left[ E (e U-E)+
\left \lbrace k_x (-a_x+a_y+k_x)-k_y (a_x+a_y)+k_y^2\right \rbrace 
\left \lbrace a_x (k_y-k_x)-a_y (k_x+k_y)+k_x^2+k_y^2\right \rbrace
\right]^2}
{k_z^2 {\tilde k_z}^2}+\cos ^2({\tilde k_z} L)}
& \text{ for } J=2\\
%%%%%%%%%%%%%%%%%%%%%%%%%%%%%%%%%%%
\frac{1}{\frac{16 \sin ^2({\tilde k_z} L) 
\left[ E (e U-E)
+ \left(k_x\,{\tilde k_x}+k_y \,{\tilde k_y} \right ) 
\left \lbrace k_y^2 \left(-3 a_x^2-2 a_x k_x+a_y^2+2 k_x^2\right)+k_x^2 \left( {\tilde k_x}^2-3 a_y^2\right)-2 a_y k_x k_y (k_x-4 a_x)-2 a_y k_y^3+k_y^4\right \rbrace
\right ]^2}
{k_z^2 {\tilde k_z}^2}
+\cos ^2({\tilde k_z} L)}
& \text{ for } J=3
\end{cases}\,.
\end{align}
%%%%%%%%%%%%%%%%
%%%%%%%%%%%%%%%
Clearly, $T=1 $ for $\cos ({\tilde k_z} L) =\pm 1$ and $\sin ({\tilde k_z} L) = 0$.
Hence we expect an oscillatory behavior with $T$ becoming unity whenever ${\tilde k_z}=
\frac{N\,\pi}{L}$ for $N \in \mathbb{Z}$. 
%%%%%%%%
We also note that there will be regions of zero transmission for large enough $L$, which coincide with the regions where $\tilde k_z$ is imaginary (or $\left [ \left(E-e\,U \right) ^2 -  
\frac{v_{\perp}^2\left( \tilde k_x^2 + \tilde k_y^2 \right)^{J}} {k_0^{2J-2}} \right ]
<0$), because $T$ then falls off as $e^{-2\,|\tilde k_z|\, L}$.

%%%%%%%%%%%%%%%%%%%%%%%%%%%
\begin{figure}[]
\subfigure[]{\includegraphics[width=.32\textwidth]{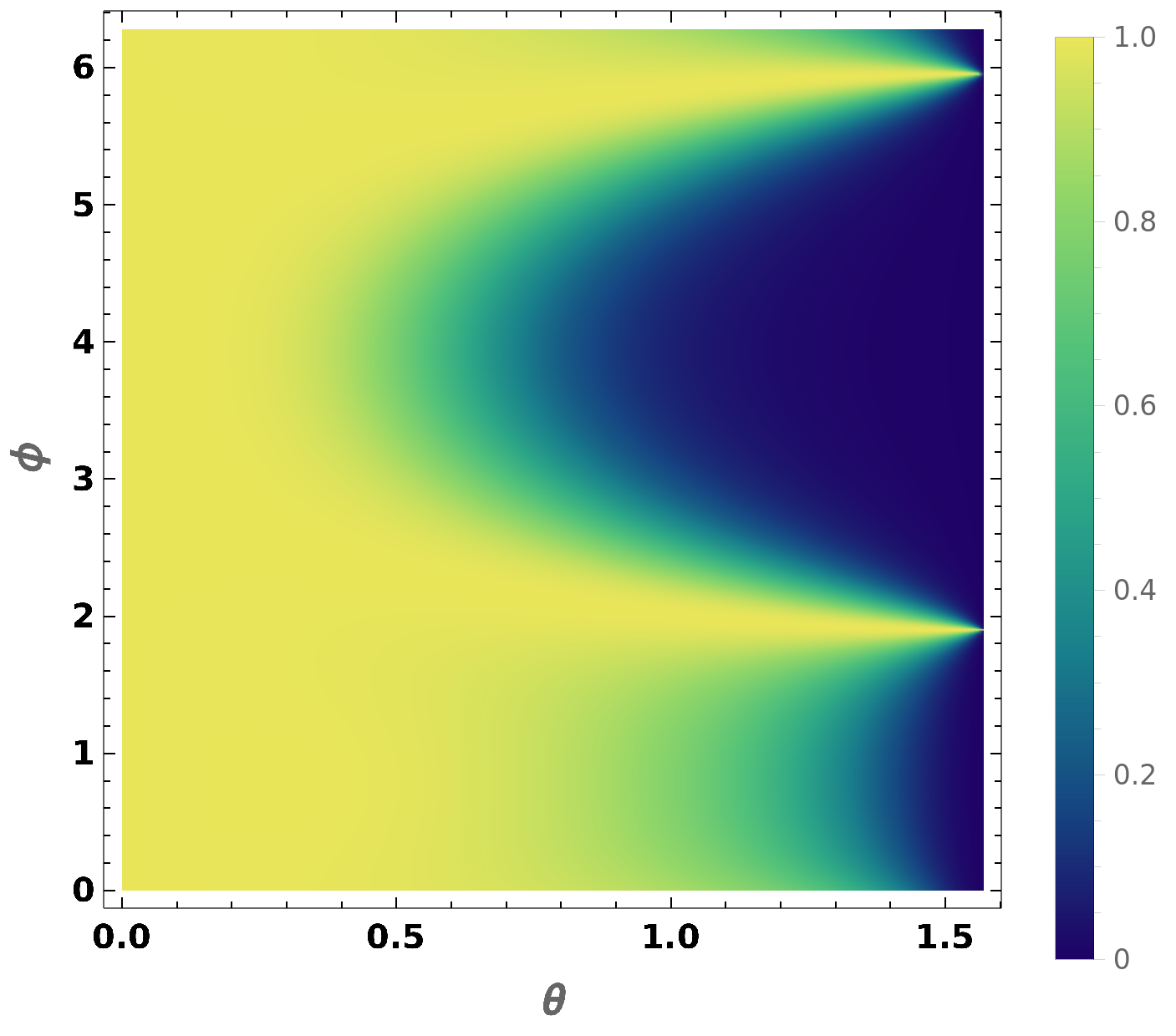}}\quad
\subfigure[]{\includegraphics[width=.32\textwidth]{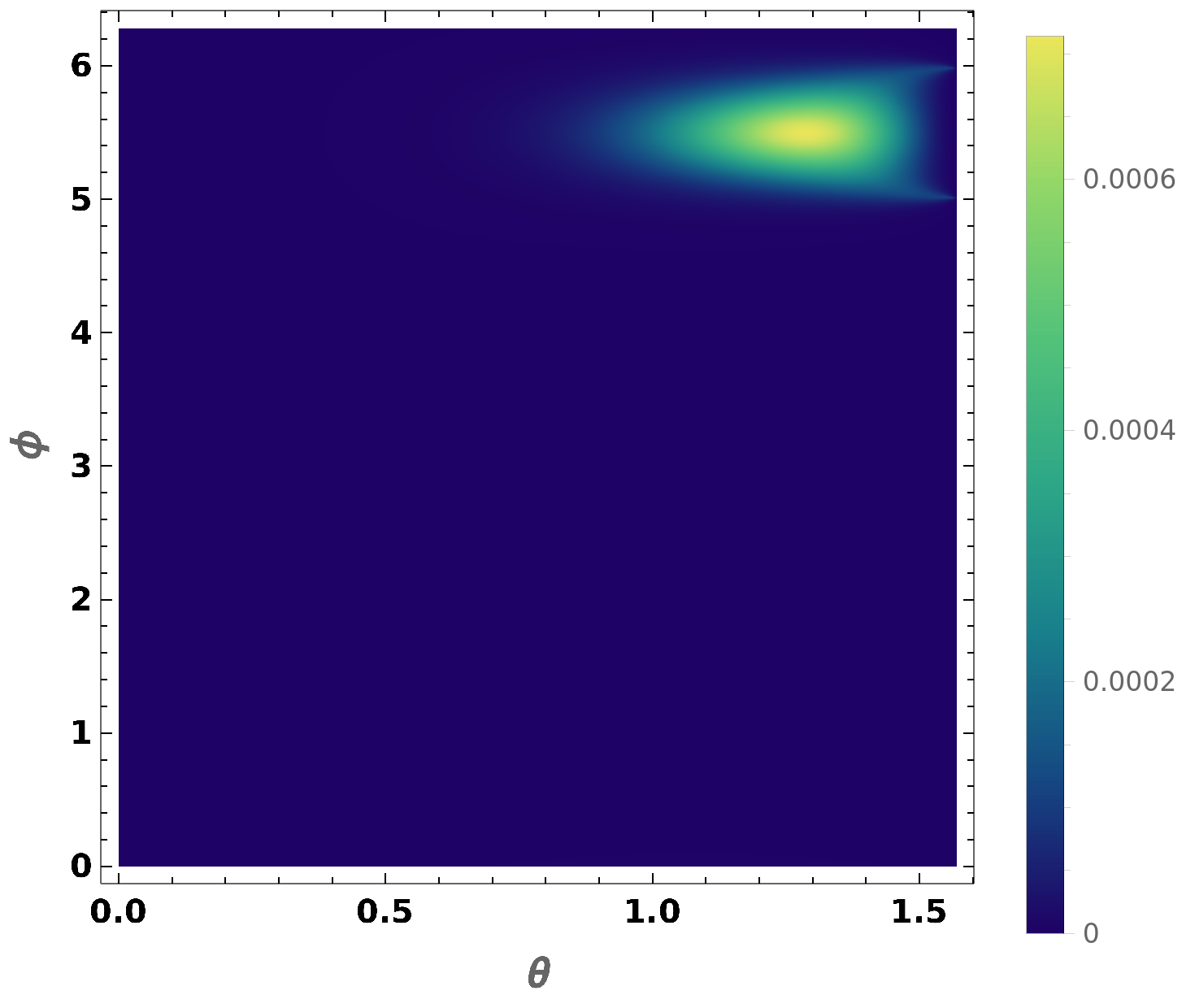}}\quad
\subfigure[]{\includegraphics[width=.32\textwidth]{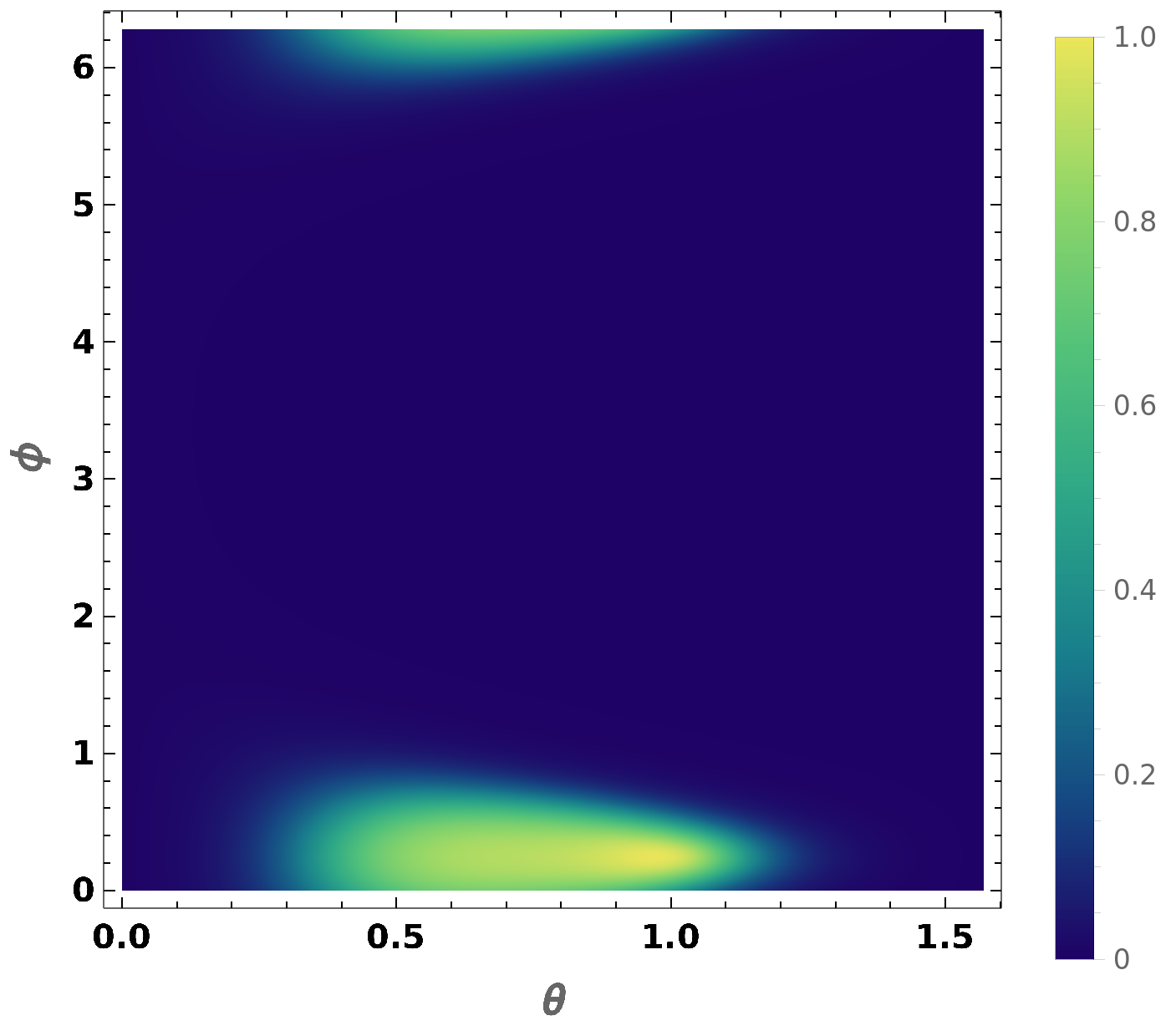}}
%\caption{$J=1$}
\subfigure[]{\includegraphics[width=.32\textwidth]{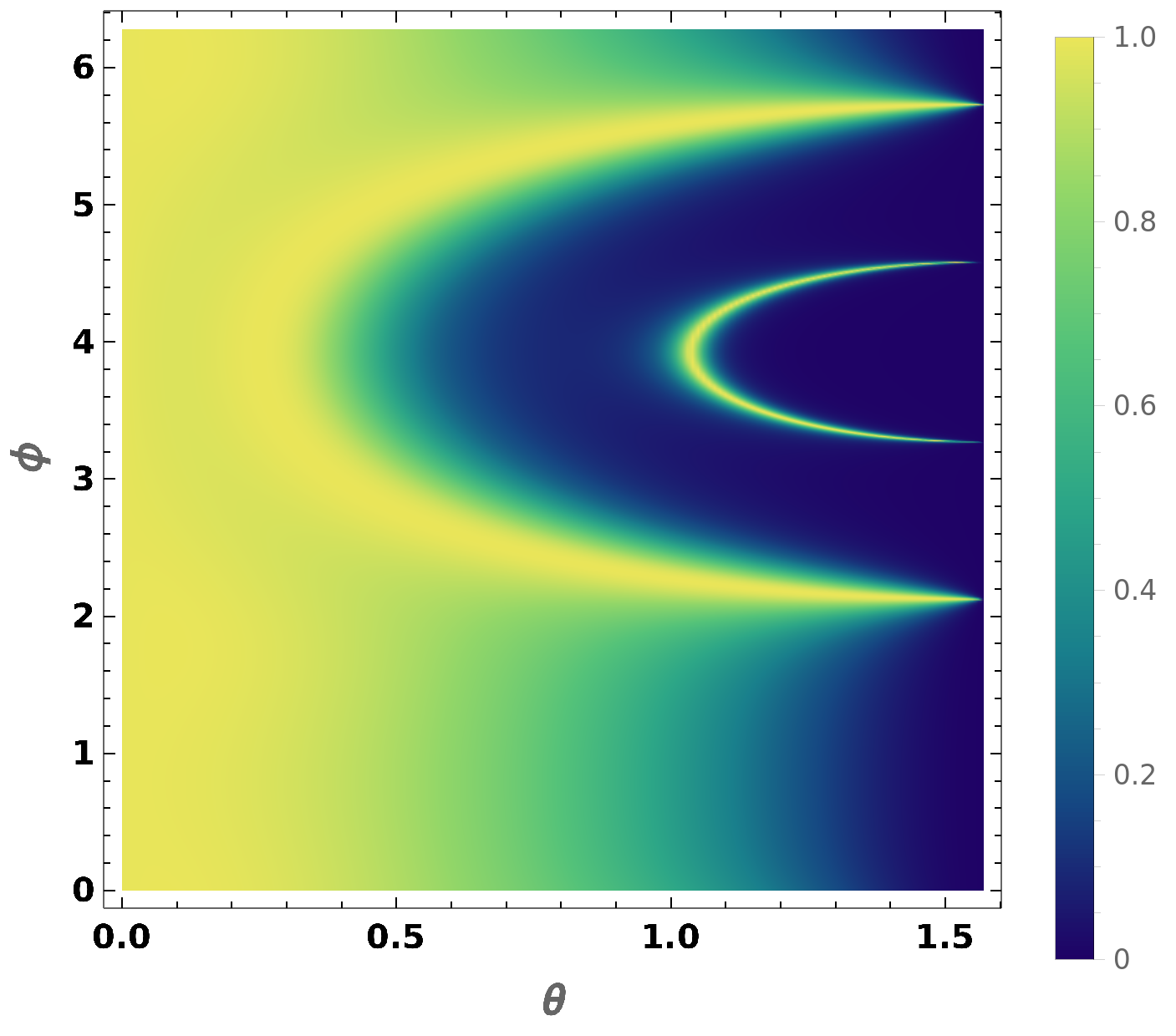}}\quad
\subfigure[]{\includegraphics[width=.32\textwidth]{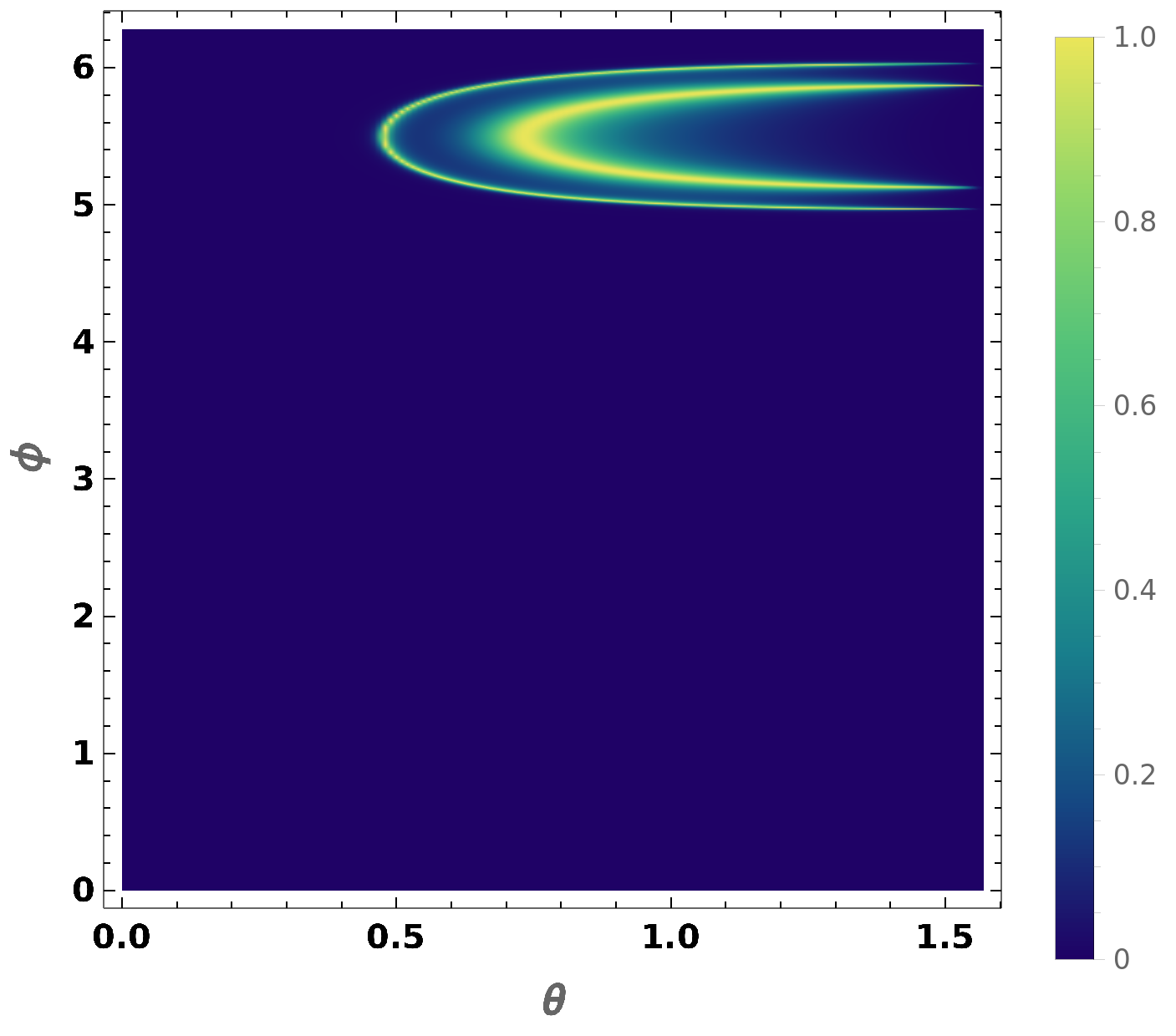}}\quad
\subfigure[]{\includegraphics[width=.32\textwidth]{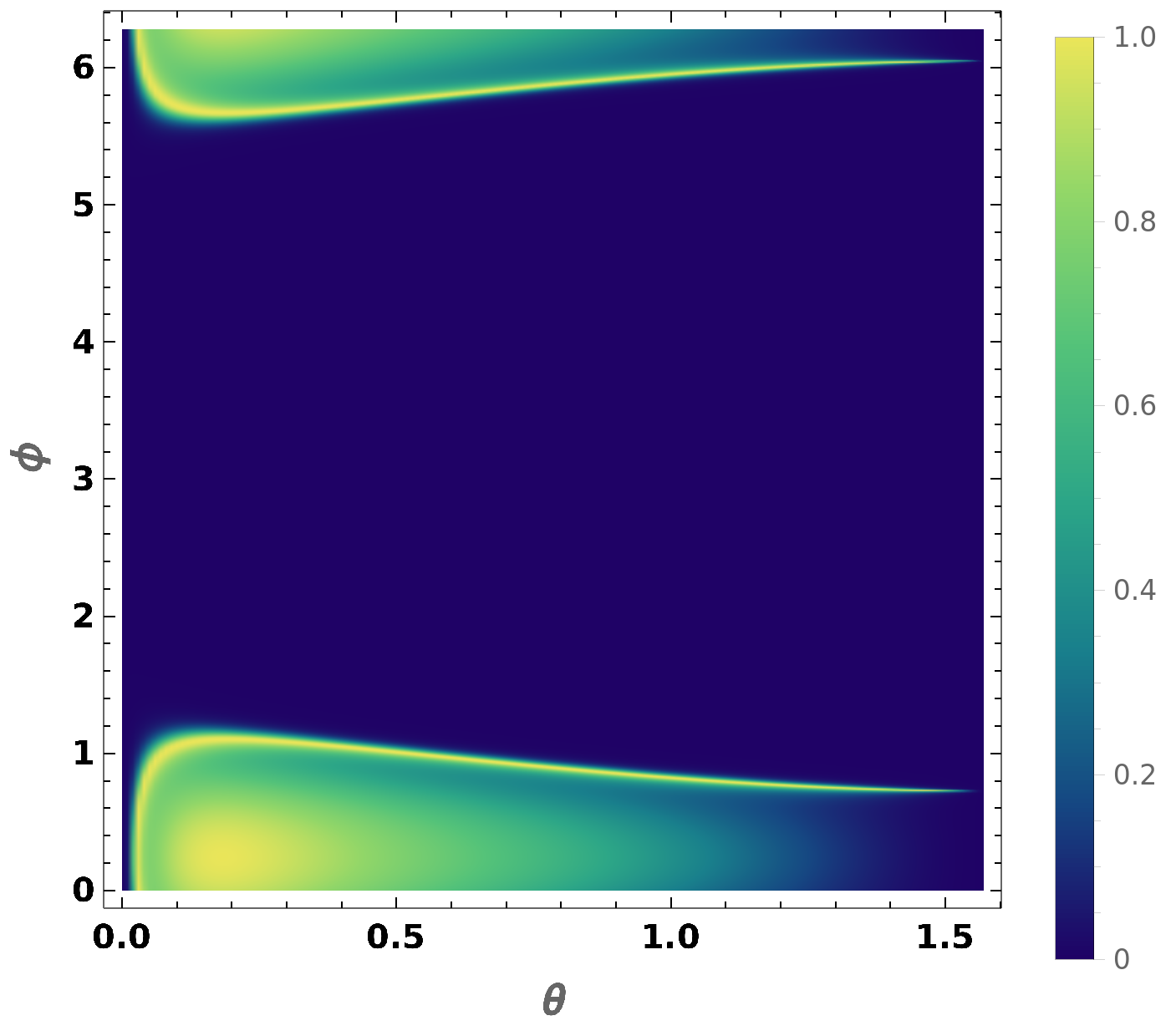}}
%\caption{$J=2$}
\subfigure[]{\includegraphics[width=.32\textwidth]{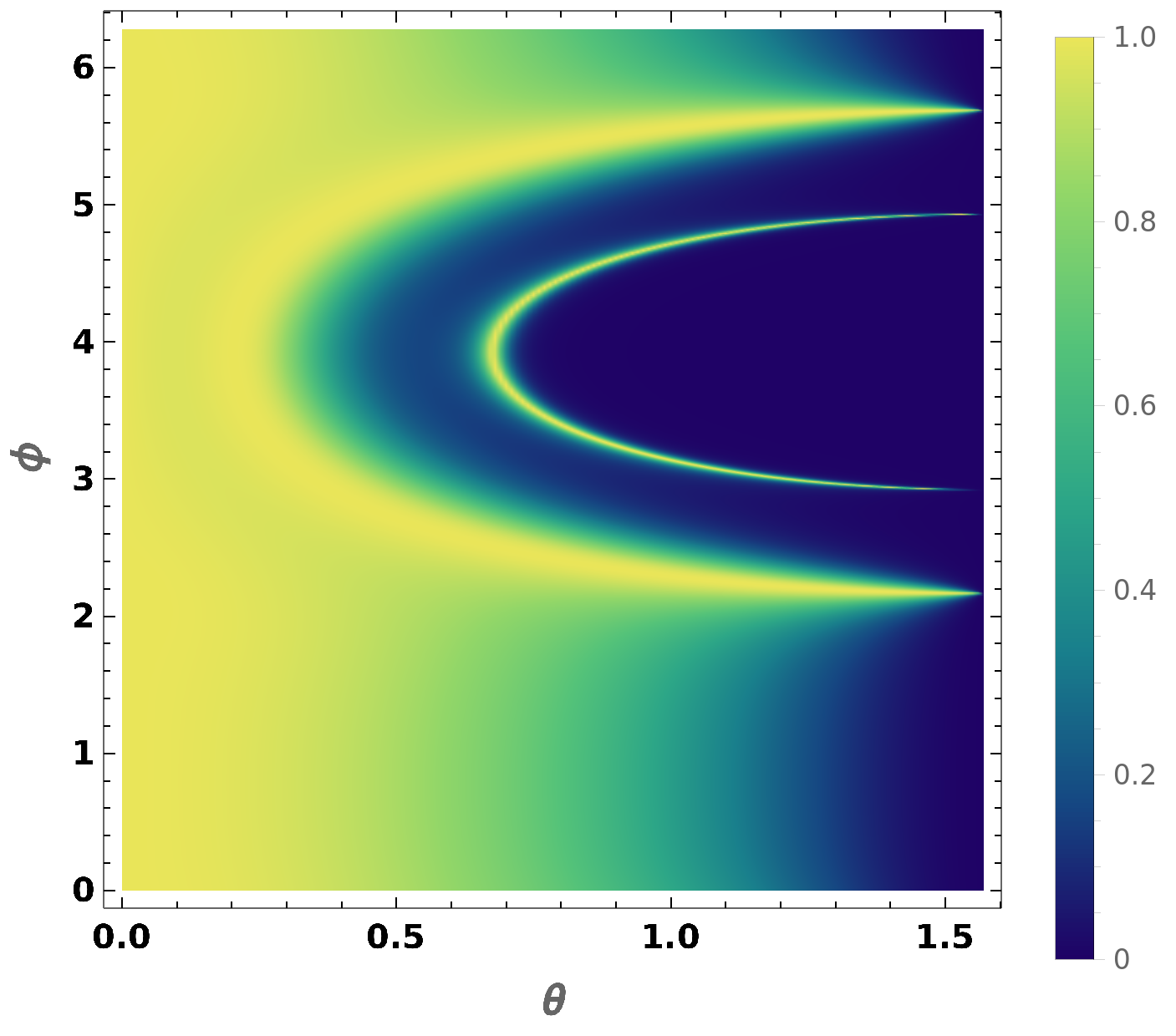}}\quad
\subfigure[]{\includegraphics[width=.32\textwidth]{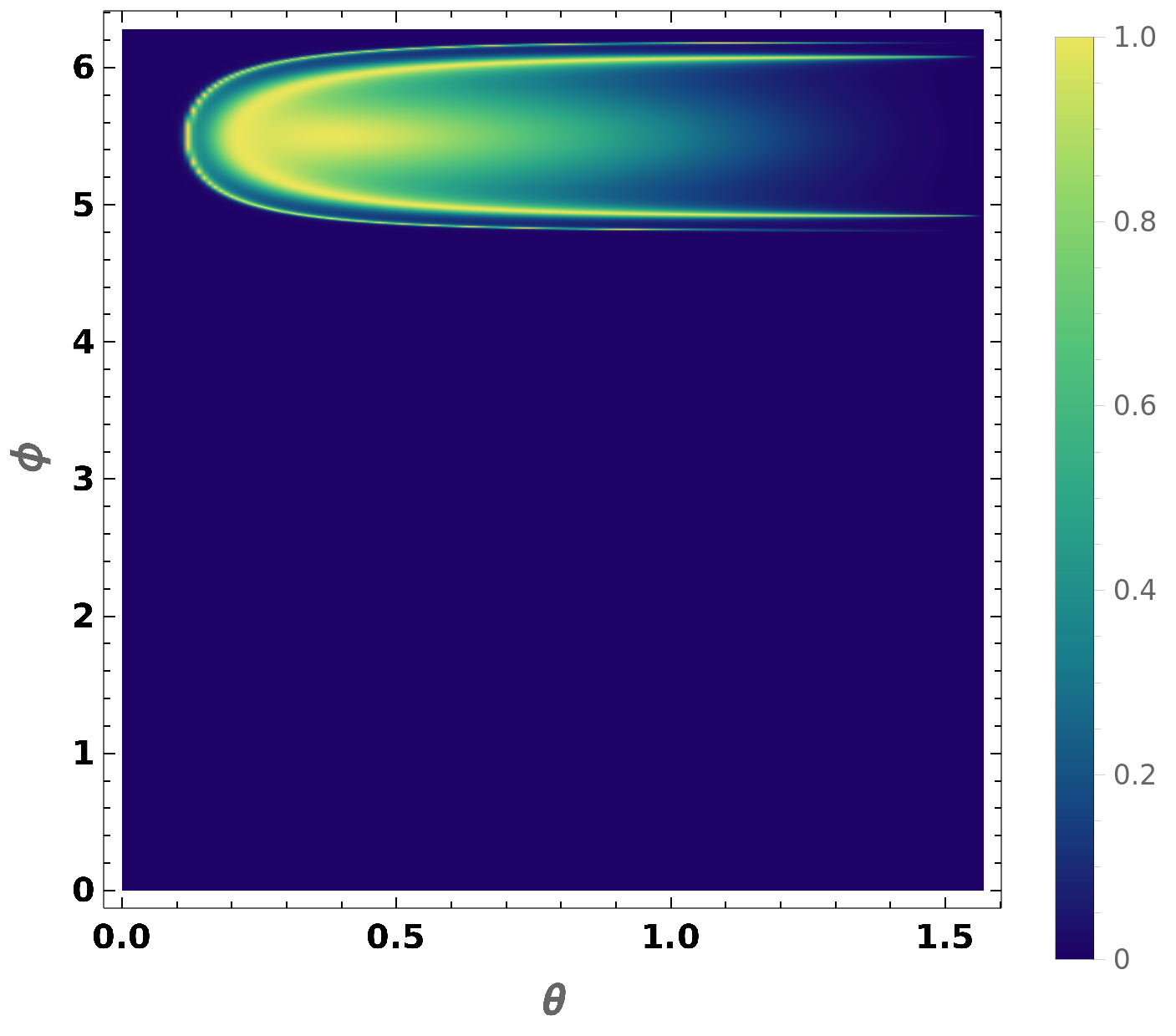}}\quad
\subfigure[]{\includegraphics[width=.32\textwidth]{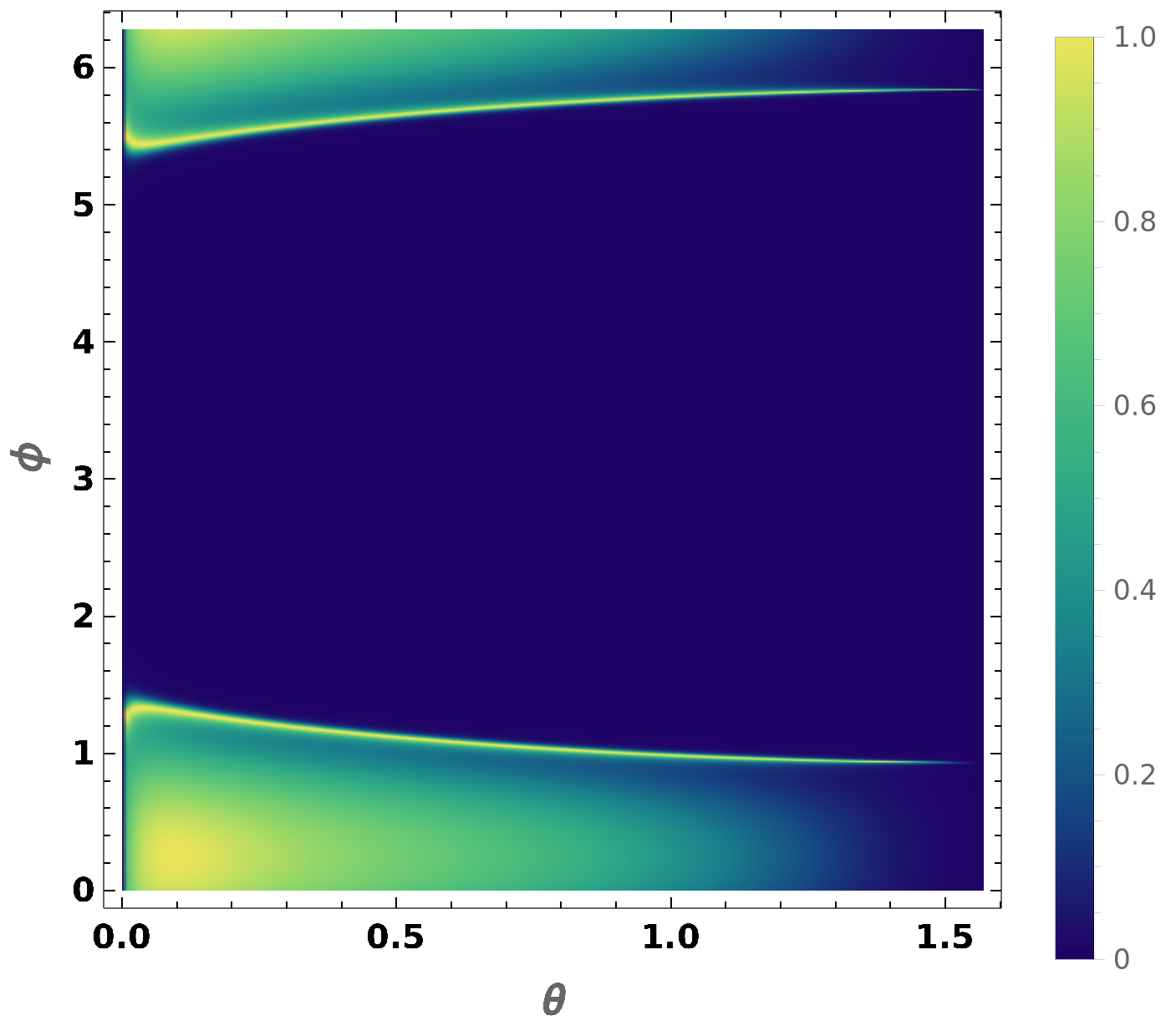}}
%\caption{$J=3$}
\caption{\label{contourforKz}
%%%%%%%%%%%%%%%%%%%%%%%%%%%%%%5
For a barrier perpendicular to $k_z$, contour-plots of the transmission coefficient $T$ as a function of the orientation of the incident beam, parameterized by the angles $(\theta, \,\phi)$: Panels (a), (d), and (g) show $T$ for $J=1,\,2,$ and $3$, respectively, with $e\,U=1$, $E=0.3$, $L=5$, and the $( a_x, \, a_y) =(0.2,\,0.2) $.
Panels (b), (e), and (h) show $T$ for
$J=1,\,2$ and $3$, respectively, with $e\,U=1$, $E=0.6$, $L=10$, and $( a_x, \, a_y) =(0.8,\,-0.8) $.
Panels (c), (f), and (i) show $T$ for
$J=1,\,2$ and $3$, respectively, with $e\,U=1$, $E=1.5$, $L=5$, and $( a_x, \, a_y) =(0.8,\,0.2) $.
}
\end{figure}
%%%%%%%%%%%%%%%%%%%%%%%%%%%

%%%%%%%%%%%%%%%%%%%%%%%%%%%
Fig.~\ref{TforKz} shows some representative plots to capture the behavior of the transmission coefficient $T$ as functions of $E$ (both for $E < e\, U$ and $E >e\, U$) and $L$, respectively, when the other parameters are held fixed at some constant values.
As expected, it shows oscillatory behavior, reaching the value $1$ whenever $\tilde k_z = \frac{N\,\pi}{L} $ (with $N \in \mathbb{Z}$).
%%%%%%%%%%%
In Figs.~\ref{TforKz}(b), (d), and (f), we find that some curves decay exponentially as functions of $L$. These are the ones for which $\tilde k_z$ become imaginary.

Fig.~\ref{contourforKz} shows the characteristic $T$ as a function of the orientation of the incident beam, parameterized by the angles $(\theta, \,\phi)$, when the other parameters are held fixed at some constant values.The choice of parameters include both the $E < e\, U$ and $E >e\, U$ cases. For these contour-plots we have used the coordinate transformations as follows:
\begin{align}
\label{eqsph}
k_x= \sqrt[J]{ \frac{k_0^{J-1}\,E\, \sin \theta}
{v_\perp}} \,\cos \phi\,, 
\quad k_y= \sqrt[J]{ \frac{k_0^{J-1}\,E\, \sin \theta}
{v_\perp}}\, \sin \phi \,,\quad
k_z=\frac{E\, \cos \theta}{v_z} \,.
\end{align}
Compared to the cases of zero magnetic field, these plots show oval-shaped contours.
Note that in the absence of magnetic fields, $T$ decreases monotonically from one as $\theta$ increases from zero to $\pi/2$, irrespective of the value of $\phi$ (as the system is isotropic with respect to a rotation in the $xy$-plane when $a_x=a_y=0$). 
Let us discuss the features seen for different $J$-values:
\begin{enumerate}

\item
$J=1$: In Fig.~\ref{contourforKz}(a), $\tilde k_z$ is real in the entire region, and shows areas where $T$ is nearly equal to one as $\sin^2 \left(  \tilde k_z L\right)$ is nearly equal to zero. We see two semi-oval-shaped regions of zero $T$. In the upper lobe, $k_z$ and $\tilde k_z$ are close to zero, while $\sin^2 \left(  \tilde k_z L\right) > 0.2$. This makes the factor $\frac{\sin^2 \left(  \tilde k_z L\right)} {k_z^2\,\tilde k_z^2}$ in the denominator of $T$ very large, driving $T$ towards zero. In the lower lobe, $k_z$ is close to zero, $0.02<\sin^2 \left(  \tilde k_z L\right) < 0.2$, and $\tilde k_z>0.45$, and all the factors conspire to make $T$ zero.
In Fig.~\ref{contourforKz}(b), $\tilde k_z$ is imaginary in the entire region, and $T$ never reaches the value of unity -- it remains close to zero for most areas (as $e^{-2\,\text{Im}\left({\tilde k_z}\right) L} \rightarrow 0 $), reaching some small nonzero values in narrow spots where the magnitude of $\text{Im}\left({\tilde k_z}\right) $ approaches zero (such that $e^{-2\,\text{Im}\left({\tilde k_z}\right) L}$ is not effectively zero). In Fig.~\ref{contourforKz}(c), $\tilde k_z$ is imaginary most of the region, except in two lobes in the uppermost and lowermost areas, within which $T$ takes values close to unity whenever $\sin^2 \left(  \tilde k_z L\right)$ is close to zero. Consequently, $T$ remains close to zero in most parts, except when the magnitude of $\text{Im}\left({\tilde k_z} \right)$ is very small or zero.

\item
$J=2$: In Fig.~\ref{contourforKz}(d), $\tilde k_z$ is real in the entire region, and shows areas where $T$ is nearly equal to one or zero depending on the value of the factor $\frac{\sin^2 \left(  \tilde k_z L\right)} {k_z^2\,\tilde k_z^2}$ in the denominator. In Fig.~\ref{contourforKz}(e), $\tilde k_z$ is imaginary in the entire region, except in an oval region towards the upper right. $T$ remains close to zero for most areas (as $e^{-2\,\text{Im}\left({\tilde k_z}\right) L} \rightarrow 0 $), reaching unity within a narrow ring within the aforementioned oval region. $T$ also shows values close to unity when the magnitude of $\text{Im}\left({\tilde k_z} \right)$ is small such that $e^{-2\,\text{Im}\left({\tilde k_z}\right) L}$ is also small.
In Fig.~\ref{contourforKz}(f), $\tilde k_z$ is imaginary most of the region, except in two lobes in the uppermost and lowermost areas, within which $T$ takes values close to unity whenever $\sin^2 \left(  \tilde k_z L\right)$ is close to zero. Consequently, $T$ remains close to zero in the middle areas, and slowly approaches unity when the magnitude of $\text{Im}\left({\tilde k_z} \right)$ becomes very small or zero.

\item
$J=3$: Figs.~\ref{contourforKz}(g), (h), and (i) show features similar to Figs.~\ref{contourforKz}(d), (e), and (f), respectively. The underlying physical interpretations are similar to those of the $J=2$ case.
\end{enumerate}
%%%%%%%%%%%%%%%
The plots indicate that although there are some small differences in the behavior of $T$, there
is no significant change in generic features for the different values of $J$. This stems from the fact that the quasiparticles for different $J$ values have the same linear dispersion along the tunneling direction when the barrier is perpendicular to $k_z$-component of the momentum.

%%%%%%%%%%%%%%%%%%%%%%%%%%%%%%%%%%%%%%%%%%%%%5 
\subsection{Conductivity and Fano factors}

%%%%%%%%%%%%%%%%%%%%%%%%%%%
\begin{figure}[]
\subfigure[]{\includegraphics[width=.45\textwidth]{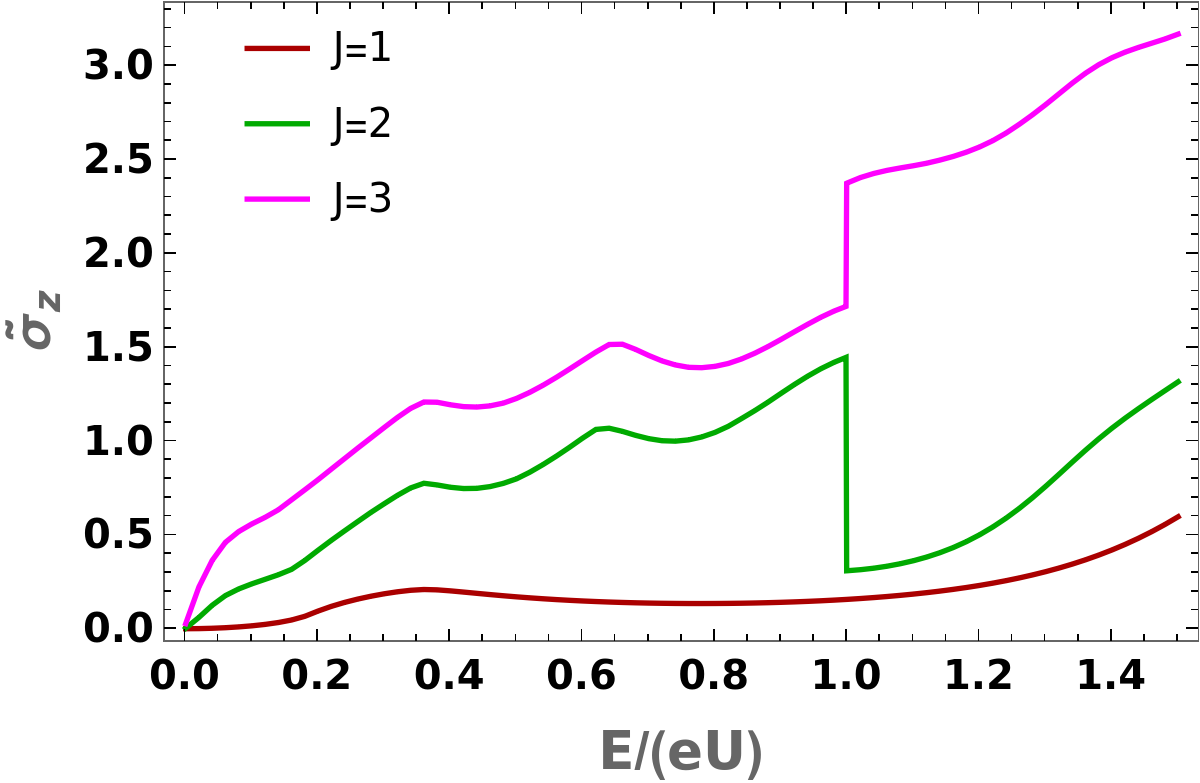}}\qquad
\subfigure[]{\includegraphics[width=.45\textwidth]{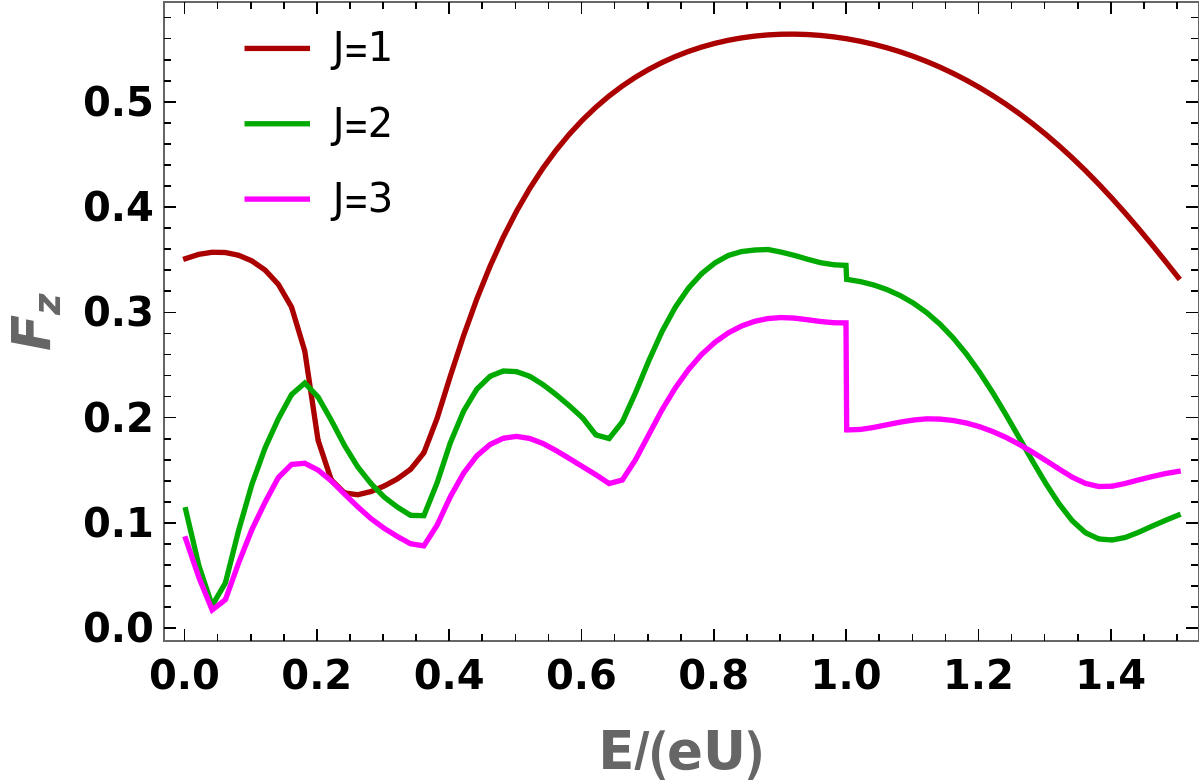}}
\caption{\label{condforKz}
Barrier perpendicular to $k_z$: Conductivity ($\tilde \sigma_z$ in units of $ \frac{ L^2 } {4\, \pi ^2 } $) and Fano factors ($F_z$) for $e\,U=1$, $L= 5$, $a_x=0.2$, and $a_y=0.25$.
}
\end{figure}
%%%%%%%%%%%%%%%%%%%%%%%%%%%

We assume $ W $ to be large enough such that $k_x$ and $k_y$ can effectively be treated as continuous variables, allowing us to perform the integrations over them to obtain the conductivity and Fano factor. 
Using $ k_z = \frac{E\, \cos \theta}{v_z}\,,
\,\, n_x= \frac{W}{2\,\pi}
\sqrt[J]{ \frac{k_0^{J-1}\,E\, \sin \theta}
{v_\perp}} \,\cos \phi\ \,,
\,\, n_y = \frac{W}{2\,\pi}
\sqrt[J]{ \frac{k_0^{J-1}\,E\, \sin \theta}
{v_\perp}}\, \sin \phi\,,
\,\, dn_x \, dn_y =
\frac{ W^2
\left \vert
\cot \theta  \left(
\frac{ E \sin \theta  \,k_0^{J-1}}{ v_\perp }\right)^{2/J} 
\right \vert \,d\theta \,d\phi}
{4 \,\pi ^2 J} $, 
in the zero-temperature limit and for a small applied voltage, the conductance is given by \cite{blanter-buttiker}:
\begin{align}
G_z(E,U,\mathbf B) & = \frac{ e^2}{h} \sum_{\mathbf n}   |t_ {J,\mathbf n}|^2 
\rightarrow \frac{  e^2} {h} \int   |t_ {J, \mathbf n}|^2 \,dn_x\,dn_y
= \frac{  e^2\,W^2 } {4 \,\pi ^2\,h\, J} 
  \int_{ \theta=0 }^{\pi/2} \int_{\phi=0}^{2\,\pi}
T \, \left \vert
\cot \theta  \left(
\frac{ E \sin \theta  \,k_0^{J-1}}{ v_\perp }\right)^{2/J} 
\right \vert \,d\theta \,d\phi \,,
\end{align}
leading to the conductivity expression:
\begin{align}
\tilde \sigma_z (E,U,\mathbf B)  & = \left( \frac{L }{W} \right)^2 
\frac{ G_z(E,U,\mathbf B) } {e^2/h}
 =   \frac{ L^2 } {4\, \pi ^2 \,J} 
  \int_{ \theta=0 }^{\pi/2} \int_{\phi=0}^{2\,\pi}
T \, \left \vert
\cot \theta  \left(
\frac{ E \sin \theta  \,k_0^{J-1}}{ v_\perp }\right)^{2/J} 
\right \vert \,d\theta \,d\phi \,.
\end{align}
The shot noise is captured by the Fano factor, which can be expressed as:
\begin{align}
F_z(E, U ,\mathbf B)  &=\frac 
{\int_{ \theta=0 }^{\pi/2} \int_{\phi=0}^{2\,\pi}
T \left( 1-T \right)\, \left \vert
\cot \theta  \left(
\frac{ E \sin \theta  \,k_0^{J-1}}{ v_\perp }\right)^{2/J} 
\right \vert \,d\theta \,d\phi  } 
{ \int_{ \theta=0 }^{\pi/2} \int_{\phi=0}^{2\,\pi}
T \, \left \vert
\cot \theta  \left(
\frac{ E \sin \theta  \,k_0^{J-1}}{ v_\perp }\right)^{2/J} 
\right \vert \,d\theta \,d\phi  } \,.
\end{align}

The results are plotted in Fig.~\ref{condforKz}, as functions of the Fermi energy, for some representative parameter values. The curves clearly show that local minima of conductivity no longer appear at $E =e\, U$ for nonzero magnetic fields, unlike the zero magnetic field cases \cite{Deng2020}. For $J>1$, we also see jumps in $ \tilde \sigma_z$ at $E =e\, U$, the sign on the jump alternating for the $J=2$ and $J=3$ cases. For $E > e\, U$, $ \tilde \sigma_z$ increases monotonically with $E$ for all $J$-values.

%%%%%%%%%%%%%%%%%%%%%%%%%%%%%%%%%%%%%%%%%%%%%%%%%%%%%%%%%%%%%%%%

%%%%%%%%%%%%%%%%%%%%%%%%%%%%%%%%%%%%%%%%%%%%%%5
\section{Barrier perpendicular to $k_x $}
\label{secperpkx}

We consider the second case where the barrier is perpendicular to $k_x$, so that the other two components $k_y$ and $k_z$ are conserved. Similar to previous case potential is expressed as 
%%%%%%%%%
\begin{equation}
U(x,y,z)=\begin{cases}U & 0\leq x\leq L\\0 & \text{otherwise}\end{cases}\,.
\end{equation}
%%%%%%%%%%%
In this case, the momentum components $k_y$ and $k_z$ are conserved.
On imposing periodic boundary conditions along these directions, we get the corresponding momentum components quantized as:
\begin{align}
k_y=\frac{2\,\pi\, n_y}{W}\,,\quad
k_z=\frac{2\,\pi \,n_z}{W}\,.
\end{align}

In this set-up, we now subject the sample to the magnetic field $\mathbf{B}  = 
 B_y\, \hat{\mathbf j}  \,\left [  \delta\left(x\right) - \delta\left( x-L \right) \right ] $, and directed perpendicular to the $zx$-plane. This can be created from a vector potential with the components:
\begin{align}
\mathbf {A}( x ) \equiv  \lbrace 0,0, a_z  \rbrace = \begin{cases} 
\lbrace 0, 0 , -B_y \rbrace &  \text{ for } 0 < x < L \\
\mathbf 0 & \text{ otherwise} \,.
\end{cases}
\label{eqvecpot}
\end{align}
The vector potential modifies the linear momentum as $k_z \rightarrow k_z-e\, a_z$, such that the effective Hamiltonians within the barrier region are given by $\mathcal{H}_J(  k_x, k_y ,  k_z - {e\, a_z} ) + e\,U$.

The momentum along $x$-direction outside the barrier region is given by:
\begin{align}
k_\eta=\pm
\sqrt{e^{\frac{2\,\pi\,\mathrm{i}\,j}{J}}
\Bigg \vert
\left[ \frac{\left( E^2-  v_z^2\, k_z^2 \right) k_0^{2J-2}} {v_{\perp}^2}\right]^{1/J}
\Bigg \vert-k_y^2} \,,\text{ where } j=1,\cdots J\,.
\end{align}
Hence, we have $2J$ possible solutions for $k_\eta$ for a given Fermi energy $E$. 
Within the barrier region, the momentum along $x$-direction is given by:
\begin{align}
\tilde k_\eta=\pm
\sqrt{e^{\frac{2\,\pi\,\mathrm{i}\,j}{J}}
\Bigg \vert
\left[ \frac{\left \lbrace 
\left( E-e\,U\right)^2-  v_z^2\left(  k_z - e\, a_z \right)^2 \right \rbrace k_0^{2J-2}} {v_{\perp}^2}\right]^{1/J}
\Bigg \vert-k_y^2} \,,\text{ where } j=1,\cdots J\,.
\end{align}
Again, we have $2J$ possible solutions for $\tilde k_\eta$ for a given set $(E, U, a_z)$. 

For this case, the analytical expressions for the transmission and reflection coefficients become unwieldy, and hence we find their values numerically and show some representative results in the next section. 

%%%%%%%%%%%%%%%%%%%%%%%%%%%%%555
\subsection{$J=2$}

The solutions $k_\eta=\pm
\sqrt{ \frac{k_0\,\sqrt{  E^2-  v_z^2\, k_z^2 } } {v_{\perp}}-k_y^2}$ give propagating modes, while $k_\eta=\pm \mathrm{i}\,\sqrt{ \frac{k_0\,\sqrt{  E^2-  v_z^2\, k_z^2 } } {v_{\perp}}+k_y^2}$ give evanescent modes. Among the evanescent modes, we only consider the physically admissible exponentially decaying solution, as the wavefunction cannot increase in an unbounded fashion as we approach $x =\pm \infty$.
%%%%%%%%%t
Within the barrier region, both the exponentially increasing and decaying solutions are allowed, and hence we need to consider all the four values of $\tilde k_\eta$.

A scattering state $\Psi_{ 2,\mathbf n}(x)$, in the mode labeled by $\mathbf n =\lbrace n_y, n_z \rbrace$, is
constructed from the following states:
\begin{align}
 \Psi_{2, \mathbf n} (x)=&  \begin{cases}   \phi_L & \text{ for } x<0  \\
 \phi_M & \text{ for } 0< x < L \\
  \phi_R &  \text{ for } x > L 
\end{cases} \,,\nonumber \\
%%%%%%%%%%%%%%%%%%%%%%%%%%%%%%%%%%%%%%%%
  \phi_L = & \,\frac{   
 \psi_{2}^+ ( k_\eta^{\text{re}},k_y,k_z)\,e^{\mathrm{i}\, k_\eta^{\text{re}} \,x} 
 + r_{2,\mathbf n}\, \psi_{2}^+ ( -k_\eta^{\text{re}},k_y,k_z)\,e^{-\mathrm{i}\, k_\eta^{\text{re}} x}}
{\sqrt{  { \mathcal{V} }( k_\eta^{\text{re}}, k_y,k_z) }}
+ r^{\prime}_{2,\mathbf n}\, \psi_{2}^+ ( -k_\eta^{\text{im}},k_y,k_z)\,
e^{-\mathrm{i}\,k_\eta^{\text{im}} x}
\, ,\nonumber \\
%%%%%%%%%%%%%%%%%%%%%%%%%%%%%%%%%%%%%%%%%%%%5
  \phi_M  = &\,\Big[  
 \alpha^+_{2,\mathbf n} \,\psi_2^+ ( \tilde k_\eta^+, k_y,\tilde{k}_z) \,
 e^{\mathrm{i}\,\tilde k_\eta^+  x } 
 + 
 \beta^+_{2,\mathbf n} \, \psi_2^+ (-\tilde k_\eta^+, k_y,\tilde{k}_z) \,
 e^{-\mathrm{i}\,\tilde k_\eta^+ x }
 +
 \alpha^{-}_{2,\mathbf n} \,\psi_2^+ ( \tilde k_\eta^-, k_y,\tilde{k}_z) \,
 e^{\mathrm{i}\,\tilde k_\eta^-  x } 
\nn &\quad  + 
 \beta^-_{2,\mathbf n} \, \psi_2^+ (-\tilde k_\eta^-, k_y,\tilde{k}_z) \,
 e^{-\mathrm{i}\,\tilde k_\eta^- x } 
\Big]   \Theta\left( E-e\,U  \right)
 \nonumber\\ & 
 %%%
 + \Big[  
 \alpha^+_{2,\mathbf n} \,\psi_2^- ( \tilde k_\eta^+, k_y,\tilde{k}_z) \,
 e^{\mathrm{i}\,\tilde k_\eta^+  x } 
 + 
 \beta^+_{2,\mathbf n} \, \psi_2^- (-\tilde k_\eta^+, k_y,\tilde{k}_z) \,
 e^{-\mathrm{i}\,\tilde k_\eta^+ x }
 +
 \alpha^{-}_{2,\mathbf n} \,\psi_2^- ( \tilde k_\eta^-, k_y,\tilde{k}_z) \,
 e^{\mathrm{i}\,\tilde k_\eta^-  x } 
\nn &\qquad  + 
 \beta^-_{2,\mathbf n} \, \psi_2^- (-\tilde k_\eta^-, k_y,\tilde{k}_z) \,
 e^{-\mathrm{i}\,\tilde k_\eta^- x } 
\Big] \,\Theta\left( e\, U-E  \right),\nonumber \\
 %%%%%%%%%%%%%%%%%%%%%%%%%%%%%%%%%%%%%%%%%%%%%%%%%%%%%%%%%%%55
 \phi_R = & \,\frac{  t_{2,\mathbf n} \,\psi_{2}^+ ( k_\eta^{\text{re}},k_y,k_z) 
\, e^{\mathrm{i}\, k_\eta^{\text{re}} \left( x-L\right)} 
 } 
{\sqrt{  { \mathcal{V} }_x (k_\eta^{\text{re}},k_y,k_z)}}
+ t^{\prime}_{2,\mathbf n}\,\psi_{2}^+ ( k_\eta^{\text{im}},k_y,k_z) 
\, e^{\mathrm{i}\, k_\eta^{\text{im}} \left( x-L\right)}
\,,\nn
%%%%%%%%%%%%%%%%%%%%%%%%%%%%%%%%%
 k_\eta^{\text{re}} = & \sqrt{ \frac{k_0\,\sqrt{  E^2-  v_z^2\, k_z^2 } } {v_{\perp}}-k_y^2}\,, \quad
k_\eta^{\text{im}} =\mathrm{i}\,
\sqrt{ \frac{k_0\,\sqrt{  E^2-  v_z^2\, k_z^2 } } {v_{\perp}}+k_y^2}\,,\quad
%%%%%%
{ \mathcal{V} }_x  (k_\eta^{\text{re}},k_y,k_z) 
=\big    |\partial_{k_\eta^{\text{re}}} 
  \mathcal{E}_{2}^+ (k_\eta^{\text{re}},k_y,k_z)\big |\,,\quad
\nn
\tilde k_z = & \, k_z - e\, a_z\,,\quad 
\tilde k_\eta^\pm = \sqrt{ \pm\frac{k_0\,\sqrt{  \left(E-e\,U\right) ^2
- v_z^2\,\tilde k_z^2 } } {v_{\perp}}-k_y^2}\,,
\end{align}
where we have used the velocity $ { \mathcal{V} }_x  (k_\eta^{\text{re}},k_y,k_z) $ to normalize the incident, reflected, and transmitted plane waves.
Here $r_{2,{\mathbf n}}$ and $t_{2,{\mathbf n}}$ are the amplitudes of the reflected and transmitted waves, respectively, while $r^{\prime}_{2,{\mathbf n}}$ and $t^{\prime}_{2,{\mathbf n}}$ are the amplitudes of the exponentially decaying modes outside the barrier.
The latter two do not contribute to the reflection or transmission coefficients.
Altogether, we have $8$ unknown parameters $(r_{2,{\mathbf n}}, \,t_{2,{\mathbf n}}, \,
r^\prime_{2,{\mathbf n}}, \,t^\prime_{2,{\mathbf n}},\,
\alpha^\pm_{2,{\mathbf n}},\, \beta^\pm_{2,{\mathbf n}})$, and to solve for these, we need $8$ equations. These $8$ equations are obtained from the imposition of the continuity conditions on the two components of the wavefunction, and their derivatives with respect to $x$, at the boundaries $x=0$ and $x=L$ of the barrier. These conditions arise from the fact that the Hamiltonian in position space has a $\partial_x^2$ operator.

Acting the Hamiltonian operator twice on the spinor $\Psi_{ 2,\mathbf n}(x)$, we find that each of its two components (let us call it $\zeta_2$) satisfies a Schrodinger-like equation:
\begin{align}
\frac{ v_{\perp}^2 
\left( \partial_x^2+ k_y^2\right)^2}{k_0^2}\,\zeta_2
+ v_z^2\left(  k_z - e\, a_z\right)^2 \zeta_2= \left(E-e\,U\right) \zeta_2\,.
\end{align}
The solution is symmetric about $k_z =0$ only for a zero $a_z$. The system is of course symmetric about $k_y=0$ (i.e. $\phi=0$ axis in terms of the parametrization in Eq.~\eqref{eqsph}) irrespective the value of $a_z$.

%%%%%%%%%%%%%%%%%%%%%%%%%%%%%%%%%%%%%%%%%%%%%%%%%

\subsection{$J=3$}

The solutions $k_\eta=\pm
\sqrt{\left[ \frac{\left( E^2-  v_z^2\, k_z^2 \right) k_0^{4}} {v_{\perp}^2}\right]^{1/3}
-k_y^2} $ give propagating modes, while
$k_\eta^\zeta=\pm\sqrt{
e^{\zeta \frac{ 2\,\pi\,\mathrm{i}}{3}}\left[ \frac{\left( E^2-  v_z^2\, k_z^2 \right) k_0^{4}} {v_{\perp}^2}\right]^{1/3}-k_y^2} $ 
( where $\zeta = \pm 1 $)
give complex modes. Again, among the complex modes, we only consider those which have physically admissible exponentially decaying components, as the wavefunction will increase in an unbounded fashion as we approach $x =\pm \infty$ if we include the exponentially diverging components.
%%%%%%%%%t
Within the barrier region, both the exponentially increasing and decaying solutions are allowed, and hence we need to consider all the six values of $\tilde k_\eta$.

A scattering state $\Psi_{ 3,\mathbf n}(x)$, in the mode labeled by $\mathbf n =\lbrace n_y, n_z \rbrace$, is
constructed from the following states:
\begin{align}
& \Psi_{ 3, \mathbf n} (x)=  \begin{cases}   \phi_L & \text{ for } x<0  \\
 \phi_M & \text{ for } 0< x < L \\
  \phi_R &  \text{ for } x > L 
\end{cases} \,,\nonumber \\
%%%%%%%%%%%%%%%%%%%%%%%%%%%%%%%%%%%%%%%%
&  \phi_L =  \,\frac{   
 \psi_{3}^+ ( k_\eta^{\text{re}},k_y,k_z)\,e^{\mathrm{i}\, k_\eta^{\text{re}} \,x} 
 + r_{3,\mathbf n}\, \psi_{ 3}^+ ( -k_\eta^{\text{re}},k_y,k_z)\,e^{-\mathrm{i}\, k_\eta^{\text{re}} x}}
{\sqrt{  { \mathcal{V} }( k_\eta^{\text{re}}, k_y,k_z) }}
%%%
+ r^{+}_{ 3,\mathbf n}\, \psi_{3}^+ ( -k_\eta^+,k_y,k_z)\,
e^{-\mathrm{i}\,k_\eta^+ x}
+ r^{-}_{ 3,\mathbf n}\, \psi_{3}^+ ( k_\eta^-,k_y,k_z)\,
e^{ \mathrm{i}\,k_\eta^- x}
\, ,\nonumber \\
%%%%%%%%%%%%%%%%%%%%%%%%%%%%%%%%%%%%%%%%%%%%5
&  \phi_M  = \Big[
 \alpha^0_{ 3,\mathbf n} \,\psi_3^+ ( \tilde k_\eta^0, k_y,\tilde{k}_z) \,
 e^{\mathrm{i}\,\tilde k_\eta^0  x } 
+  \beta^0_{ 3,\mathbf n} \,\psi_3^+ ( -\tilde k_\eta^0, k_y,\tilde{k}_z) \,
 e^{ -\mathrm{i}\,\tilde k_\eta^0  x }  
+ \alpha^+_{ 3,\mathbf n} \,\psi_3^+ ( \tilde k_\eta^+, k_y,\tilde{k}_z) \,
 e^{\mathrm{i}\,\tilde k_\eta^+  x } 
 + 
 \beta^+_{ 3,\mathbf n} \, \psi_3^+ (-\tilde k_\eta^+, k_y,\tilde{k}_z) \,
 e^{-\mathrm{i}\,\tilde k_\eta^+ x }\nn
&\hspace{1.5 cm} +
 \alpha^{-}_{ 3,\mathbf n} \,\psi_3^+ ( \tilde k_\eta^-, k_y,\tilde{k}_z) \,
 e^{\mathrm{i}\,\tilde k_\eta^-  x } 
+ 
 \beta^-_{ 3,\mathbf n} \, \psi_3^+ (-\tilde k_\eta^-, k_y,\tilde{k}_z) \,
 e^{-\mathrm{i}\,\tilde k_\eta^- x } 
\Big]  \, \Theta\left( E-e\,U  \right)
 \nonumber\\ & 
 %%%
\hspace{0.75 cm} + \Big[
 \alpha^0_{ 3,\mathbf n} \,\psi_3^- ( \tilde k_\eta^0, k_y,\tilde{k}_z) \,
 e^{\mathrm{i}\,\tilde k_\eta^0  x } 
+  \beta^0_{ 3,\mathbf n} \,\psi_3^- ( -\tilde k_\eta^0, k_y,\tilde{k}_z) \,
 e^{ -\mathrm{i}\,\tilde k_\eta^0  x }  
+ \alpha^+_{ 3,\mathbf n} \,\psi_3^- ( \tilde k_\eta^+, k_y,\tilde{k}_z) \,
 e^{\mathrm{i}\,\tilde k_\eta^+  x } 
 + 
 \beta^+_{ 3,\mathbf n} \, \psi_3^- (-\tilde k_\eta^+, k_y,\tilde{k}_z) \,
 e^{-\mathrm{i}\,\tilde k_\eta^+ x }\nn
&\hspace{1.5 cm} +
 \alpha^{-}_{ 3,\mathbf n} \,\psi_3^- ( \tilde k_\eta^-, k_y,\tilde{k}_z) \,
 e^{\mathrm{i}\,\tilde k_\eta^-  x } 
+ 
 \beta^-_{ 3,\mathbf n} \, \psi_3^- (-\tilde k_\eta^-, k_y,\tilde{k}_z) \,
 e^{-\mathrm{i}\,\tilde k_\eta^- x } 
\Big] \,\Theta\left( e\, U-E  \right),\nonumber \\
 %%%%%%%%%%%%%%%%%%%%%%%%%%%%%%%%%%%%%%%%%%%%%%%%%%%%%%%%%%%55
 \phi_R = & \,\frac{  t_{3,\mathbf n} \,\psi_{2}^+ ( k_\eta^{\text{re}},k_y,k_z) 
\, e^{\mathrm{i}\, k_\eta^{\text{re}} \left( x-L\right)} 
 } 
{\sqrt{  { \mathcal{V} }_x (k_\eta^{\text{re}},k_y,k_z)}}
%%%
+ t^{+}_{ 3,\mathbf n}\, \psi_{3}^+ ( k_\eta^+,k_y,k_z)\,
e^{\mathrm{i}\,k_\eta^+ \left( x-L\right)}
+ t^{-}_{ 3,\mathbf n}\, \psi_{3}^+ ( -k_\eta^-,k_y,k_z)\,
e^{ -\mathrm{i}\,k_\eta^- \left( x-L\right)}
\,,\nn
%%%%%%%%%%%%%%%%%%%%%%%%%%%%%%%%%
& k_\eta^{\text{re}} =  \sqrt{\left[ \frac{\left( E^2-  v_z^2\, k_z^2 \right) k_0^{4}} {v_{\perp}^2}\right]^{1/3}
-k_y^2}\,, \quad
k_\eta^\pm= \sqrt{
e^{\pm \frac{ 2\,\pi\,\mathrm{i}}{3}}\left[ \frac{\left( E^2-  v_z^2\, k_z^2 \right) k_0^{4}} {v_{\perp}^2}\right]^{1/3}-k_y^2}\,,\nn
%%%%%%
& { \mathcal{V} }_x  (k_\eta^{\text{re}},k_y,k_z) 
=  \big    |\partial_{k_\eta^{\text{re}}} 
  \mathcal{E}_{3}^+ (k_\eta^{\text{re}},k_y,k_z)\big |\,,\quad
\tilde k_z =   k_z -e\, a_z \,,  
\nn
%%%
& \tilde k_\eta^0 = \sqrt{
\Bigg \vert \frac{\left \lbrace \left( E -e\,U\right )^2-  
\tilde k_z ^2 \right \rbrace k_0^{4}} {v_{\perp}^2}
\Bigg \vert^{1/3}-k_y^2}\,,\quad
%%%
\tilde k_\eta^\pm = \sqrt{
e^{\pm \frac{ 2\,\pi\,\mathrm{i}}{3}}
\Bigg \vert \frac{\left \lbrace \left( E -e\,U\right )^2-  
v_z^2 \,\tilde k_z ^2 \right \rbrace k_0^{4}} {v_{\perp}^2}
\Bigg \vert^{1/3}-k_y^2} \,,
\end{align}
where we have used the velocity $ { \mathcal{V} }_x  (k_\eta^{\text{re}},k_y,k_z) $ to normalize the incident, reflected, and transmitted plane waves.
In writing the wavefunction in the regions outside the barrier, we have used the fact that $\text{Im}\left[ k_\eta^+\right] >0 $ and $\text{Im}\left[ k_\eta^-\right] < 0 $.
There is only one reflection channel (with amplitude $r_{3,{\mathbf n}}$) and one transmission channel (with amplitude $t_{3,{\mathbf n}}$) as the parts containing the complex solutions $k_\eta^\pm$ (namely, $r^{\pm}_{3,{\mathbf n}}$ and $t^{\pm}_{3,{\mathbf n}}$) do not contribute to the probability current. 
Altogether, we have $12$ unknown parameters $(r_{3,{\mathbf n}}, \,t_{3,{\mathbf n}},\,
r^\pm_{3,{\mathbf n}}, \,t^\pm_{3,{\mathbf n}},\,
\alpha^0_{3,{\mathbf n}},\, \beta^0_{3,{\mathbf n}}, \,\alpha^\pm_{3,{\mathbf n}},\, \beta^\pm_{3,{\mathbf n}})$, and we need $12$ equations to determine these. These $12$ equations are obtained from the imposition of the continuity conditions on the two components of the wavefunction, and their first and second derivatives with respect to $x$, at the boundaries $x=0$ and $x=L$ of the barrier. These conditions arise from the fact that the Hamiltonian in position space has a $\partial_x^3$ operator.

Acting the Hamiltonian operator twice on the spinor $\Psi_{ 3,\mathbf n}(x)$, we find that each of its two components (let us call it $\zeta_3$) satisfies a Schrodinger-like equation:
\begin{align}
\frac{ v_{\perp}^2 
\left( \partial_x^2+ k_y^2\right)^3}{k_0^4}\,\zeta_3
+ v_z^2\left(  k_z - e\, a_z\right)^2 \zeta_3 = \left(E-e\,U\right) \zeta_3\,.
\end{align}
The solution is symmetric about $k_z =0$ only for a zero $a_z$. The system is of course symmetric about $k_y=0$ (i.e. $\phi=0$ axis in terms of the parametrization in Eq.~\eqref{eqsph}) irrespective the value of $a_z$.

%%%%%%%%%%%%%%%%%%%%%%%%%%%%%%%%%%%%%%%%%%%%%%%%%%%%%%%%%%

%%%%%%%%%%%%%%%%%%%%%%%%%%%%%%%%%%%%%
\subsection{Transmission coefficients}

%%%%%%%%%%%%%%%%%%%%%%%%%%%
\begin{figure}[]
\subfigure[]{\includegraphics[width=.47\textwidth]{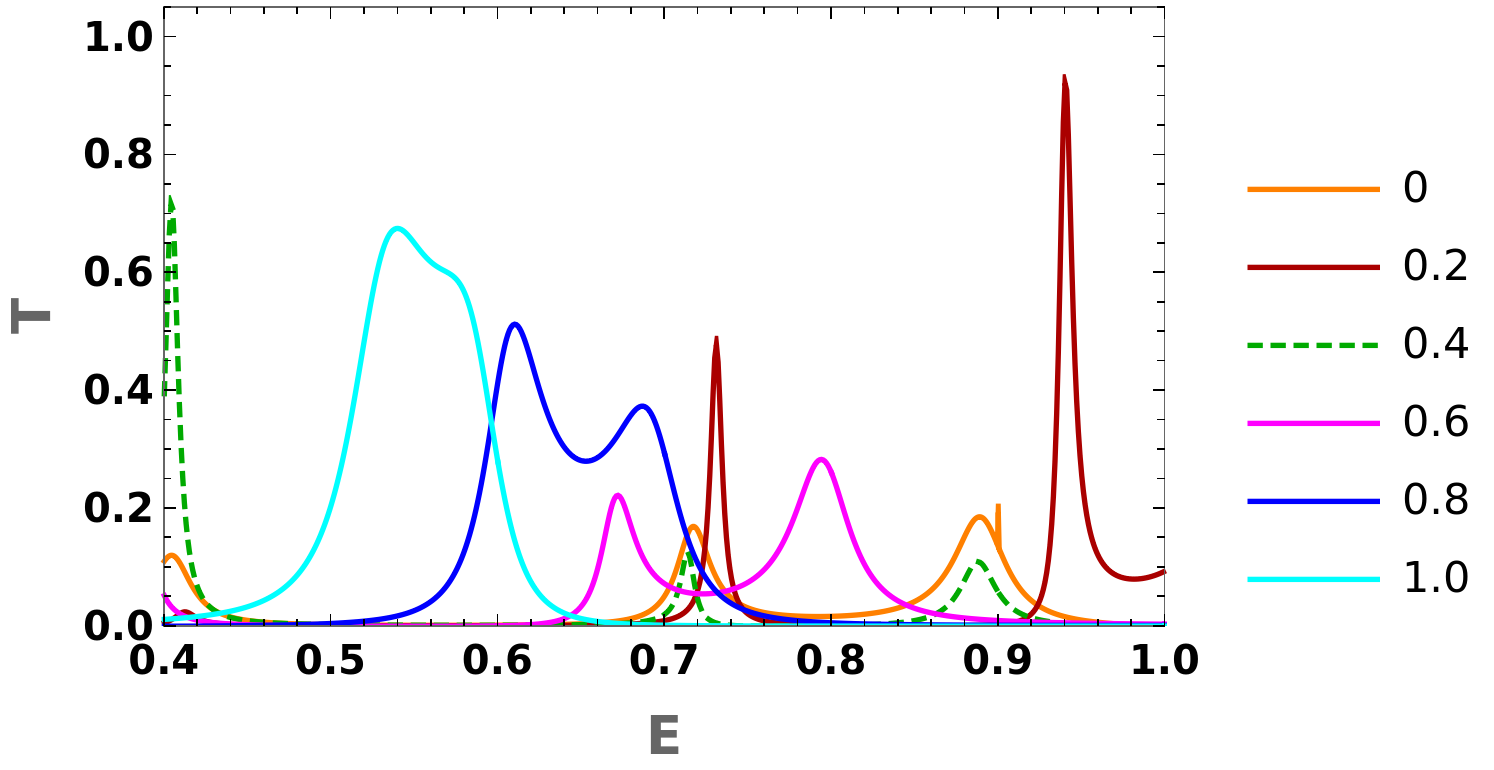}}\qquad
\subfigure[]{\includegraphics[width=.47\textwidth]{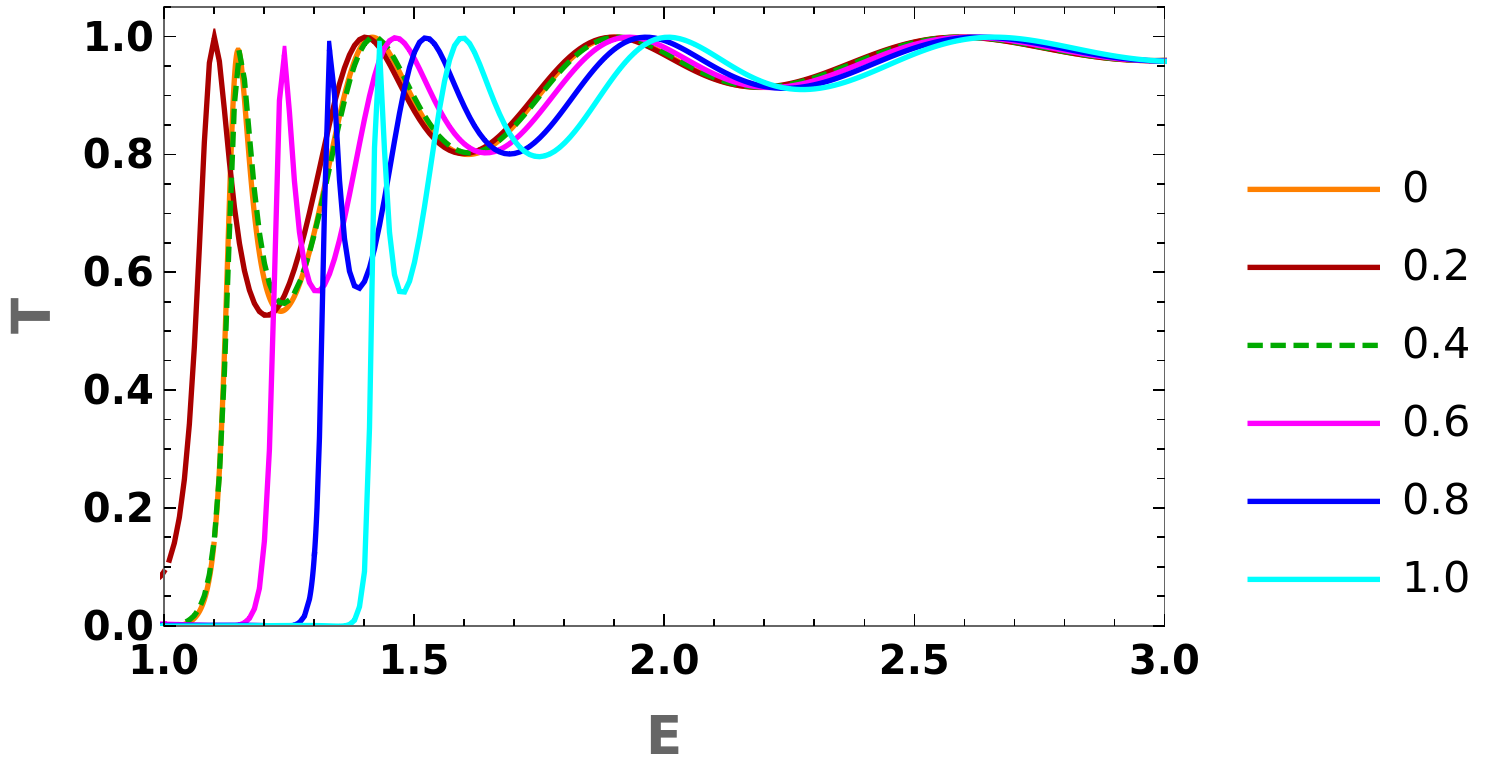}}\\
\subfigure[]{\includegraphics[width=.47\textwidth]{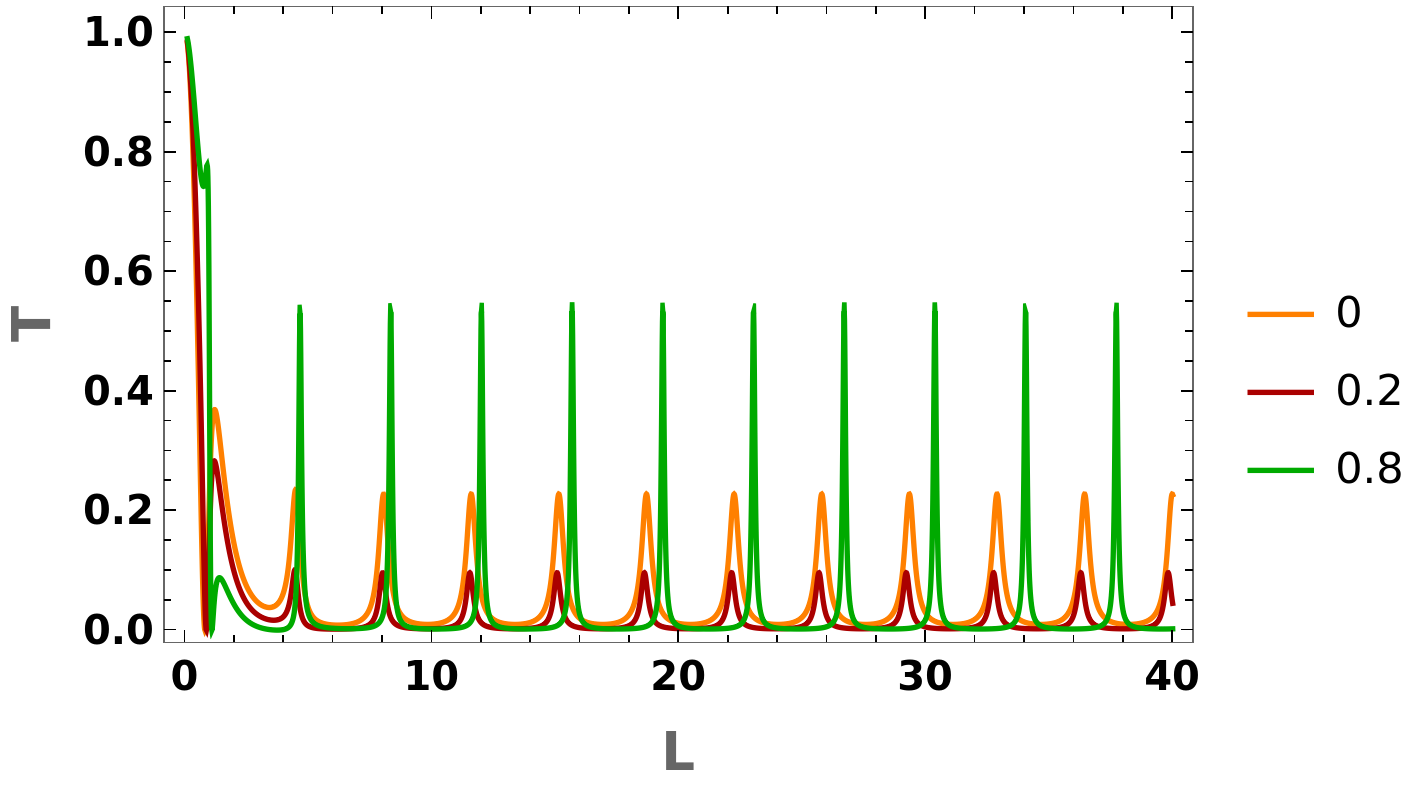}}\qquad
\subfigure[]{\includegraphics[width=.47\textwidth]{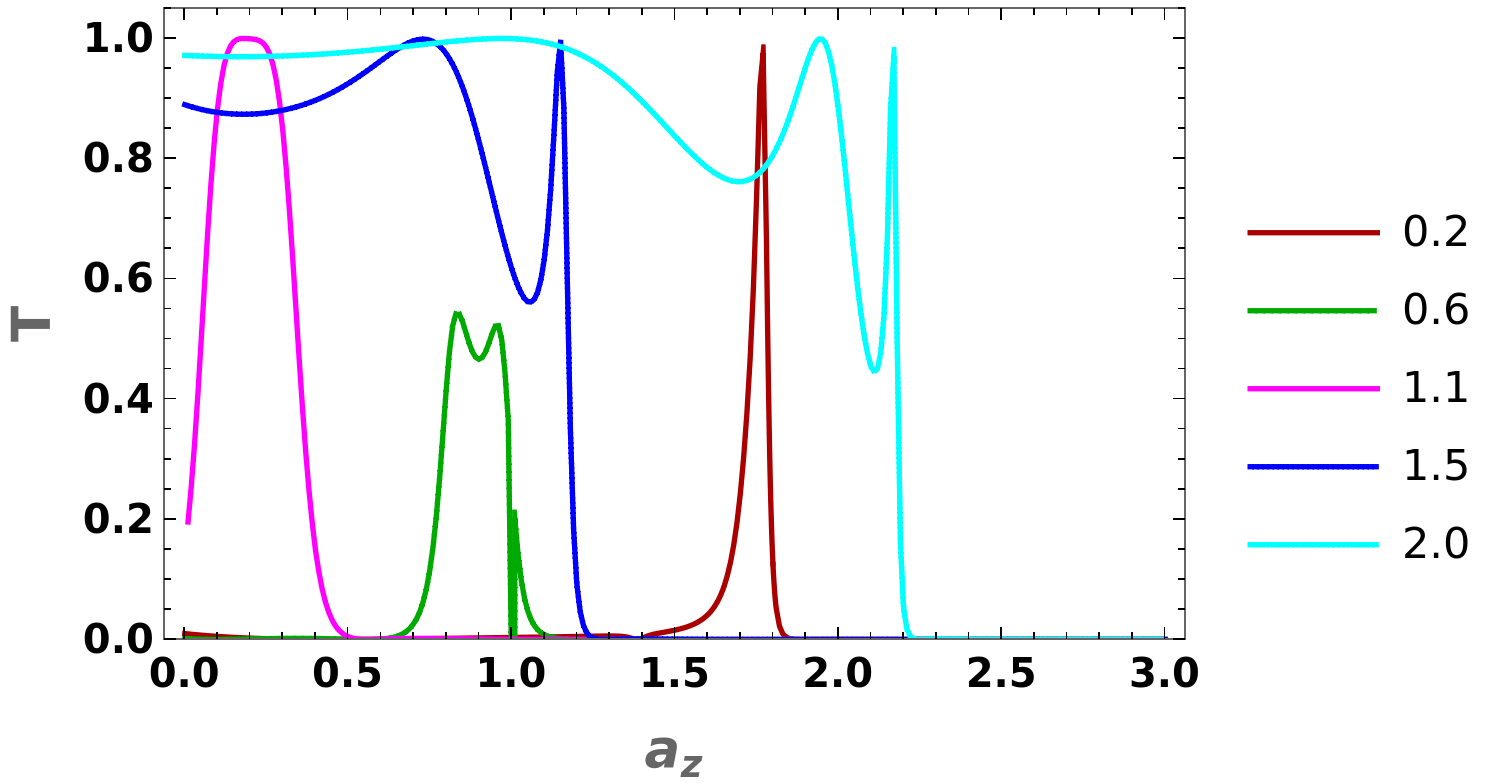}}
\subfigure[]{\includegraphics[width=.47\textwidth]{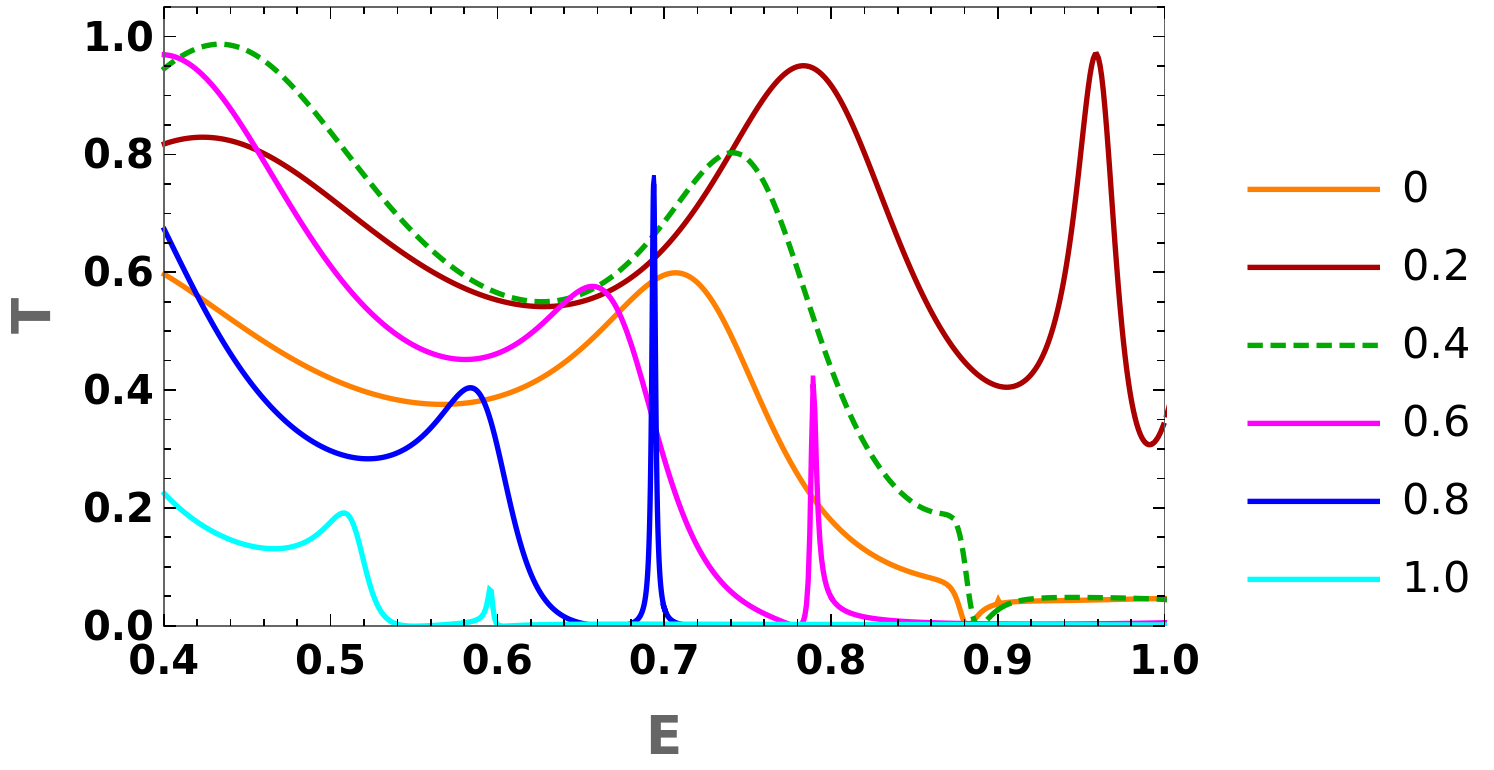}}\qquad
\subfigure[]{\includegraphics[width=.47\textwidth]{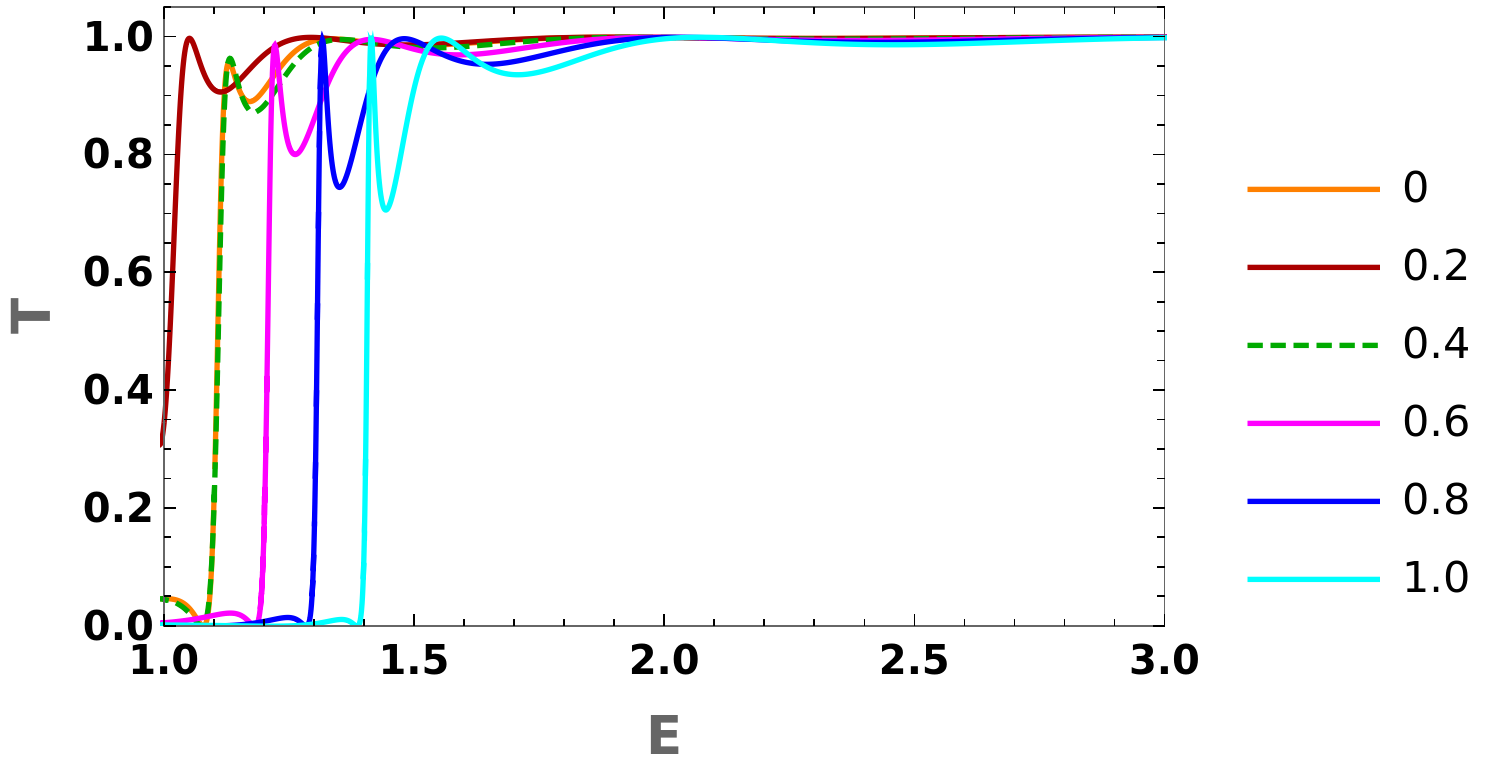}}\\
\subfigure[]{\includegraphics[width=.47\textwidth]{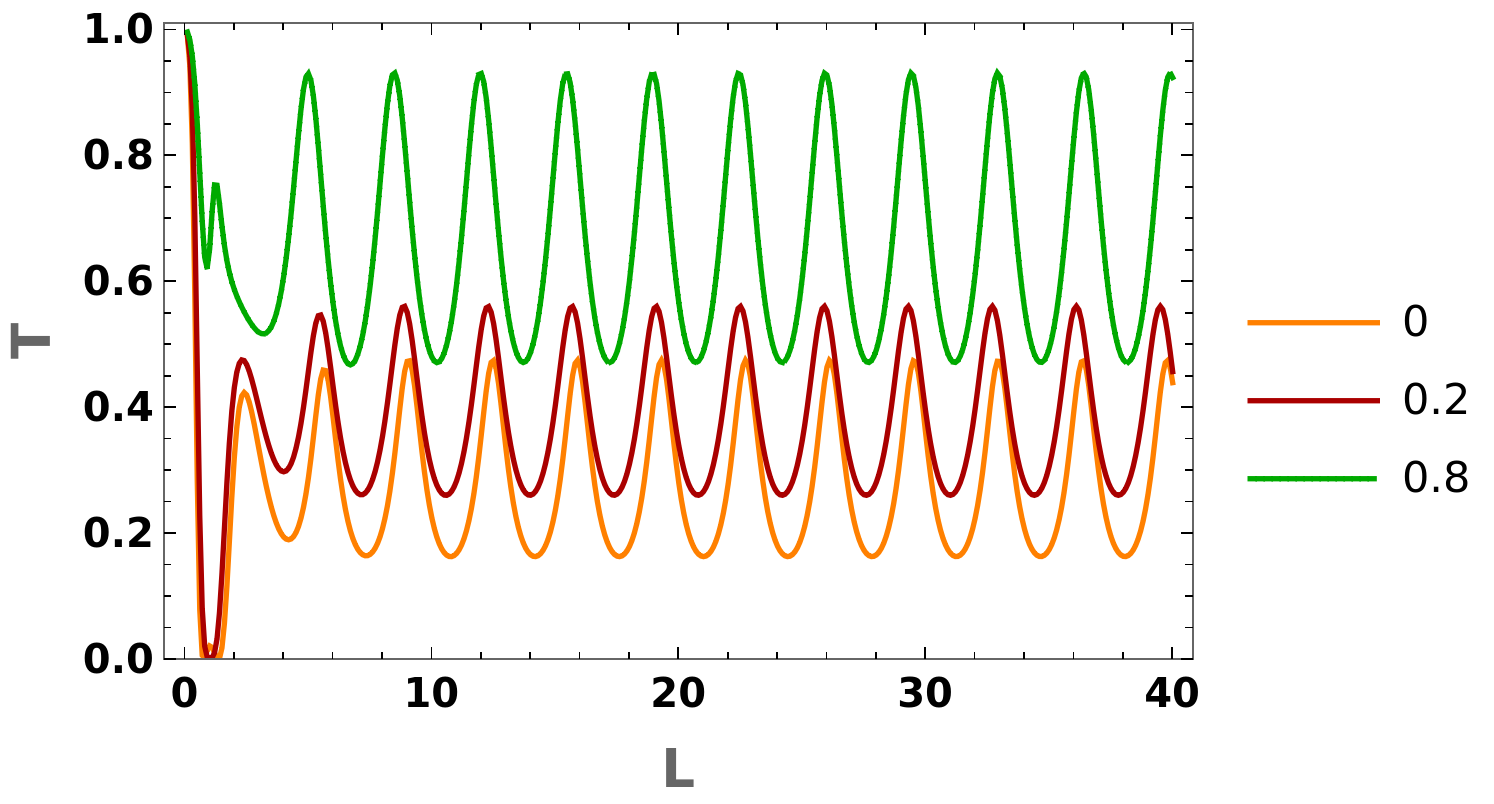}}\qquad
\subfigure[]{\includegraphics[width=.47\textwidth]{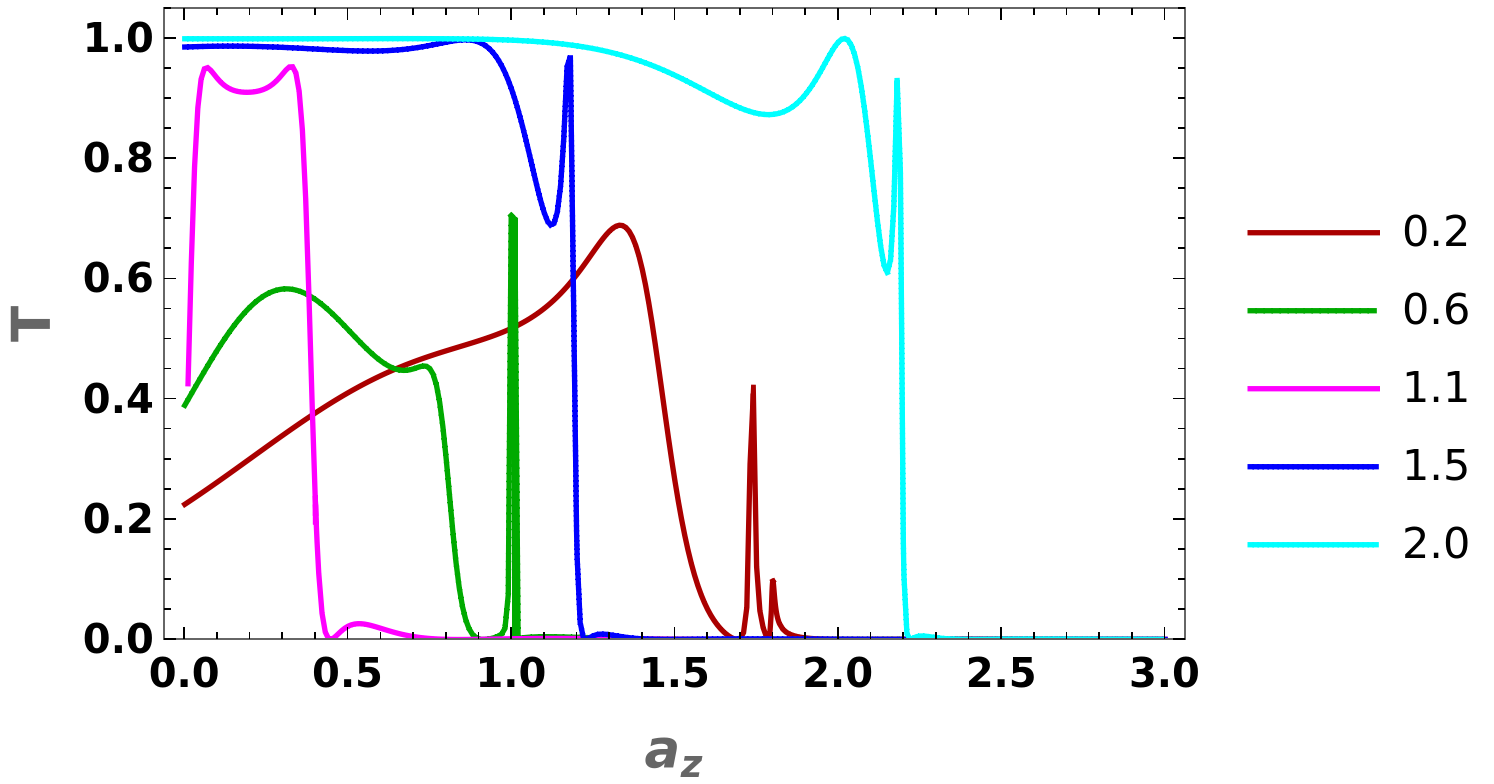}}
%\caption{$J=3$}
\caption{\label{TforKx}
Barrier perpendicular to $k_x$:
Panels (a) and (b) (for $J=2$), (e) and (f) ($J=3$) show the transmission coefficient $T$ as a function of the Fermi energy $E$, with $e\,U =1$, $L=10$, $k_y =0.1$, $k_z=0.2$, and the $a_z $-values indicated in the plot-legends.
Panels (c) and (g) show $T$ as a function of the barrier length $L$ for
$J=2$ and $3$, respectively, with $e\,U=1$, $E=0.2$, $k_y =0.1$, $k_z=0.2$, and the $a_z $-values indicated in the plot-legends.
Panels (d) and (h) show $T$ as a function of the barrier length $a_z$ for
$J=2$ and $3$, respectively, with $e\,U =1$, $L=10$, $k_y =0.1$, $k_z=0.2$, and the $E $-values indicated in the plot-legends.
}
\end{figure}
%%%%%%%%%%%%%%%%%%%%%%%%%%%

%%%%%%%%%%%%%%%%%%%%%%%%%%%
\begin{figure}[]
\subfigure[]{\includegraphics[width=.32\textwidth]{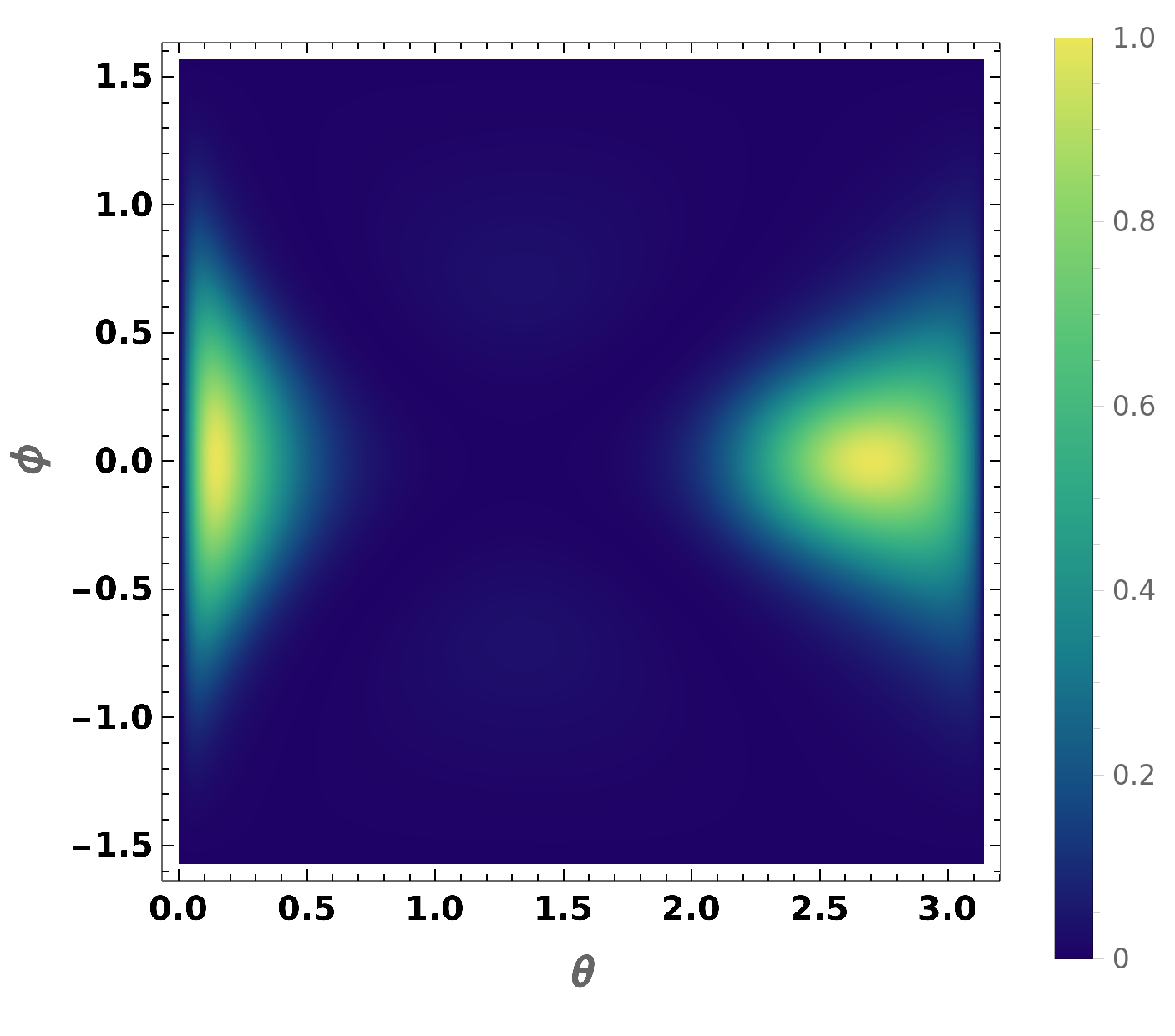}}\quad
\subfigure[]{\includegraphics[width=.32\textwidth]{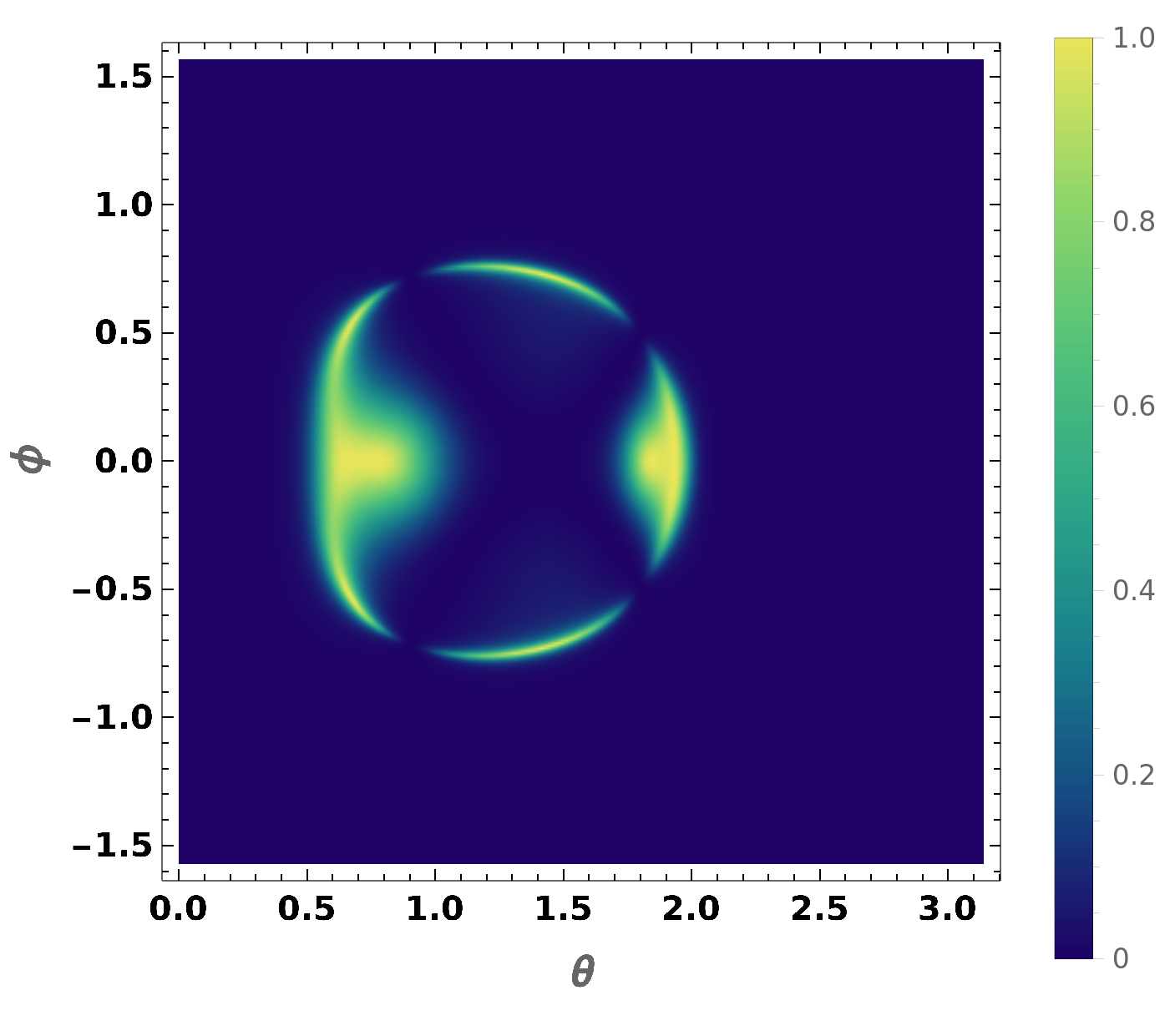}}\quad
\subfigure[]{\includegraphics[width=.32\textwidth]{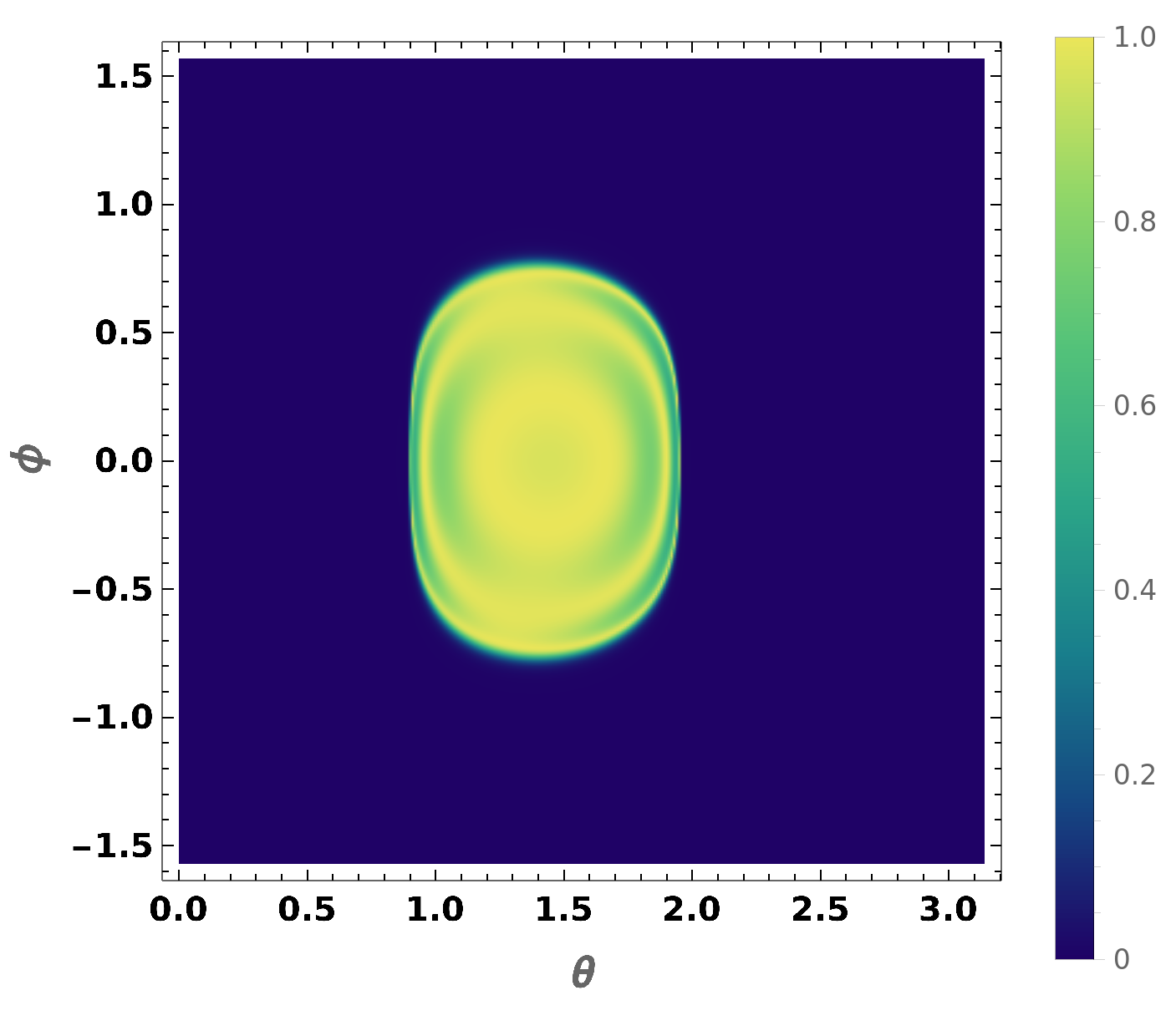}}\\
\subfigure[]{\includegraphics[width=.32\textwidth]{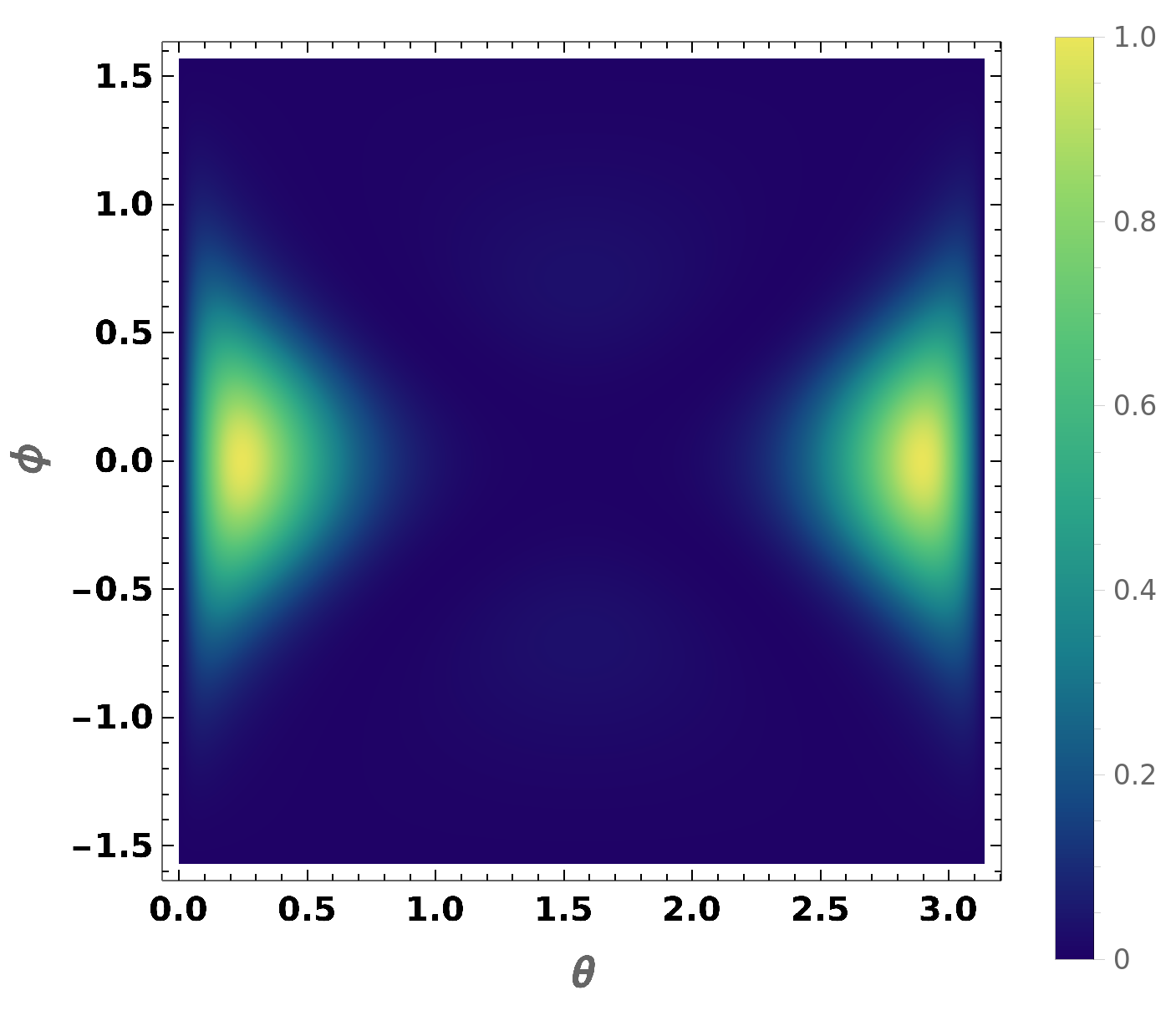}}\quad
\subfigure[]{\includegraphics[width=.32\textwidth]{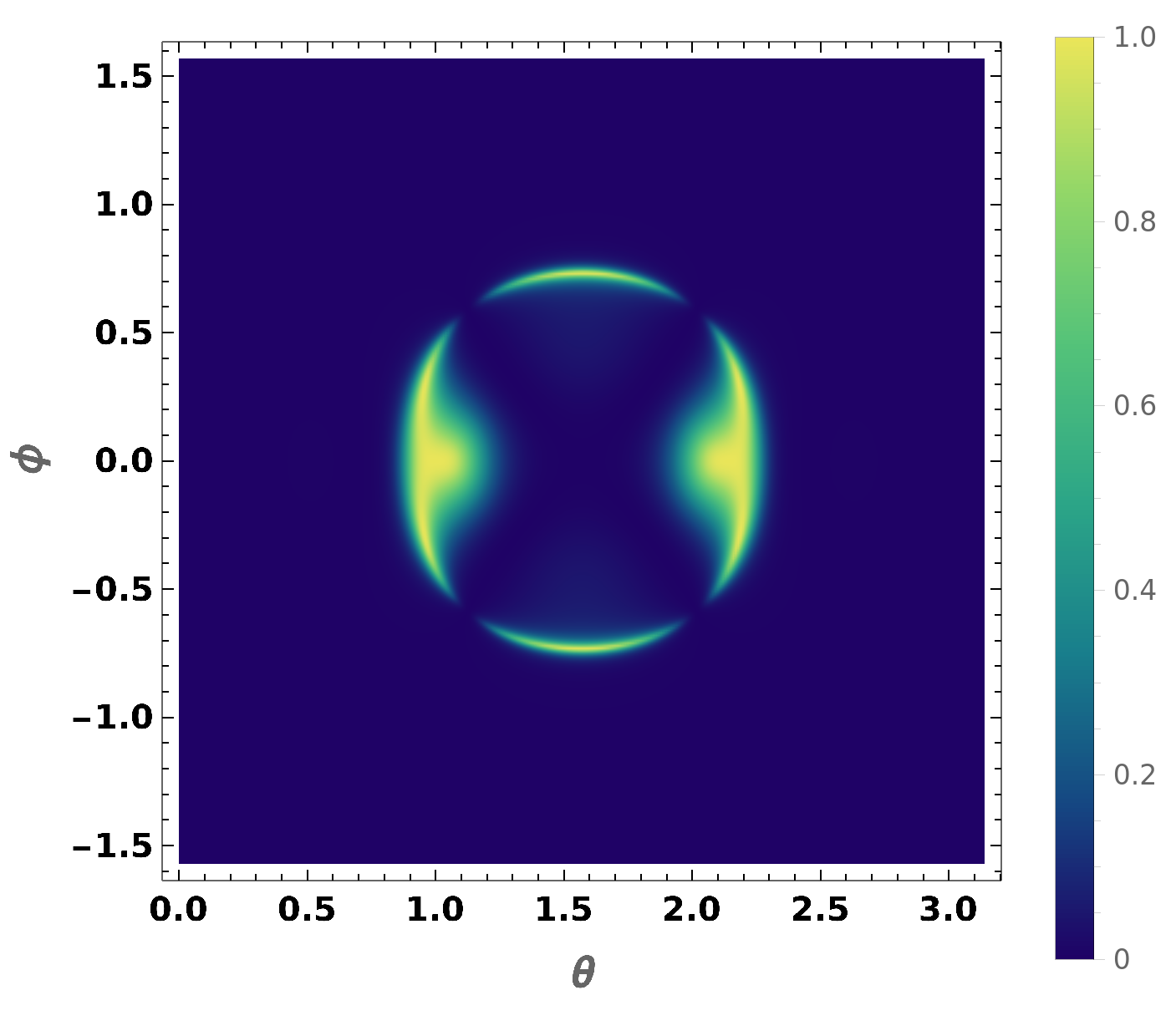}}\quad
\subfigure[]{\includegraphics[width=.32\textwidth]{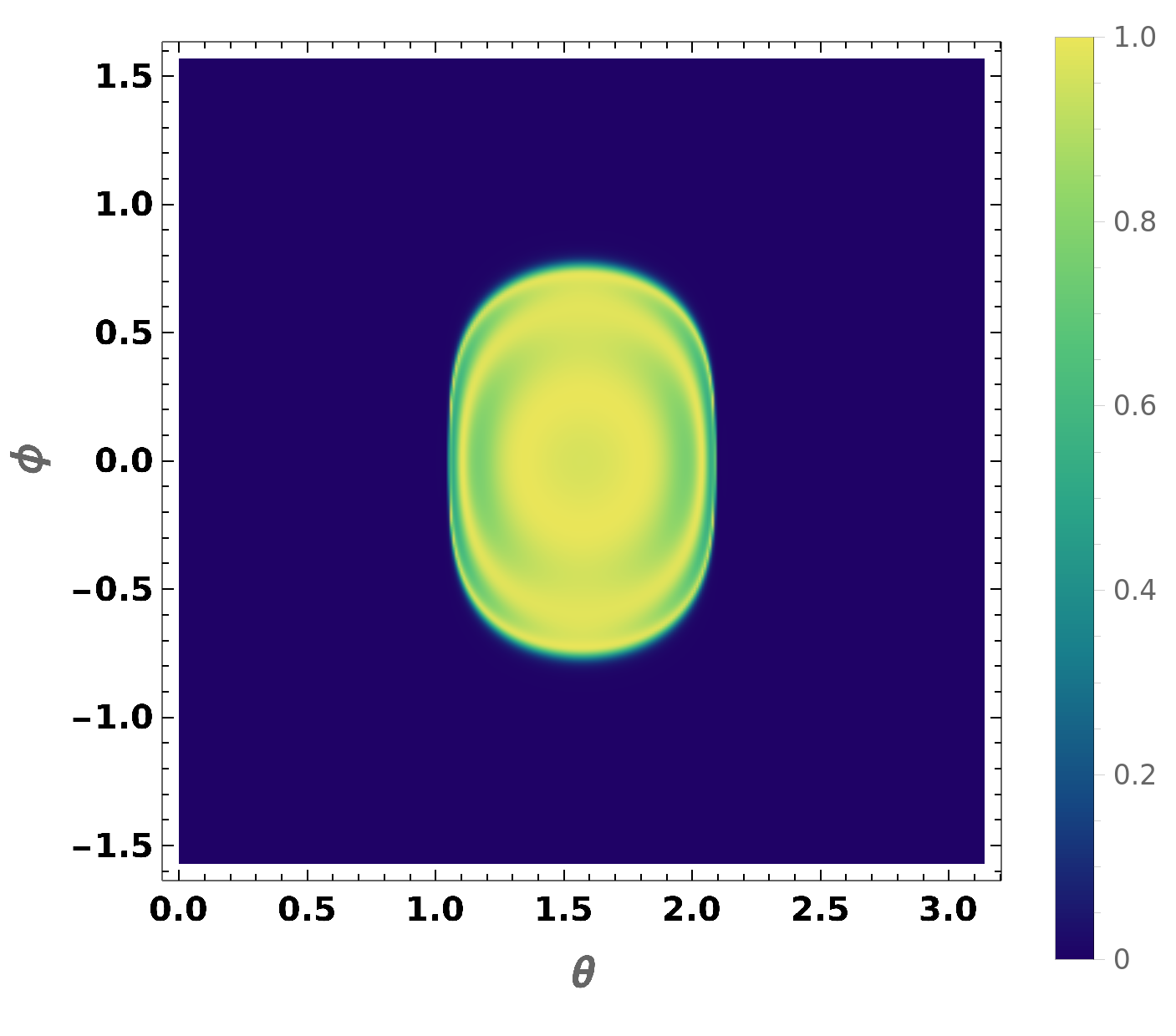}}\\
%\caption{$J=2$}
\subfigure[]{\includegraphics[width=.32\textwidth]{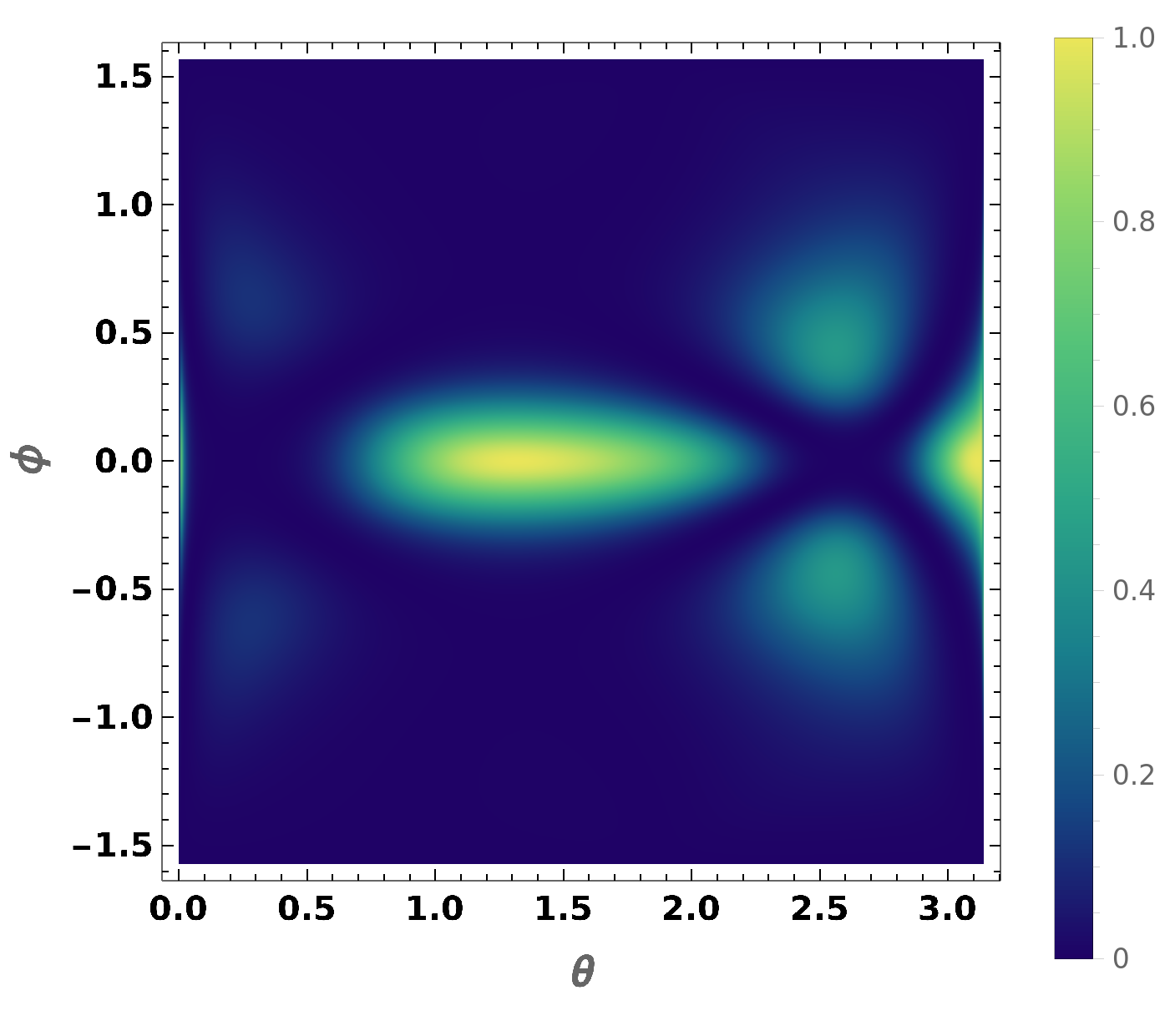}}\quad
\subfigure[]{\includegraphics[width=.32\textwidth]{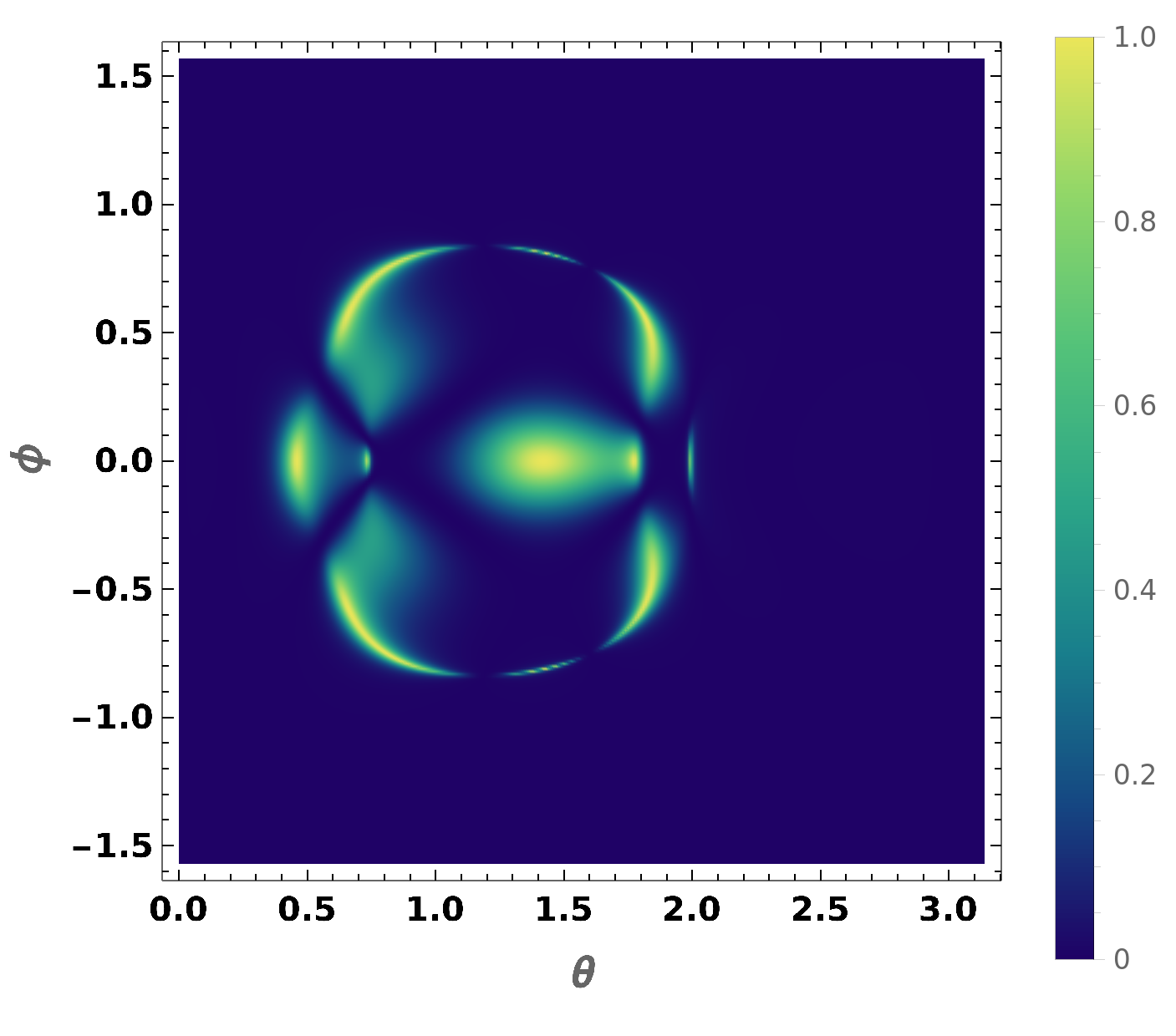}}\quad
\subfigure[]{\includegraphics[width=.32\textwidth]{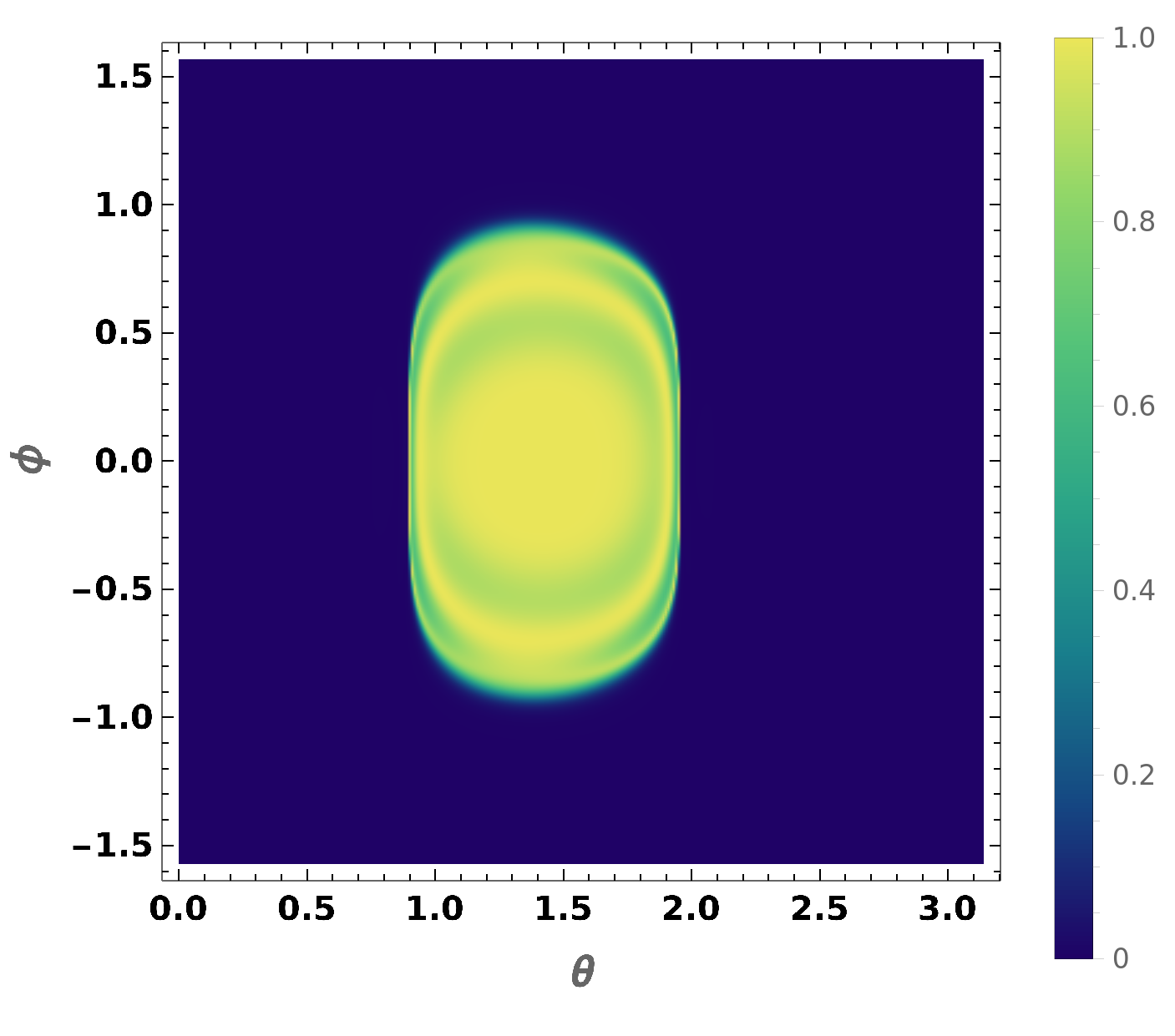}}\\
\subfigure[]{\includegraphics[width=.32\textwidth]{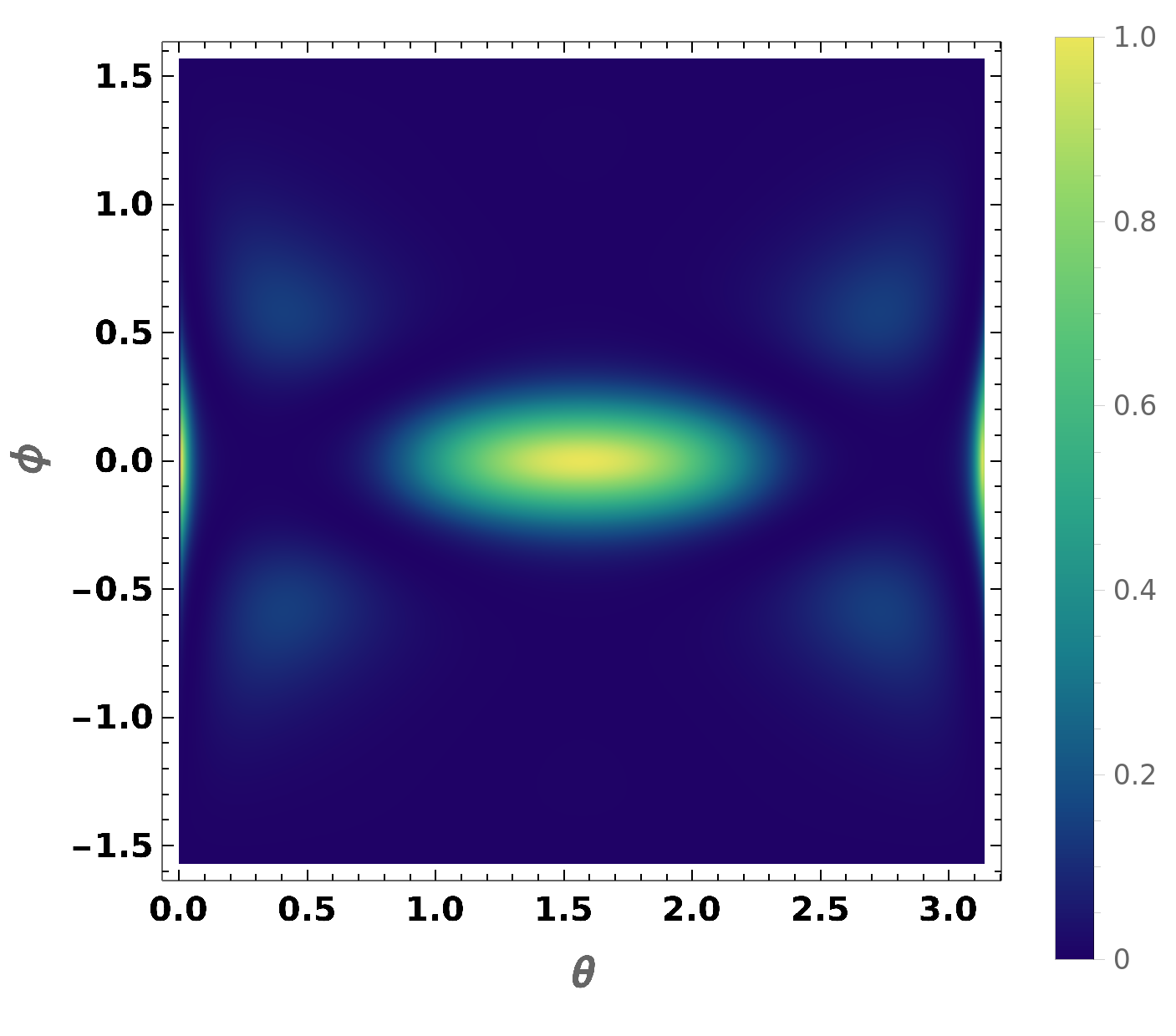}}\quad
\subfigure[]{\includegraphics[width=.32\textwidth]{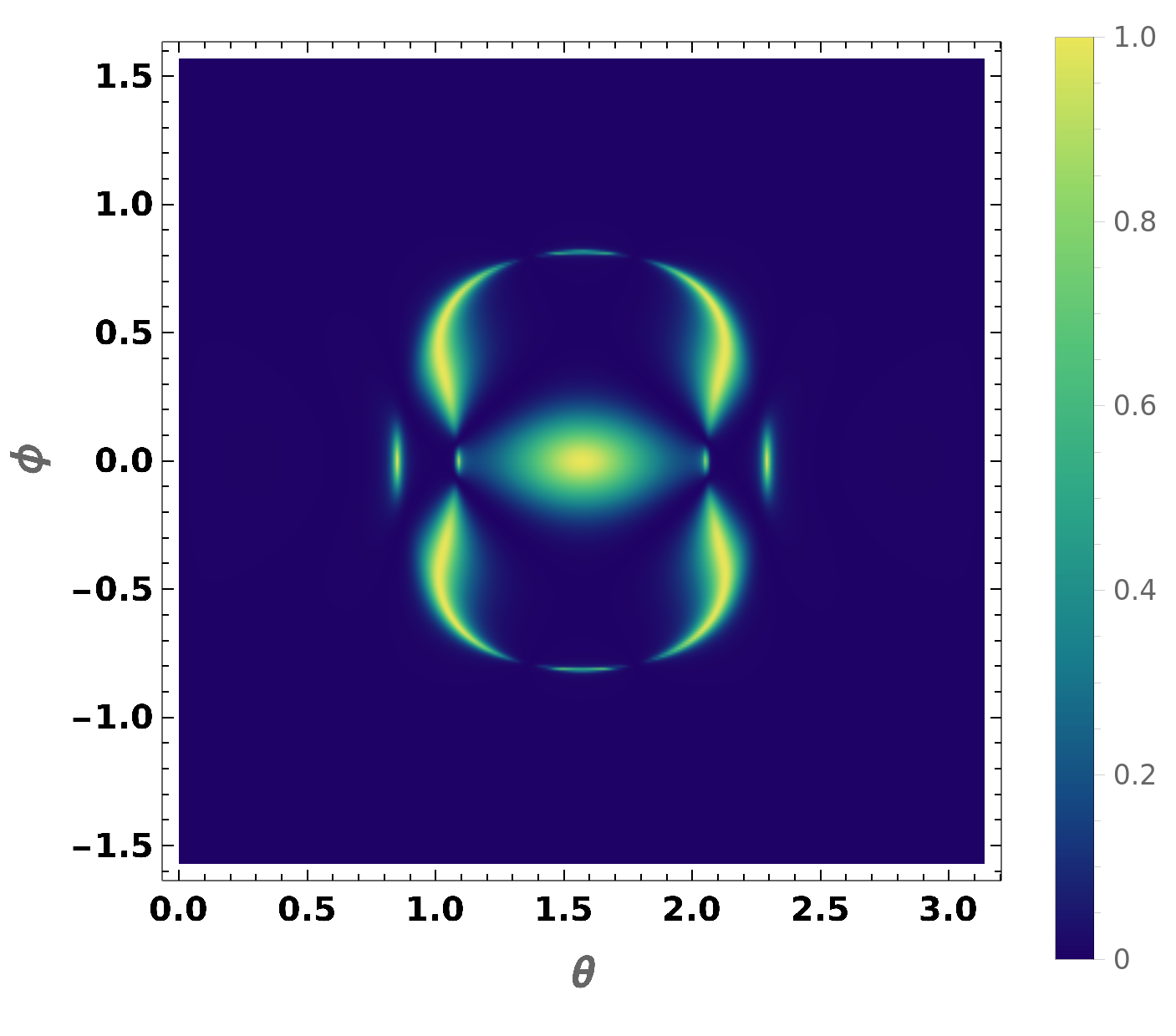}}\quad
\subfigure[]{\includegraphics[width=.32\textwidth]{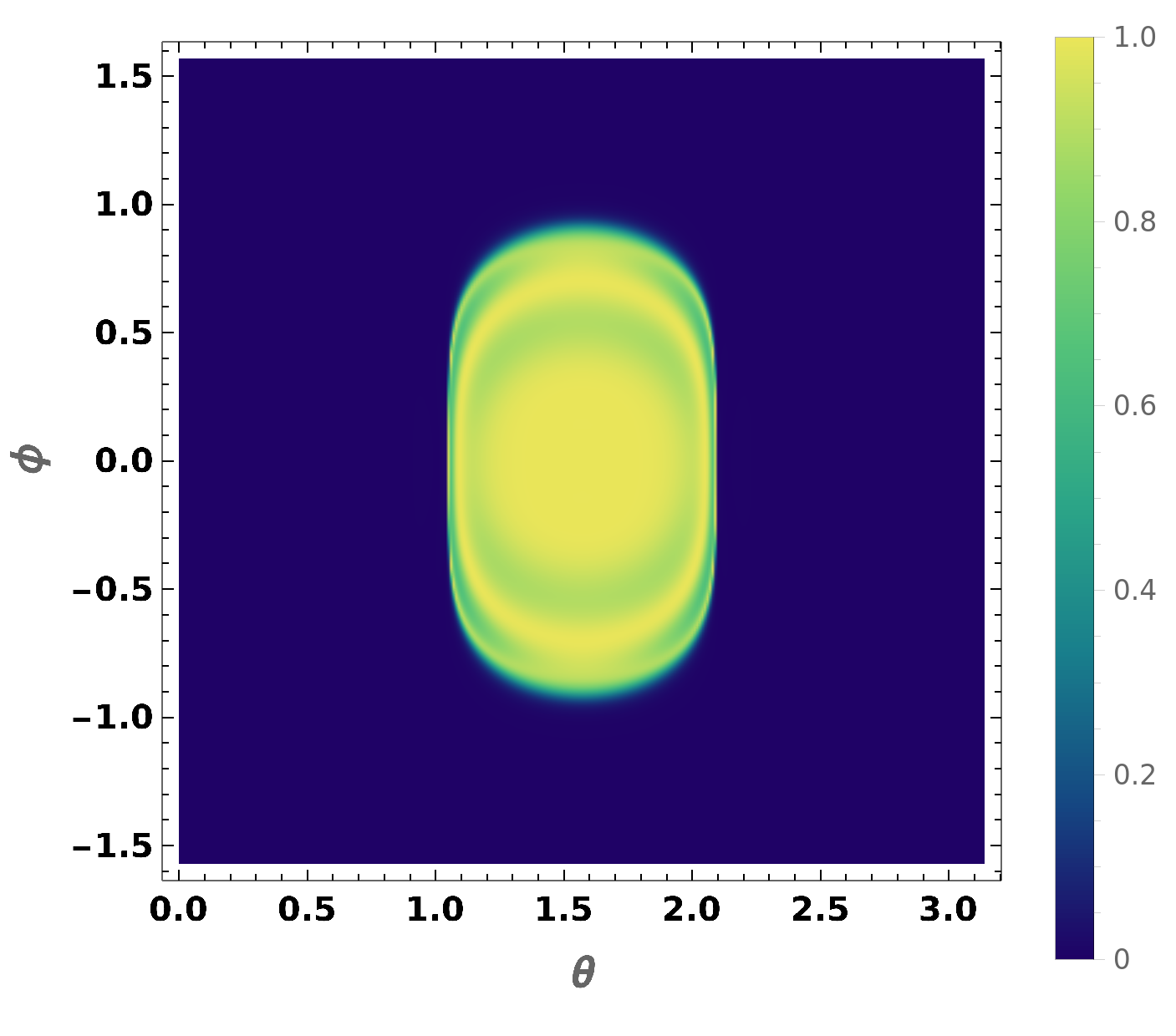}}\\
%\caption{$J=3$}
\caption{\label{contourforKx}
For a barrier perpendicular to $k_x$,
contour-plots of the transmission coefficient $T$ as a function of the orientation of the incident beam, parameterized by the angles $(\theta, \,\phi)$: Panels (a), (b), (c), (d), (e), and (f) show $T$ for $J= 2$, with $e\,U=1$, $L=10$, and $(E,\,a_z)=(0.2,\,0.5)$, $(0.6,\,0.3) $, $(2.0,\,0.5)$, $(0.2,\,0)$, $(0.6,\,0)$, and $(2.0,\,0)$, respectively.
Panels (g), (h), (i), (j), (k), and (l) show $T$ for
$J= 3$ for the same sequence of parameters as for $J=2$.
}
\end{figure}
%%%%%%%%%%%%%%%%%%%%%%%%%%%
 
 As before, the value of the transmission coefficient $T$ is obtained by taking the square of the absolute value of the corresponding transmission amplitude. 
For the contour-plots, we use the same parametrization as in Eq.~\eqref{eqsph}.
The features are heavily dependent on the $J$-value of the system, unlike the earlier case of a barrier perpendicular to the $k_z$-axis. This is due to the fact that the dispersion along the transmission direction now goes as $k_x^{2J}$.

Fig.~\ref{TforKx} shows some representative plots to capture the features of $T$ as functions of $E$ (both for $E < e\, U$ and $E >e\, U$), $L$, and $a_z$, respectively, when the other parameters are held fixed at some constant values.
Just like Fig.~\ref{TforKx}, $T$ shows oscillatory behavior, as a function of $E$ or $L$,
but unlike the previous case, does not generically reach the value of unity within the oscillatory cycles. This is true for both zero and nonzero values of $a_z$.

The contour-plots in Fig.~\ref{contourforKx} include the corresponding zero magnetic field cases for the sake of comparison. They clearly show that the effect of introducing a nonzero vector potential $a_z$ is to make $T$ asymmetric about the $\theta =\frac{\pi}{2}$ axis (which corresponds to $k_z=0$).

In zero magnetic field, we always get $T < 1$ for $J=2$ for $E<e\,U$ at normal incidence ($\theta =\frac{\pi}{2}$, $\phi =0$). This feature persists in the presence of the magnetic field. Also, $T = 1$ always at normal incidence for $J=1,\, 3$ for zero magnetic field. Nonzero magnetic field can shift the position of maximum $T$ value from normal incidence to different orientations (for example, in Fig.~\ref{contourforKx}(h), $T=0.786071 $ at normal incidence). The $E> e\, U$ features show that the specific incident angles that exhibit absence of reflection can be tuned by the vector potentials and a very small range of perfect transmission angles can be selected in the transverse plane. Moreover, by changing the difference between $E$ and $e\,U$, the radius of perfect transmission points can be adjusted as well.

%%%%%%%%%%%%%%%%%%%%%%%%%%%%%%%%%%%%%%%%%%%%%%%%
\subsection{Conductivity and Fano factors}

%%%%%%%%%%%%%%%%%%%%%%%%%%%
\begin{figure}[]
\subfigure[]{\includegraphics[width=.45\textwidth]{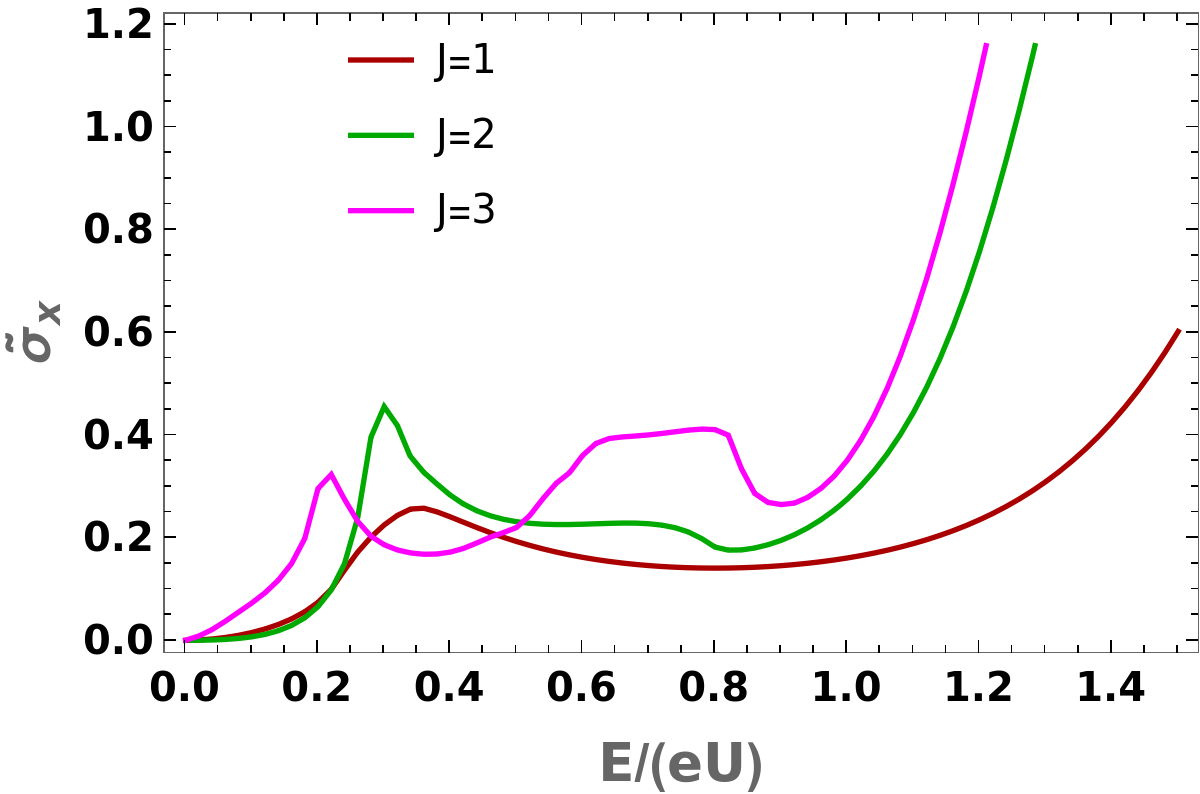}}\qquad
\subfigure[]{\includegraphics[width=.45\textwidth]{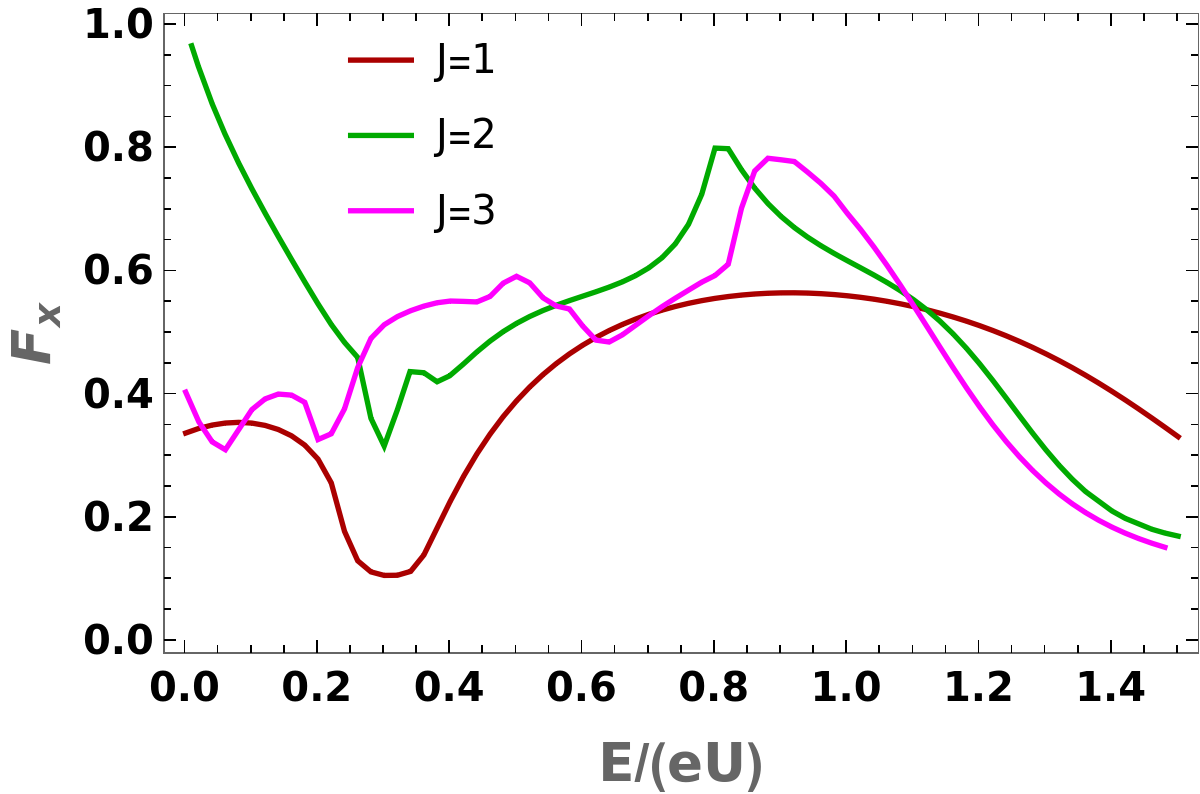}}
\caption{\label{condforKx}
Barrier perpendicular to $k_x$: Conductivity ($\tilde \sigma_x$ in units of $\frac{ L^2 } {4 \,\pi ^2} $) and Fano factors ($F_x $) as functions of the Fermi energy $E$, for
$e\,U =1$, $L=5$, and $ a_z=0.2$. 
}
\end{figure}
%%%%%%%%%%%%%%%%%%%%%%%%%%%

We assume $ W $ to be large enough such that $k_x$ and $k_y$ can effectively be treated as continuous variables, allowing us to perform the integrations over them to obtain the conductivity and Fano factor. 
Using $ k_x= 
\sqrt[J]{ \frac{k_0^{J-1}\,E\, \sin \theta}
{v_\perp}} \,\cos \phi\ \,,
\,\, n_y = \frac{W}{2\,\pi}
\sqrt[J]{ \frac{k_0^{J-1}\,E\, \sin \theta}
{v_\perp}}\, \sin \phi\,,
%%%
\,\,n_z = \frac{ W\,E\, \cos \theta}{ 2\,\pi\,v_z}\,,
\,\, dn_y \, dn_z =
\frac{ W^2\, E\,
\left \vert
\sin \theta  \cos \phi  \left(
\frac{ E \sin \theta \,
 k_0^{J-1}}{ v_{\perp}} \right)^{1/J}\right \vert \,d\theta \,d\phi}
{4 \,\pi ^2 \,v_z} $, 
in the zero-temperature limit and for a small applied voltage, the conductance is given by \cite{blanter-buttiker}:
\begin{align}
G_x(E,U,\mathbf B) & = \frac{ e^2}{h} \sum_{\mathbf n}   |t_ {J,\mathbf n}|^2 
\rightarrow \frac{  e^2} {h} \int   |t_ {J, \mathbf n}|^2 \,dn_y\,dn_z
= \frac{  e^2\,W^2 \, E} {4 \,\pi ^2\,h \,v_z } 
  \int_{ \theta=0 }^{\pi} \int_{\phi= -\pi/2}^{\pi/2}
T \, \left \vert
\sin \theta  \cos \phi  \left(
\frac{ E \sin \theta \,
 k_0^{J-1}}{ v_{\perp}} \right)^{1/J}\right \vert
  \,d\theta \,d\phi \,,
\end{align}
leading to the conductivity expression:
\begin{align}
\tilde \sigma_x (E,U,\mathbf B)  & = \left( \frac{L }{W} \right)^2 
\frac{ G_x(E,U,\mathbf B) } {e^2/h}
 =   \frac{ L^2\,E } {4 \,\pi ^2\, v_z} 
  \int_{ \theta=0 }^{\pi} \int_{\phi= -\pi/2}^{\pi/2}
T \, \left \vert
\sin \theta  \cos \phi  \left(
\frac{ E \sin \theta \,
 k_0^{J-1}}{ v_{\perp}} \right)^{1/J}\right \vert
  \,d\theta \,d\phi\,.
\end{align}
The Fano factor, quantitatively describing the shot noise, can be expressed as:
\begin{align}
F_x(E, U ,\mathbf B)  &=\frac 
{\int_{ \theta=0 }^{\pi} \int_{\phi= -\pi/2}^{\pi/2}
T \left(1-T\right)\, \left \vert
\sin \theta  \cos \phi  \left(
\frac{ E \sin \theta \,
 k_0^{J-1}}{ v_{\perp}} \right)^{1/J}\right \vert
  \,d\theta \,d\phi  } 
{ \int_{ \theta=0 }^{\pi} \int_{\phi= -\pi/2}^{\pi/2}
T \, \left \vert
\sin \theta  \cos \phi  \left(
\frac{ E \sin \theta \,
 k_0^{J-1}}{ v_{\perp}} \right)^{1/J}\right \vert
  \,d\theta \,d\phi  } \,.
\end{align}

The results are plotted in Fig.~\ref{condforKx}, as functions of the Fermi energy, for some representative parameter values. Again, the curves clearly show that local minima of conductivity no longer appear at $E =e\, U$ for nonzero magnetic fields, unlike the zero magnetic field cases \cite{Deng2020}. Unlike the case of the barrier perpendicular to the $k_z$-case, we do not see jumps in $ \tilde \sigma_x $ at $E =e\, U$. For $E < e\, U$, $ \tilde \sigma_x$ shows a non-monotonic behavior.
For $E > e\, U$, $ \tilde \sigma_x$ increases monotonically with $E$ for all $J$-values.

%%%%%%%%%%%%%%%%%%%%
\section{Summary and outlook}
\label{secsum}

In this paper, we have computed the transmission coefficients of the multi-Weyl semimetals with anisotropic dispersions and nonzero Chern numbers $J$. The $J=2$ and $J=3$ cases are the higher winding-number generalizations of the well-studied Weyl semimetals with $J=1$. The transmission coefficients have been calculated in the presence of both scalar and vector potentials, existing uniformly in a bounded region. The patterns found clearly serve as fingerprints of the corresponding semimetal, resulting from their distinct dispersion relations. Similar computations were done for the case of Weyl fermions in Ref.~\cite{mansoor}, and for the pseudospin-1 and pseudospin-3/2 semimetals with linear dispersion in Ref.~\cite{ips-tunnel-linear}. Comparing with those features, one can easily see that the characteristics for these higher-$J$ cases differ considerably
other kinds of semimetals. 
The conductivities and Fano factors obtained for some representative parameter values also serve as another set of fingerprints to identify the different types of semimetals. 
Most importantly, depending on whether the propagation direction is along the linear dispersion direction or nonlinear dispersion directions, the $J>1$ systems give us two independent sets of transport characteristics.
An important practical application of our theoretical calculations is that the results will help us find the perfect transmission regions by tuning the Fermi level and/or the magnetic fields, which can then potentially be used in generating localized transmission in the bulk of the semimetals (e.g. in electro-optic applications).

A future direction will study these transport properties in the presence of disorder, as was done in the case of Weyl \cite{emil2} and double-Weyl \cite{emil} nodes in the absence of any magnetic field. Computation of thermopower in the presence of a quantizing magnetic field is another avenue for future studies, as was done for the 2d double-Weyl case in Ref.~\cite{ips-kush}.
Lastly, if this exercise is carried out in the presence of interactions, it will show whether these
can destroy the quantization of various physical quantities in the topological phases \cite{PhysRevResearch.2.013069,kozii,Mandal_2020}, or whether new strongly correlated phases can emerge \cite{ips-seb,MoonXuKimBalents,rahul-sid,ipsita-rahul,ips-qbt-sc} where quasiparticle description of transport breaks down \cite{ips-subir,ipsc2,ips-mem-mat}.

%%%%%%%%%%%%%%%%%%
\section{Acknowledgements}
We thank Surajit Basak for participating in the initial stages of the project.

%%%%%%%%%%%%%%%%%
\bibliography{biblio}
\end{document}